\makeatletter
\def\input@path{{classes/}{sections/}{figs/}{figs/code}}
\makeatother

\documentclass[10pt,conference]{IEEEtran}
\usepackage{pbalance}
\usepackage{cite}
\usepackage{amsmath,amssymb,amsfonts,amsthm}
\usepackage{algorithmic}
\usepackage{graphicx}
\usepackage{textcomp}
\usepackage{xcolor}
\usepackage{makecell}
\usepackage{listings}
\usepackage{textcomp}
\usepackage{bm}
\usepackage[caption=false,font=footnotesize]{subfig}
\usepackage{kbordermatrix}
\usepackage{gensymb}
\usepackage{proof}
\usepackage{scalerel}

\usepackage{tikz}
\usetikzlibrary{calc}
\usetikzlibrary{quantikz2}
\usetikzlibrary{tikzmark}
\usetikzmarklibrary{listings}

\graphicspath{ {./figs/} }
\providecommand{\lngroveroraclestart}{3}
\providecommand{\lngroveroracleend}{5}
\providecommand{\lngroveroraclexpr}{5}
\providecommand{\lngroverbtrans}{10}

\providecommand{\lngroverqlit}{15}

\providecommand{\lngrovermeasure}{18}

\providecommand{\lngroversign}{9}
\providecommand{\lngroveriterdef}{8}
\providecommand{\lngroverpipeonestart}{9}
\providecommand{\lngroverpipeoneend}{10}
\providecommand{\lngroverpipetwostart}{15}
\providecommand{\lngroverpipetwoend}{18}
\providecommand{\lngroverqpudecorator}{7}
\providecommand{\lngroverhardcodestart}{7}
\providecommand{\lngroverhardcodeend}{18}
\providecommand{\lngroveroracledef}{4}
\providecommand{\lngroveriterpipestart}{15}
\providecommand{\lngroveriterpipeend}{17}

\providecommand{\lnsuperdensebellstart}{15}
\providecommand{\lnsuperdensebellend}{18}
\providecommand{\lnsuperdensebellpair}{6}

\providecommand{\lnsuperdenseflip}{9}
\providecommand{\lnsuperdensez}{11}

\providecommand{\lnpreludezero}{2}
\providecommand{\lnpreludeone}{3}
\providecommand{\lnpreludesymstart}{4}
\providecommand{\lnpreludesymend}{7}
\providecommand{\lnpreludebasisstart}{10}
\providecommand{\lnpreludebasisend}{16}
\providecommand{\lnpreludei}{5}
\providecommand{\lnpreludej}{7}

\providecommand{\lnpreludetwomeasure}{15}

\providecommand{\lnpreludetwofourierrecstart}{7}

\providecommand{\lnpreludetwofourierinductstart}{8}
\providecommand{\lnpreludetwofourierrecend}{9}

\providecommand{\lnpreludetwofourierinductend}{9}
\providecommand{\lnpreludetwomeasuremacro}{3}
\providecommand{\lnpreludetwoflipmacro}{2}
\providecommand{\lnpreludetwoflipvar}{14}
\providecommand{\lnpreludetwobasecase}{7}

\providecommand{\lngrovermetaforstart}{10}
\providecommand{\lngrovermetaforend}{11}

\providecommand{\lnqpescanstart}{8}
\providecommand{\lnqpescanend}{12}
\providecommand{\lnqpemeas}{13}
\providecommand{\lnqpeloop}{12}
\providecommand{\lnqpeinstant}{9}

\providecommand{\lnqpekernelcall}{16}

\providecommand{\lnqpeuserangle}{4}
\providecommand{\lnqpeuserfinalcalc}{17}

\providecommand{\lnqpeuserangleexp}{14}
\providecommand{\lnqpeuserJ}{11}
\providecommand{\lnqpeuserrevdec}{12}

\providecommand{\lnteleportcondone}{10}
\providecommand{\lnteleportcondtwostart}{11}

\providecommand{\lnteleportcondtwoend}{12}

\providecommand{\lnbvwrapperfunc}{3}
\providecommand{\lnbvfdef}{5}
\providecommand{\lnbvkerneldef}{9}
\providecommand{\lnbvand}{6}
\providecommand{\lnbvpipestart}{10}
\providecommand{\lnbvpipeend}{12}
\providecommand{\lnbvsecretdef}{16}

\providecommand{\lnperiodinit}{8}
\providecommand{\lnperiodpipestart}{8}
\providecommand{\lnperiodpipeend}{11}
\providecommand{\lnperiodblackboxdef}{23}
\providecommand{\lnperiodxor}{9}
\providecommand{\lnperioddiscard}{10}
\providecommand{\lnperiodmeas}{11}
\providecommand{\lnperiodpoststart}{13}
\providecommand{\lnperiodpostend}{20}
\providecommand{\lnperiodlcm}{19}

\providecommand{\lnshorinplace}{19}

\providecommand{\lnshorprecstart}{7}
\providecommand{\lnshorprecend}{8}
\providecommand{\lnshorinitone}{12}
\providecommand{\lnshorreversible}{15}
\providecommand{\lnshormult}{16}

\providecommand{\lnshorcfrac}{24}
\providecommand{\lnshorconv}{25}
\providecommand{\lnshorpoststart}{23}
\providecommand{\lnshorpostend}{30}

\newenvironment{codeannotations}[1]{\providecommand\thecode{}\renewcommand\thecode{#1}\input{\thecode-lines}\begin{tikzpicture}[remember picture,overlay,shorten >= -3pt,shorten <= -3pt,thick,color=blue!50!gray]}{\end{tikzpicture}}

\newcommand{\codeannotation}[7][top right]{
    \node (\thecode#2 top left) at ($(pic cs:line-\thecode-\number\numexpr\csname ln\thecode#2start\endcsname-1\relax-start)+#3$) {};
    \node (\thecode#2 bottom right) at ($(pic cs:line-\thecode-\number\numexpr\csname ln\thecode#2end\endcsname+#5\relax-start)+#4$) {};
    \path let \p{tl}=(\thecode#2 top left), \p{br}=(\thecode#2 bottom right) in node (\thecode#2 bottom left) at (\x{tl},\y{br}) {} node (\thecode#2 top right) at (\x{br},\y{tl}) {};
    \node (\thecode#2 top middle) at ($(\thecode#2 top left)!0.5!(\thecode#2 top right)$) {};
    \node (\thecode#2 top middle right) at ($(\thecode#2 top left)!0.75!(\thecode#2 top right)$) {};
    \draw[] (\thecode#2 top left) rectangle (\thecode#2 bottom right);
    \node[anchor=west,align=left] (\thecode#2Note) at ($(\thecode#2 #1)+#6$) {\small #7};
    \draw[->] (\thecode#2Note) -- (\thecode#2 #1);
}

\newcommand{\gigaorange}{orange!70!black}
\lstdefinestyle{base}{language=Python,emph={@qpu,@classical},morekeywords={func,cfunc,qfunc,rev_func,rev_qfunc,qubit,bit,basis,std,pm,id,fourier,exp,discard,discardz,bell,sym,revolve,reverse,xor_reduce,and_reduce,sign,inplace,xor,measure,flip,rotate,prep,ij,@reversible,cfrac},basicstyle=\ttfamily,emphstyle=\bfseries\color{\gigaorange},keepspaces=true,stringstyle=\color{blue},commentstyle=\color{gray!60!black},showstringspaces=false,upquote=true,literate={~}{{\raise.17ex\hbox{$\scriptstyle\sim$}}}{1} {^}{{\raise.60ex\hbox{$\scriptstyle\wedge$}}}{1},aboveskip=0pt,belowskip=0pt}
\lstdefinestyle{s}{style=base,basicstyle=\ttfamily\small}
\lstdefinestyle{xs}{style=base,basicstyle=\ttfamily\footnotesize}
\lstdefinestyle{num}{style=base,frame=single,rulecolor=\color{gray},numbers=left,numberstyle=\sffamily\footnotesize\color{gray},numbersep=4.5pt}
\lstdefinestyle{nums}{style=num,basicstyle=\ttfamily\small}
\lstdefinestyle{numxs}{style=num,basicstyle=\ttfamily\footnotesize}
\lstdefinestyle{numxxs}{style=num,basicstyle=\ttfamily\scriptsize}
\lstdefinestyle{numxxxs}{style=num,basicstyle=\ttfamily\tiny}
\DeclareRobustCommand\circleit[1]{\tikz[baseline=(num.base)]{\node[thick,black,shape=circle,draw,inner sep=0.2ex] (num) {\textbf{#1}};}}

\newcommand{\varbv}{\mathrm{bv}}

\newcommand{\formname}{Mini-Qwerty}

\newcommand{\babyket}[1]{\ensuremath{|#1\rangle}}
\newcommand{\babybra}[1]{\ensuremath{\langle#1|}}

\newcommand{\desc}[1]{\textit{#1}}
\newcommand{\alt}{\,\mid\,}
\newcommand{\nvar}{n}
\newcommand{\mvar}{m}
\newcommand{\tiltvar}{\theta}
\newcommand{\typevar}{\tau}
\newcommand{\regvar}{r}
\newcommand{\exprvar}{e}
\newcommand{\basisvar}{b}
\newcommand{\termvar}{t}
\newcommand{\varvar}{x}
\newcommand{\qlitvar}{q\ell}
\newcommand{\qavar}{qa}
\newcommand{\indexvar}{i}
\newcommand{\bvvar}{bv}
\newcommand{\vavar}{va}
\newcommand{\qrefvar}[1]{q_{#1}}
\newcommand{\valvar}{v}
\newcommand{\propervalvar}{w}
\newcommand{\regelemvar}{re}
\newcommand{\qstatevarraw}{\psi}
\newcommand{\qstatevar}{\babyket{\qstatevarraw}\!}
\newcommand{\qstatevaralt}{\babyket{\qstatevarraw{'}}\!}

\newcommand{\probvar}{p}
\newcommand{\funcvalvar}{fv}
\newcommand{\revfuncvalvar}{rv}
\newcommand{\funckindvar}{\gamma}
\newcommand{\irrevtok}{\textit{irrev}}
\newcommand{\revtok}{\textit{rev}}
\newcommand{\pipetok}{\parallel}
\newcommand{\unittok}{\!\texttt{\textbf{[]}}\!}
\newcommand{\zerotok}{0}
\newcommand{\onetok}{1}
\newcommand{\rawunboldbitensortok}{\texttt{*}}
\newcommand{\rawbitensortok}{\textbf{\rawunboldbitensortok}}
\newcommand{\bitensortok}[2]{#1\,\rawbitensortok\,#2}
\newcommand{\measuretok}[1]{#1\texttt{.\textbf{measure}}}
\newcommand{\discardtok}{\texttt{\textbf{discard}}}
\newcommand{\rawadjtok}{{\sim}}
\newcommand{\adjtok}[1]{\rawadjtok{}#1}
\newcommand{\rawtranstok}{\texttt{>{}>}}
\newcommand{\transtok}[2]{#1 \,\rawtranstok{}\, #2}
\newcommand{\ifelsetok}[3]{#1 \ \texttt{\textbf{if}}\  #2 \ \texttt{\textbf{else}}\  #3}
\newcommand{\lambdatok}[3]{\lambda #1\texttt{\textbf{:}}#2 \texttt{\textbf{.}} #3}
\newcommand{\revlambdatok}[3]{\lambda^{\textit{rev}} #1\texttt{\textbf{:}}#2 \texttt{\textbf{.}} #3}
\newcommand{\qlittok}[1]{\text{\textcolor{blue}{\texttt{\textquotesingle #1\textquotesingle}}}}
\newcommand{\superpostok}[2]{#1 \,\texttt{\textbf{+}}\, #2}
\newcommand{\tilttok}[2]{#1 \ \texttt{@}\  #2}
\newcommand{\basislittok}[1]{\texttt{\textbf{\{}}\bvvar_1,\bvvar_2,\ldots,\bvvar_{#1}\texttt{\textbf{\}}}}

\newcommand{\basislittokvecs}[1]{\texttt{\textbf{\{}}#1\texttt{\textbf{\}}}}
\newcommand{\basislittokstd}{\texttt{\textbf{\{}}\qlittok{0},\qlittok{1}\texttt{\textbf{\}}}}
\newcommand{\unpacktok}[3]{\texttt{\textbf{let}}\ \texttt{\textbf{(}}\varvar_1,\varvar_2,\ldots,\varvar_{#1}\texttt{\textbf{)}}\ \texttt{\textbf{=}}\ #2\ \texttt{\textbf{in}}\:#3}
\newcommand{\unittype}{\texttt{\textbf{unit}}}
\newcommand{\bittype}{\texttt{\textbf{bit}}}
\newcommand{\qubittype}{\texttt{\textbf{qubit}}}
\newcommand{\basistype}{\texttt{\textbf{basis}}}
\newcommand{\arrowtok}[3]{{#1 \xrightarrow{#2} #3}}
\newcommand{\broadcasttok}[1]{\!\texttt{[}#1\texttt{]}\!}
\newcommand{\broadcasted}[2]{#1\broadcasttok{#2}}

\newcommand{\ouremptyset}{\varnothing}
\newcommand{\hilbertspace}{\mathcal{H}_2}
\newcommand{\rawpermutation}{\pi}
\newcommand{\rawaltpermutation}{\pi'}
\newcommand{\rawaltaltpermutation}{\pi''}
\newcommand{\permutationof}[1]{\rawpermutation{}(#1)}
\newcommand{\altpermutationof}[1]{\rawaltpermutation{}(#1)}
\newcommand{\altaltpermutationof}[1]{\rawaltaltpermutation{}(#1)}
\newcommand{\invpermutationof}[1]{\rawpermutation^{-1}(#1)}
\newcommand{\qctxunion}{\sqcup}
\newcommand{\bigqctxunion}{\bigsqcup}
\newcommand{\emptyveclist}{\cdot}
\newcommand{\emptyindexlist}{\cdot}

\newcommand{\emptytyctx}{\cdot}
\newcommand{\tyctx}{\Gamma}
\newcommand{\emptyqctx}{\ouremptyset}
\newcommand{\qctx}{\Delta}
\newcommand{\typebindtok}{:}
\newcommand{\typebind}[2]{#1 \typebindtok #2}
\newcommand{\typectxbind}[2]{#1 {\typebindtok} #2}
\newcommand{\typerel}[4]{#1 \vdash_{#2} \typebind{#3}{#4}}
\newcommand{\typerelwrap}[4]{\begin{gathered}#1 \vdash_{#2} #3 \assumbrk{} \typebindtok #4\end{gathered}}
\newcommand{\typerelaltwrap}[4]{\begin{gathered}#1 \vdash_{#2} \assumbrk{} \typebind{#3}{#4}\end{gathered}}
\newcommand{\orthorel}[2]{\braket{#1}{#2} = 0}
\newcommand{\rawsubtyperel}{<:}
\newcommand{\subtyperel}[2]{#1 \rawsubtyperel #2}
\newcommand{\freevars}[1]{\mathrm{FV}(#1)}
\newcommand{\statepair}[2]{\left[#1;#2\right]}
\newcommand{\smallstepraw}{\longrightarrow}
\newcommand{\smallstep}[4]{\statepair{#1}{#2} \smallstepraw \statepair{#3}{#4}}
\newcommand{\smallstepwrap}[4]{\begin{aligned}& \textstyle\statepair{#1}{#2} \\ &\smallstepraw \textstyle \statepair{#3}{#4}\end{aligned}}
\newcommand{\smallstepprob}[5]{\statepair{#1}{#2} \xlongrightarrow{#3} \statepair{#4}{#5}}
\newcommand{\smallstepprobwrap}[5]{\begin{aligned}& \textstyle\statepair{#1}{#2} \\ &\xlongrightarrow{#3} \textstyle \statepair{#4}{#5}\end{aligned}}
\newcommand{\multismallstep}[4]{\statepair{#1}{#2} \smallstepraw^{\ast} \statepair{#3}{#4}}
\newcommand{\multismallstepwrap}[4]{\begin{aligned}& \textstyle\statepair{#1}{#2} \\ &\smallstepraw^{\ast} \textstyle \statepair{#3}{#4}\end{aligned}}

\newcommand{\qdim}[1]{\vert #1 \vert}
\newcommand{\bspan}[1]{\mathrm{span}(#1)}
\newcommand{\atomposany}[2]{\Xi{#1}[#2]}

\newcommand{\atomposanynegmany}[2]{\Xi{\neg \vavar_1,\vavar_2,\ldots,\vavar_{#1}}[#2]}
\newcommand{\atomposanysup}[3]{\Xi{#1}[#3]^{#2}}
\newcommand{\atompos}[2]{\atomposany{\qlittok{#1}}{#2}}
\newcommand{\atomposneg}[2]{\atomposany{\neg\qlittok{#1}}{#2}}
\newcommand{\atomposnegneg}[3]{\atomposany{\neg\qlittok{#1},\qlittok{#2}}{#3}}

\newcommand{\atomposcard}[2]{\vert \atompos{#1}{#2} \vert}
\newcommand{\eqrel}[2]{#1 = #2}
\newcommand{\neqrel}[2]{#1 \ne #2}
\newcommand{\defas}{:=}
\newcommand{\vecsf}[1]{\mathbb{V}[#1]}
\newcommand{\vecf}[1]{\ket{#1}}
\newcommand{\basisvecidx}[2]{\babyket{{#1}^{(#2)}}\!}
\newcommand{\basisvecidxbra}[2]{\babybra{{#1}^{(#2)}}}
\newcommand{\intbit}[3]{\mathbb{B}^{#1,#2}_{#3}}

\newcommand{\substitute}[3]{[#1 \mapsto #2]#3}

\newcommand{\rangesubstitute}[5]{[#1 \mapsto #2]_{#3}^{#4}#5}
\newcommand{\btransunitary}[2]{U_{#1 \rightarrow #2}}
\newcommand{\indexvec}{\vec{i}}

\newcommand{\indexveccoord}[1]{i_{#1}}

\newcommand{\qstatevarproj}[1]{\babyket{{\qstatevarraw}_{#1}}}
\newcommand{\probof}[1]{\braket{{\qstatevarraw}_{#1}}{{\qstatevarraw}_{#1}}}
\newcommand{\qstatevarhat}[1]{\babyket{\hat{\qstatevarraw}_{#1}}\!}

\newcommand{\predunitary}[3]{{#1}^{#2}_{#3}}
\newcommand{\permunitary}{U_{\rawpermutation{}}}
\newcommand{\altpermunitary}{U_{\rawaltpermutation{}}}
\newcommand{\altaltpermunitary}{U_{\rawaltaltpermutation{}}}

\DeclareMathOperator{\vspan}{span}
\DeclareMathOperator*{\bigplus}{\scalerel*{\rawunboldbitensortok}{\sum}}

\newtheorem{theorem}{Theorem}
\newtheorem{lemma}{Lemma}
\renewcommand\descriptionlabel[1]{\hspace\labelsep\normalfont #1}

\begin{document}
\bstctlcite{IEEEexample:BSTcontrol}

\title{Qwerty: A Basis-Oriented Quantum \\ Programming Language}

\author{%
\IEEEauthorblockN{%
    Austin J. Adams\IEEEauthorrefmark{1},
    Sharjeel Khan\IEEEauthorrefmark{3},
    Arjun S. Bhamra\IEEEauthorrefmark{1},
    Ryan R. Abusaada\IEEEauthorrefmark{1}, \\
    Jeffrey S. Young\IEEEauthorrefmark{4},
    and Thomas M. Conte\IEEEauthorrefmark{1}\IEEEauthorrefmark{2}}
\IEEEauthorblockA{%
    \IEEEauthorrefmark{1}School of Computer Science, \IEEEauthorrefmark{2}School of Electrical \& Computer Engineering, \\
    \IEEEauthorrefmark{4}Partnership for an Advanced Computing Environment \\
    Georgia Institute of Technology, Atlanta, GA, USA}
\IEEEauthorblockA{%
    \IEEEauthorrefmark{3}Google, Mountain View, CA, USA}
\IEEEauthorblockA{\textit{Email: aja@gatech.edu}}}

\maketitle
\thispagestyle{plain}
\pagestyle{plain}

\begin{abstract}
Quantum computers have leaped from the theoretical realm into a race
to large-scale implementations. This is due to the promise of
revolutionary speedups, where achieving such speedup requires designing
an algorithm that harnesses the structure of a problem using quantum
mechanics. Yet many quantum programming languages today require programmers to
reason at a low level of physics notation and quantum gate circuitry. This
presents a significant barrier to entry for programmers who have not yet built
up an intuition about quantum gate semantics, and it can prove to be tedious
even for those who have. In this paper, we present Qwerty, a new quantum
programming language that allows programmers to manipulate qubits more
expressively than gates and trace programs without bra--ket notation. Due to
its novel basis type and easy interoperability with Python, Qwerty is a
powerful framework for high-level quantum--classical computation.

\end{abstract}


\section{Introduction}
Quantum computers have evolved from the realm of theoretical to an active and
highly competitive commercial race to large-scale implementations. The force behind this is
the promise of revolutionary speedups for
important problems (e.g., unstructured search~\cite{grover_fast_1996}, factoring large
integers~\cite{shor_polynomial-time_1999}, etc.).
Achieving such speedup is non-trivial and requires discovering algorithms that
leverage both insight about the problem structure and intuition about quantum
mechanics~\cite{shor_why_2003,aaronson_how_2022}.
Even if a programmer has strong intuition about a particular quantum algorithm,
realizing the algorithm often
requires an expertise in linear algebra and quantum physics notation coupled with a mastery of quantum
gate engineering~\cite{cobb_towards_2022,di_matteo_abstraction_2024,furntratt_towards_2024}.
This significant, abrupt gap between algorithmic intuition and low-level
quantum gate realization creates a barrier to entry for the budding quantum programmer~\cite{meyer_todays_2022,johansson_shut_2018,singh_review_2015}.

\textit{\textit{Qwerty}} is a new quantum programming language whose novel constructs such as literals for qubit states and the \lstinline!basis! type allow programmers to create and trace quantum programs without needing to understand bra--ket notation or gate-based semantics.  Qwerty is embedded in Python, allowing both
intuitive circuit synthesis from classical logic and
easy interoperability
between classical and quantum code, making Qwerty a robust framework for mixed
quantum--classical computation.

Qwerty differs from many other quantum languages in three aspects:
\begin{enumerate}
\item \textbf{Programs are expressed through
      basis translations} instead of
      low-level circuitry;
\item Quantum program behavior can be \textbf{traced intuitively through language constructs} without physics notation; and
\item \textbf{Hybrid quantum/classical programs are written in a popular programming language}, both
      on qubits and on classical bits, either inside or outside qubit lifetime.
\end{enumerate}
Figure~\ref{fig:diffuser} illustrates these differences by comparing the diffusion step of Grover's search algorithm in circuitry versus in Qwerty.  The Qwerty code directly expresses, ``replace $\ket{+}\!\ket{+}\!\ket{+}\!\ket{+}$ with $-\ket{+}\!\ket{+}\!\ket{+}\!\ket{+}$,'' whereas the circuit implementation requires significantly more explanation and expertise to understand.

\begin{figure}
\centering
\input{diffuser}
\caption{The diffuser from Grover's algorithm in Qwerty versus a traditional quantum circuit}\label{fig:diffuser}
\end{figure}

This paper introduces Qwerty with Grover's algorithm as a ``hello world'' program in Section~\ref{sec:hello-world}.
Subsequent sections describe programmer-friendly Qwerty features in more detail, beginning with qubit initialization (Section~\ref{sec:qlit}), then state evolution (Section~\ref{sec:evol}), metaprogramming (Section~\ref{sec:meta}), running operations in subspaces (Section~\ref{sec:pred}), and classical operations (Section~\ref{sec:classical}). Section~\ref{sec:ex} elaborates on these features by describing and tracing through several example Qwerty programs. A discussion of the differences between Qwerty and existing quantum programming languages is presented in Section~\ref{sec:relwork}.
Appendix~\ref{app:form} rigorously defines \formname{}, a subset of Qwerty, and proves its safety and universality.

\section{Hello World: Grover's in Qwerty}\label{sec:hello-world}

To introduce Qwerty, we begin with implementing Grover's well-known quantum search algorithm~\cite{grover_fast_1996,grover_quantum_1997,nielsen_quantum_2010,rieffel_quantum_2014,mermin_quantum_2007}.
Whereas a classical brute-force search through $2^N$ possibilities requires $O(2^N)$ queries to an
$N$-bit black box oracle, Grover's algorithm needs only $O(\sqrt{2^N})$
queries.
The Qwerty code for Grover's algorithm begins on lines~\lngroveroraclestart-\lngroveroracleend{} of Fig.~\ref{fig:grover} by
defining an example oracle.
For simplicity, the example oracle marks the bits $1010$ as the answer. As the
C-style syntax on line~\lngroveroraclexpr{} of Fig.~\ref{fig:grover} demonstrates,
Qwerty programmers express classical oracles with classical code instead of
quantum gates~\cite{green_quipper_2013,abhari_scaffold_2012,li_verified_2022,seidel_qrisp_2024}.

Grover's algorithm consists of many iterations, each of which rotates the state
away from the space of wrong answers toward the space of correct answers~\cite{nielsen_quantum_2010,mermin_quantum_2007}.
Line~\lngroveriterdef{} of Fig.~\ref{fig:grover} defines a single Grover
iteration as a Qwerty function named \lstinline!grover_iter!. (Applying the
\lstinline!@qpu! decorator to a function as seen on
line~\lngroverqpudecorator{} of Fig.~\ref{fig:grover} indicates that the
function is written in Qwerty, not Python.) Because the syntax
\lstinline!x | f | g! represents $g(f(x))$ in Qwerty, lines
\lngroverpipeonestart{}-\lngroverpipeoneend{} of Fig.~\ref{fig:grover} pass
\lstinline!q! first through an embedding of the oracle
(\lstinline!oracle.sign!) and then the \textit{Grover diffuser}
(\lstinline!'pppp' >> -'pppp'!). The syntax \lstinline!oracle.sign! produces a
quantum form of the classical function \lstinline!oracle! needed by Grover's
algorithm, which avoids the need to understand or prepare a $\ket{-}$ ancilla.
(Section~\ref{sec:bv} describes \lstinline!.sign! in more detail.)

The Grover diffuser on line~\lngroverbtrans{} of Fig.~\ref{fig:grover} is
written as a \textit{basis translation}, the fundamental state evolution
primitive in Qwerty. The basis translation \lstinline!'pppp' >> -'pppp'! allows
the programmer to write the Grover diffuser by describing how the state should
change, i.e., that  $\ket{+}^{\otimes 4}$ should be replaced with
${-}\ket{+}^{\otimes 4}$, instead of by picking specific gates that accomplish
this transformation.

The procedure for Grover's algorithm begins by preparing a superposition of all
possibilities. In the example in Fig.~\ref{fig:grover}, we have decided to
perform the search across all four-bit values; thus, line~\lngroverqlit{} of Fig.~\ref{fig:grover} prepares a superposition of all four-bit values using the \textit{qubit literal} \lstinline!'pppp'!.
Because \lstinline!'p'! is an equal superposition of both bits \lstinline!'0'! and \lstinline!'1'!, a four-fold \lstinline!'p'! state represents all combinations of four single-bit values.
In traditional quantum programming, this preparation of a $\ket{+}^{\otimes 4}$
state would be performed by writing a \textbf{\texttt{for}} loop with Hadamard gates,
but qubit literals are more succinct and express programmer intention more
explicitly.

On lines~\lngroverpipetwostart{}-\lngroverpipetwoend{} of
Fig.~\ref{fig:grover}, the pipeline in \lstinline!grover()! pipes this initial
\lstinline!'pppp'! state through three Grover iterations and into a four-qubit
measurement in the computational basis. (Note the number of iterations must be
chosen carefully because otherwise one may rotate either too far past the space
of correct answers or not far enough from the space of wrong
answers~\cite{nielsen_quantum_2010}.) The final line of Fig.~\ref{fig:grover}
is ordinary Python code and calls the Qwerty function \lstinline!grover()!
with typical Python function call syntax, demonstrating the seamless integration of Qwerty with Python.

\begin{figure}
\centering
\begin{lstlisting}[name=grover,style=nums,xleftmargin=5mm,linewidth=0.98\linewidth]
from qwerty import *

@classical
def oracle(x: bit[4]) -> bit:
  return x[0] & ~x[1] & x[2] & ~x[3]

@qpu
def grover_iter(q: qubit[4]):
  return (q | oracle.sign
            | 'pppp' >> -'pppp')

@qpu
def grover():
  return (
    'pppp' | grover_iter
           | grover_iter
           | grover_iter
           | measure**4)

print(grover())
\end{lstlisting}%
\caption{Grover's algorithm~\cite{nielsen_quantum_2010} implemented in Qwerty}\label{fig:grover}
\end{figure}

\section{Initialization: Qubit Literals}\label{sec:qlit}
Most quantum programs begin with allocating qubits and then initializing them.
Qwerty combines these two steps, providing \textit{qubit literals}, analogous to
string literals in classical programming languages (e.g., \lstinline!"howdy"!).
To stress this similarity, Qubit literals are written with the same syntax as classical string literals. For example, \lstinline!'0'! and \lstinline!'1'! represent both vectors of the computational basis, $\ket{0}$ and $\ket{1}$. Similarly to how \lstinline!"yee"+"haw"! typically equals \lstinline!"yeehaw"! for classical string literals, Qwerty defines \lstinline!'01'! as equivalent to \lstinline!'0'*'1'!, that is, \lstinline!*! represents the tensor product in Qwerty.
\begin{codeannotations}{grover}%
\codeannotation[top middle right]{oracle}{(-2pt,-3pt)}{(195pt,6pt)}{1}{(10pt,15pt)}{Classical oracle}
\codeannotation{btrans}{(75pt,-3pt)}{(167pt,-4pt)}{0}{(0,15pt)}{Basis translation}
\codeannotation{qlit}{(20pt,-2pt)}{(55pt,-4pt)}{0}{(20pt,10pt)}{Qubit literal}
\end{codeannotations}%

\subsection{Tilt}
Scalar factors, specifically phase factors of $e^{i\theta}$ where
$\theta \in \mathbb R$, are crucial ingredients in the description of qubit
states. But because complex numbers can be confusing to newcomers, Qwerty
represents multiplication by a phase, e.g. $e^{i\pi/4}\ket{1}$, with the syntax \lstinline!'1'@45!.
Qwerty's \lstinline!@! syntax is called \textit{tilt}
because \lstinline!'1'@45! evokes a rotation
\lstinline!'1'!{\Large$\circlearrowleft$}\lstinline!45!.
To emphasize the rotation metaphor, this paper often draws tilted states such as
\lstinline!'1'@45! as visually rotated states instead, as in
\rotatebox[origin=c]{45}{\lstinline!'1'!}. The syntax \lstinline!-'1'! is
syntactic sugar for \lstinline!'1'@180!, i.e.,
\rotatebox[origin=c]{180}{\lstinline!'1'!}.

\begin{figure}
\centering
\subfloat[Superposition literal]{%
\includegraphics[width=1.1in]{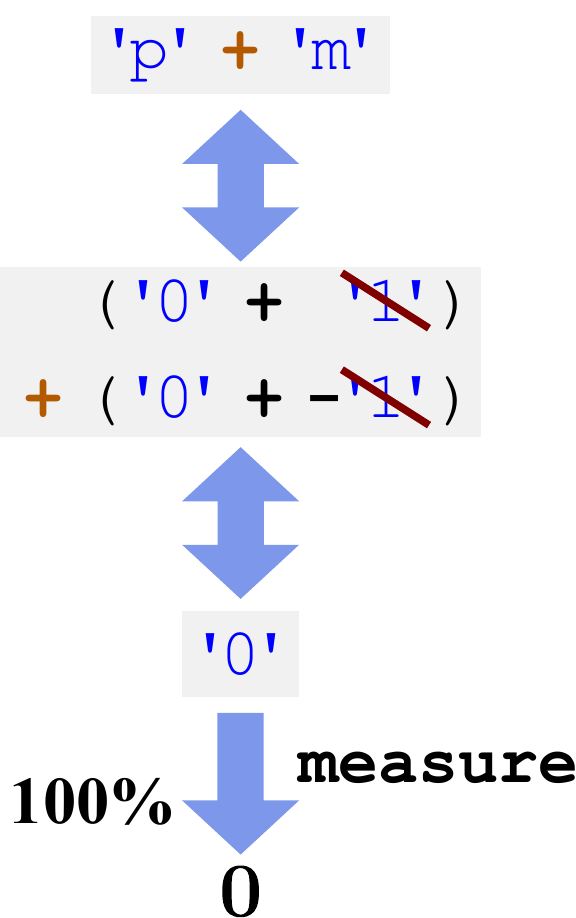}
\label{fig:mixed:pure}}
\hfill
\subfloat[Ensemble literal]{%
\includegraphics[width=2in]{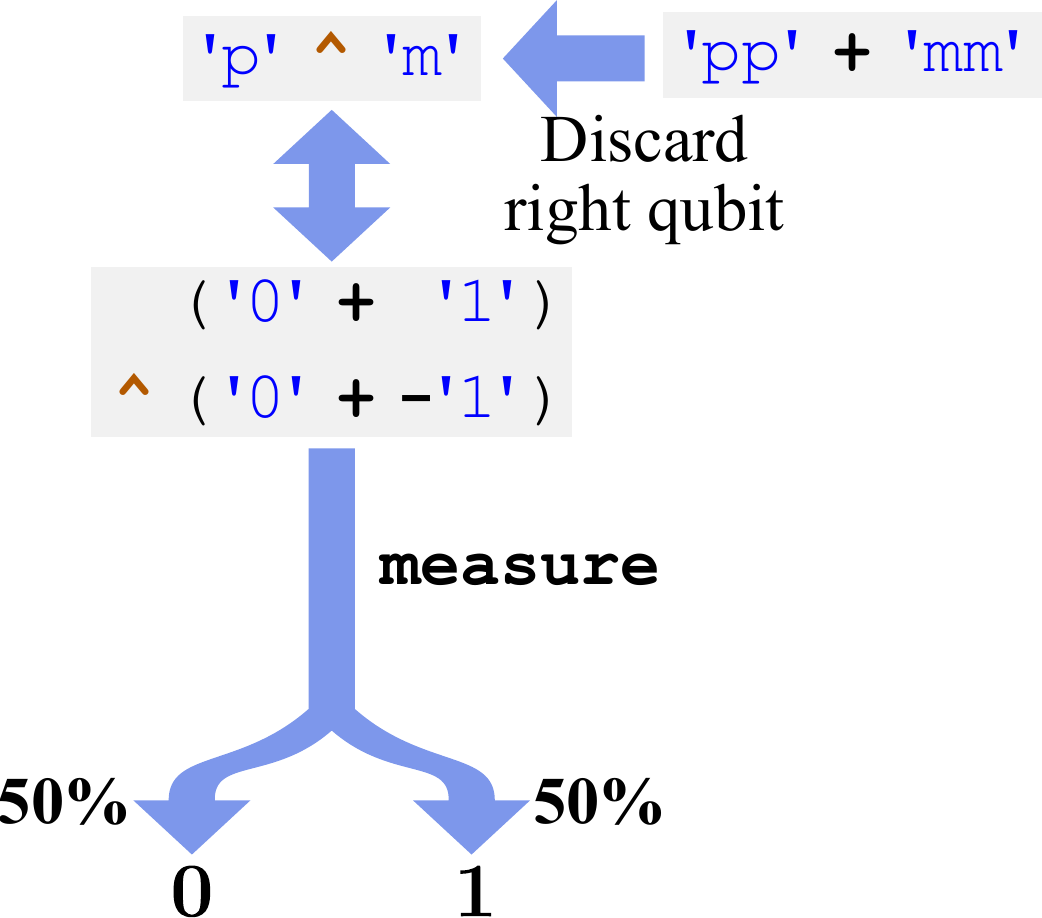}
\label{fig:mixed:mixed}}
\caption{Comparison of how the superposition literal \lstinline!('p' + 'm')!, a pure state, and the ensemble literal \lstinline!('p' ^ 'm')!, a mixed state, respond to measurement}
\label{fig:mixed}
\end{figure}

\subsection{Superposition and Ensemble Literals}\label{sec:ensemble-lit}
Since superposition is a crucial ingredient of quantum algorithms and allows
expressing qubit states in full generality, Qwerty includes a
\textit{superposition literal}. For
example, \lstinline!('0' + '1')! represents the $\ket{+}$ state
\lstinline!'p'!.
The most common use of the superposition literal in
Qwerty code is to prepare a Bell state with \lstinline!('00' + '11')! as seen
on line~\lnsuperdensebellpair{} of Fig.~\ref{fig:superdense}, but superposition
literals also allow tracing Qwerty programs without bra--ket notation. To make
this tracing feasible, the superposition literal also supports specifying
probabilities, e.g., \lstinline!(0.75*'0' + 0.25*'1')!. 

In the language of quantum mechanics, superposition literals represent pure
states, but it can be useful to describe mixed states when tracing Qwerty
programs --- that is, a classical probability distribution of pure states. For example, if the right qubit of the entangled state \lstinline!('pp' + 'mm')!, were discarded, one could describe the remaining left qubit using
the Qwerty \textit{ensemble literal} \lstinline!('p' ^ 'm')!. Fig.~\ref{fig:mixed} shows
how the mixed state \lstinline!('p' ^ 'm')! differs from the
pure superposition state \lstinline!('p' + 'm')!: states in a pure superposition
can destructively interfere (Fig.~\ref{fig:mixed:pure}),
whereas states in an ensemble (Fig.~\ref{fig:mixed:mixed}) cannot. For both
superposition and ensemble literals, the Qwerty type checker guarantees that
probabilities sum to $1$ and that all state operands are orthogonal.

\begin{figure}
\centering
\subfloat[Invalid program (implicitly discards \texttt{b})]{%
\begin{minipage}{2.7in}
\begin{lstlisting}[name=discardinvalid,style=s]
@qpu
def invalid():
    a, b = '01' + '10'
    return a | measure
\end{lstlisting}
\end{minipage}
\label{fig:discard:invalid}}
\hfil
\subfloat[Valid program (explicitly discards \texttt{b})]{%
\begin{minipage}{2.7in}
\begin{lstlisting}[name=discardvalid,style=s]
@qpu
def valid():
    a, b = '01' + '10'
    return a * b | measure * discard
\end{lstlisting}
\end{minipage}
\label{fig:discard:valid}}
\caption{Demonstration of how qubits must be discarded explicitly in Qwerty. The syntax \lstinline!a, b = q! assigns the left qubit of the two-qubit register \lstinline!q! to \lstinline!a! and the right qubit to \lstinline!b!.}
\label{fig:discard}
\end{figure}

\subsection{The \textbf{\texttt{qubit}} Type}\label{sec:qubit-type}
The type of a qubit literal of length $N$ is \lstinline!qubit[N]!. To
avoid violating the no cloning theorem~\cite{nielsen_quantum_2010},
\lstinline!qubit! is a linear
type~\cite{selinger_lambda_2006,paykin_qwire_2017,yuan_twist_2022}, meaning
that any value of type \lstinline!qubit! must be used exactly once. Programmers
can discard \lstinline!qubit!s only explicitly, as shown in
Fig.~\ref{fig:discard}.
\begin{codeannotations}{superdense}%
\codeannotation{bellpair}{(90pt,-2pt)}{(153pt,-4pt)}{0}{(0,20pt)}{Entangled state \\ \small (Bell pair)}
\end{codeannotations}%

\begin{figure}
\centering
\begin{lstlisting}[name=superdense,style=nums,xleftmargin=5mm,linewidth=0.98\linewidth]
def superdense_coding(payload: bit[2]):
  bit0, bit1 = payload

  @qpu
  def kernel() -> bit[2]:
    alice, bob = '00' + '11'

    sent_to_bob = (
      alice | ({'0'>>'1', '1'>>'0'}
               if bit0 else id)
            | ('1' >> -'1'
               if bit1 else id))

    return (sent_to_bob * bob
            | {'00' +  '11',
               '00' + -'11',
               '10' +  '01',
               '01' + -'10'}.measure)

  recovered_payload = kernel()
  return recovered_payload
\end{lstlisting}
\caption{Qwerty code for superdense coding, which allows Alice to send two classical bits to Bob by sending him one qubit, assuming an entangled state had already been shared between Alice and Bob~\cite{nielsen_quantum_2010}}\label{fig:superdense}
\end{figure}

\section{Evolution: Basis Translations}\label{sec:evol}
After initializing its qubits, a typical quantum program evolves qubits to a
useful state. Qwerty programmers achieve these state changes using
\textit{basis translations}.
One
can view basis translations as a substitution rule replacing one basis with
another. For example, \lstinline!{'0','1'} >> {'m','p'}! simultaneously
replaces the state \lstinline!'0'! with \lstinline!'m'! and the state
\lstinline!'1'! with \lstinline!'p'!. In general, a basis translation is
written $b_\text{in}$\lstinline! >> !$b_\text{out}$, where each $b$ is a basis
literal, described in the next section.
(Traditionally, a basis translation could be called
a ``change of basis,'' but this term may be confused with the linear algebra
operation that changes only the representation of a vector, not its value.)

\subsection{Basis Literals}\label{sec:blit}
In Qwerty, a basis is expressed as a \textit{basis literal}, an ordered list of
basis vectors, each written as a qubit literal (or a superposition literal). For example,
\lstinline!{'00','01','10','11'}! is a basis literal representing the two-qubit
computational basis. One could equivalently write
\lstinline!{'0','1'}*{'0','1'}! (the tensor product of two single-qubit
computational bases) or \lstinline!{'0','1'}**2! (the two-fold tensor product of
the single-qubit computational basis). Since a basis in Qwerty represents an
orthonormal basis, any two basis vectors in a basis literal must be
orthogonal. For instance, \lstinline!{'00',-'00','01'}! is an invalid basis because \lstinline!'00'! and \lstinline!-'00'! are not orthogonal.

\subsection{Type Checking}\label{sec:btrans-typecheck}
Type checking is necessary for basis translations because syntax checking alone
cannot catch all cases of nonsensical basis translations. For example, the
basis translation \lstinline!{'0'} >> {'1','0'}! is syntactically valid but
could not be interpreted as an element-wise replacement rule.
To guarantee that basis
translations represent unitary operators, the Qwerty type checker verifies that
both basis operands of a basis translation share the same span. (Here, \textit{span} is
meant in the typical linear algebraic sense~\cite{axler_linear_2023}.)
In other words, for any basis translation $b_\text{in}$\lstinline! >> !$b_\text{out}$, the Qwerty type checker requires that any state that can be described as a superposition of the vectors in the input basis $b_\text{in}$ must also be possible to write as a superposition of the vectors in the output basis $b_\text{out}$, and vice versa. The basis translation
\lstinline!{'0','1'@90} >> {'1','0'}! would pass type checking, for
example, since any state that can be written as a superposition of
\lstinline!'0'! and \rotatebox[origin=c]{90}{\lstinline!'1'!} can also be
written as a superposition of \lstinline!'1'! and \lstinline!'0'!.

\subsection{Measurement}\label{sec:meas}
Because some quantum algorithms leverage measurement in different bases, Qwerty
defines measurement in terms of a basis. For example,
\lstinline!{'p','m'}.measure! is a one-qubit $X$-measurement that returns one
classical bit. Measuring in more exotic bases is possible, such as measuring in
the $N$-qubit Fourier basis with \lstinline!fourier[[N]].measure! or measuring in
the Bell basis as seen on lines~\lnsuperdensebellstart{}-\lnsuperdensebellend{}
of Fig.~\ref{fig:superdense}.
Ultimately, any $b$\lstinline!.measure! compiles
to a basis translation
$b$\lstinline! >> !\allowbreak\lstinline!{'0','1'}**N!
followed by
\lstinline!({'0','1'}**N).measure!.

\subsection{Syntactic Sugar}
We briefly define some basis translation syntactic sugar because it is used in most of
the example code in this paper. If a basis translation involves bases with only
one vector, such as \lstinline!{'1'} >> {-'1'}!, the Qwerty compiler allows
writing \lstinline!'1' >> -'1'! instead, as seen on line~\lnsuperdensez{} of
Fig.~\ref{fig:superdense}. To make a basis translation more closely resemble a
substitution rule, a basis translation such as \lstinline!{'0','1'} >> {'1','0'}!
may be written as \lstinline!{'0'>>'1', '1'>>'0'}! instead, as seen
on line~\lnsuperdenseflip{} of Fig.~\ref{fig:superdense}.

\subsection{Mathematical View}
In mathematical terms, a basis translation
$\texttt{\{}\varbv_1,\varbv_2,\allowbreak\ldots,\varbv_m\texttt{\} >{}> \{}\varbv_1',\varbv_2',\ldots,\varbv_m'\texttt{\}}$
represents the following unitary matrix:
\begin{gather*}
\kbordermatrix{
                   & \bm{\babyket{\varbv_1}} & \bm{\babyket{\varbv_2}} & \cdots & \bm{\babyket{\varbv_m}} & {\color{darkgray}\babyket{\varbv^\perp}} \\
   \bm{\babyket{\varbv_1'}} &          \bm{1}&              0 & \cdots &              0 &                  0 \\
   \bm{\babyket{\varbv_2'}} &              0 &          \bm{1}&        &              0 &                  0 \\
                   &         \vdots &                & \ddots &                &             \vdots \\
   \bm{\babyket{\varbv_m'}} &              0 &              0 &        &          \bm{1}&                  0 \\
{\color{darkgray}\babyket{\varbv^\perp}} &              0 &              0 & \cdots &              0 &              {\color{darkgray}\bm{1}}\\
}
\end{gather*}
If all $\varbv_i$ and $\varbv_j'$ are computational basis vectors, then
a basis translation can be viewed as a classical permutation of
computational basis vectors.
If not, then the basis translation can be viewed as a classical permutation conjugated with changes of basis.
Alternatively, the matrix above can be viewed as a singular value decomposition where all singular
values are~1~\cite{axler_linear_2023}.
Above, ${\color{darkgray}\babyket{\varbv^\perp}}$ denotes vectors orthogonal to both bases; these
orthogonal vectors pass through the basis translation unchanged.
(For example, $\babyket{00}$ passes through \lstinline!'11' >> -'11'! unchanged because $\babyket{00}$ is orthogonal to $\babyket{11}$.)
This unitary representing a basis translation can be written in bra--ket notation as
\begin{gather*}
\sum_{j=1}^m \babyket{\varbv_j'}\!\babybra{\varbv_j} + {\color{darkgray}\sum_k\babyket{\varbv^\perp_k}\!\babybra{\varbv^\perp_k}}.
\end{gather*}
The vectors ${\color{darkgray}\babyket{\varbv^\perp_k}}$ are an orthonormal basis for
$\vspan(\varbv_1,\allowbreak\varbv_2,\ldots,\varbv_m)^\perp$,
the orthogonal complement of the space spanned by both basis operands.

\section{Metaprogramming: Macros and Prelude}\label{sec:meta}
Needing to write bases explicitly could become tedious. For example,
\lstinline!{'0','1'}.measure! is a verbose way to write an ordinary
measurement. To mitigate this, the Qwerty runtime has a \textit{prelude}
holding abbreviated aliases for common bases and operations. The dialect of
Qwerty that understands these aliases is called
\textit{metaQwerty}. The compilation process for a function marked with
\lstinline!@qpu! begins with expanding metaQwerty to Qwerty, after which type
checking and later circuit synthesis occur.
Expressing quantum code with metaprogramming is consistent with the
respective roles of classical computers and quantum accelerators in computation:
the classical computer is tasked with constructing quantum instructions to send to
quantum hardware, which is specialized to preserve and manipulate its delicate
quantum
system~\cite{qcl,selinger_lambda_2006,paykin_qwire_2017,amy_sized_2019}.

\begin{figure}
\centering
\begin{lstlisting}[name=prelude,style=nums,xleftmargin=5mm,linewidth=0.98\linewidth]
# Symbols in qubit literals
'0'.sym = __SYM_STD0__()
'1'.sym = __SYM_STD1__()
'p'.sym = '0' + '1'
'i'.sym = '0' + '1'@90
'm'.sym = '0' + '1'@180
'j'.sym = '0' + '1'@270

# Common bases
std = {'0', '1'}
pm = {'p', 'm'}
ij = {'i', 'j'}
bell = {'00' +  '11',
        '00' + -'11',
        '10' +  '01',
        '01' + -'10'}
\end{lstlisting}
\caption{The first portion of the default Qwerty prelude}\label{fig:prelude}
\end{figure}

Fig.~\ref{fig:prelude} shows the beginning of the default metaQwerty prelude.
Lines~\lnpreludezero{} and \lnpreludeone{} of Fig.~\ref{fig:prelude} define the computational basis
vectors using intrinsics, and then
lines~\lnpreludesymstart{}-\lnpreludesymend{} define other common single-qubit
states as superpositions of \lstinline!'0'! and \lstinline!'1!'. The
\lstinline!'i'! and \lstinline!'j'! states defined on lines~\lnpreludei{} and
\lnpreludej{} of Fig.~\ref{fig:prelude} represent the $Y$ eigenstates
$\ket{+i}$ and $\ket{-i}$, respectively.
With these qubit literal symbols defined, then
lines~\lnpreludebasisstart{}-\lnpreludebasisend{} of Fig.~\ref{fig:prelude}
define some common bases. With these \textit{basis aliases}, then a programmer
can write e.g. \lstinline!pm.measure**2! instead of
\lstinline!{'pp','pm','mp','mm'}.measure!.

In general, any \lstinline[morekeywords={foo}]!x.foo! syntax in Qwerty code is using
a metaQwerty macro named \textit{foo}. For example, the definition of the macro
\lstinline!.measure! is found in the prelude on
line~\lnpreludetwomeasuremacro{} of Fig.~\ref{fig:preludetwo}. The prelude also
defines abbreviated names for programming convenience, such as
line~\lnpreludetwomeasure{} of Fig.~\ref{fig:preludetwo}, which allows an
$N$-qubit measurement in the computational basis to be written as
\lstinline!measure**N! instead of \lstinline!(std**N).measure! (e.g.,
line~\lngrovermeasure{} of Fig.~\ref{fig:grover}). Similarly,
line~\lnpreludetwoflipmacro{} of Fig.~\ref{fig:preludetwo} enables programmers to
write \lstinline!std.flip! to perform a bit flip instead of the more verbose
\lstinline!{'0','1'}>>{'1','0'}!. (Line~\lnpreludetwoflipvar{} of Fig.~\ref{fig:preludetwo} shortens the common \lstinline!std! case further to just \lstinline!flip!.)

\begin{figure}
\centering
\includegraphics[width=0.9\linewidth]{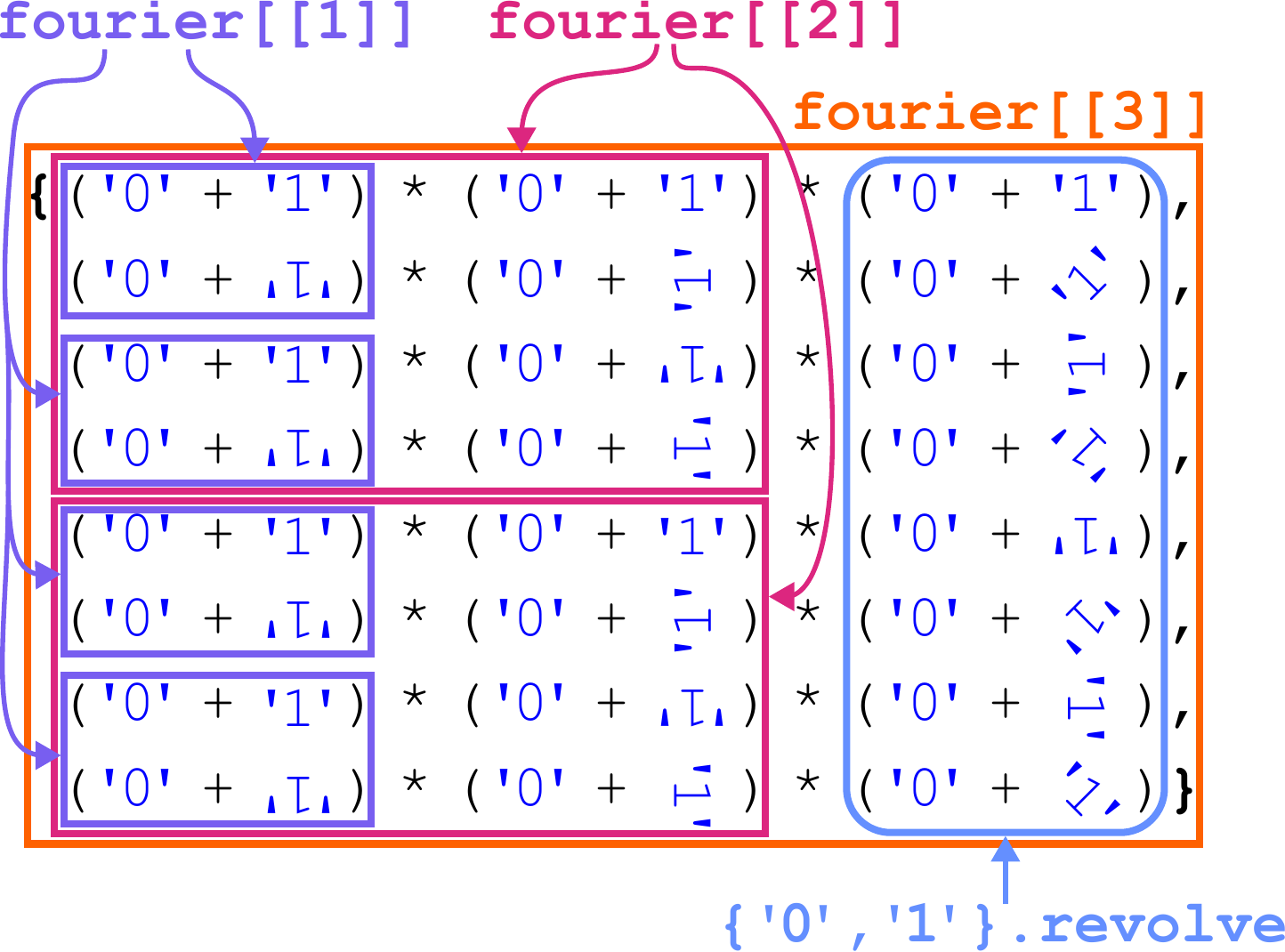}
\caption{Drawing of the recursive structure of the Fourier basis. This is a Qwerty reinterpretation of Equation (5.4) of Nielsen and Chuang~\cite{nielsen_quantum_2010}.}\label{fig:fourier-nested}
\end{figure}

Lines~\lnpreludetwofourierrecstart{}-\lnpreludetwofourierrecend{} of Fig.~\ref{fig:preludetwo} define the $N$-qubit
Fourier basis, which is vital for quantum algorithms such as period finding
(Section~\ref{sec:period}). As seen in Fig.~\ref{fig:fourier-nested}, the Fourier
basis has a recursive structure: the $N$-qubit fourier basis can be constructed
by concatenating the full list of $(N{-}1)$-qubit Fourier basis
vectors with itself and then extending each resulting vector with a
superposition of \lstinline!'0'! and a revolving \lstinline!'1'!.
The revolving \lstinline!'1'! is accomplished by
gradually increasing the tilt of \lstinline!'1'! in units of $360/2^N$ degrees; the rightmost box in
Fig.~\ref{fig:fourier-nested} shows the list of these superpositions when $N{=}3$. Qwerty represents
this list of superpositions efficiently as \lstinline!{'0','1'}.revolve!, a
\textit{basis generator}. Basis generators allow constructing bases in more
complex ways than taking the tensor product of smaller bases. The knowledge of
basis structure in the compiler provided by basis generators allows efficient
synthesis of basis translations as e.g. efficient quantum Fourier transform
circuits.
Lines~\lnpreludetwofourierinductstart{}-\lnpreludetwofourierinductend{} of Fig.~\ref{fig:preludetwo}
define the $N$-qubit Fourier basis recursively by using the
\lstinline!//! operator to apply the basis generator \lstinline!std.revolve!
to the basis \lstinline!fourier[[N-1]]!. Line~\lnpreludetwobasecase{} of
Fig.~\ref{fig:preludetwo} defines the single-qubit Fourier basis, the base case
for the recursive definition.

The automatic expansion of metaQwerty to Qwerty also enables parametrically polymorphic
programming, that is, defining code in terms of e.g. the length of a quantum
register. For example, in contrast to how lines
\lngroverhardcodestart{}-\lngroverhardcodeend{} of Fig.~\ref{fig:grover}
hard code the algorithm to operate on four qubits, Fig.~\ref{fig:grovermeta}
implements Grover's algorithm in terms of a \textit{dimension variable} $N$.
The Qwerty compiler can infer dimension variables such as $N$, meaning they are
usually not specified explicitly; for example, if one calls
\lstinline!grover2(oracle)! when \lstinline!oracle! is the function defined on
line~\lngroveroracledef{} of Fig.~\ref{fig:grover}, the Qwerty compiler would infer that
$N=4$. The expression \lstinline!'p'**N! would then expand to \lstinline!'p'**4!,
a four-fold tensor product equivalent to \lstinline!'pppp'!.
Similarly, according to the number of iterations \lstinline!num_iter! value
passed by the caller of \lstinline!grover2()!, the \lstinline!for! construct on
lines~\lngrovermetaforstart{}-\lngrovermetaforend{} would expand to a pipeline
of \lstinline!grover_iter! calls as seen on
lines~\lngroveriterpipestart{}-\lngroveriterpipeend{} of Fig.~\ref{fig:grover}.

\begin{figure}
\centering
\begin{lstlisting}[name=preludetwo,style=nums,xleftmargin=5mm,linewidth=0.98\linewidth]
# Basis macros
{bv1, bv2}.flip = {bv1, bv2} >> {bv2, bv1}
b.measure = __MEASURE__(b)
{bv1, bv2}.revolve = __REVOLVE__(bv1, bv2)

# More complicated bases
fourier[[1]] = pm
fourier[[N]] = (fourier[[N-1]]
                // std.revolve)

# Built-in functions
id: qfunc = lambda q: q
discard = __DISCARD__()
flip = std.flip
measure = std.measure
\end{lstlisting}
\caption{The second portion of the default Qwerty prelude}\label{fig:preludetwo}
\end{figure}

\begin{figure}
\centering
\begin{lstlisting}[name=grovermeta,style=nums,xleftmargin=5mm,linewidth=0.98\linewidth]
def grover2(oracle, num_iter):
  @qpu[[N]]
  def grover_iter(q):
    return (q | oracle.sign
              | 'p'**N >> -'p'**N)

  @qpu[[N]]
  def kernel():
    return (
      'p'**N | (grover_iter
                for i in range(num_iter))
             | measure**N)

  return kernel()
\end{lstlisting}
\caption{A polymorphic version of the Grover's algorithm code from Fig.~\ref{fig:grover}}\label{fig:grovermeta}
\end{figure}

\section{Predication: Basis Patterns}\label{sec:pred}
While evolving the whole state space of a qubit is useful,
quantum algorithms often require
performing some state evolution only in a proper subspace. In circuit-oriented programming, this is typically performed by adding controls on gates or functions, which makes them execute only when all control qubits are $\ket{1}$. Qwerty generalizes controls, allowing programmers to run functions only on states where a \textit{basis pattern} is satisfied. These predications are written such that they resemble classical conditional expressions. For example, \lstinline!(flip if '1_' else id)! will flip the right qubit if the left is a \lstinline!'1'!; otherwise, the identity \lstinline!id! is applied to the right qubit. Thus, \lstinline!'10' | (flip if '1_' else id)! yields \lstinline!'11'!, and \lstinline!'00' | (flip if '1_' else id)! yields an unchanged state \lstinline!'00'!. The \textit{target} symbol \lstinline!'_'! is specific to basis patterns.

The previous example is equivalent to a traditional controlled-NOT gate, but
more powerful basis patterns than the trivial \lstinline!{'1_'}! pattern are
possible. For instance, \lstinline!(pm >> std if {'p_p', 'm_m'} else id)!
performs the following mapping:
\begin{center}
\setlength{\tabcolsep}{1.9pt}
\begin{tabular}{rlcrl}
    \textcolor{blue}{\texttt{\textquotesingle{}\textbf{p}\underline{p}\textbf{p}\textquotesingle{}}}
    &$\mapsto$
    \textcolor{blue}{\texttt{\textquotesingle{}\textbf{p}\underline{0}\textbf{p}\textquotesingle{}}}
    & \qquad\qquad &
    \textcolor{blue}{\texttt{\textquotesingle{}ppm\textquotesingle{}}}
    &$\mapsto$
    \textcolor{blue}{\texttt{\textquotesingle{}ppm\textquotesingle{}}} \\
    \textcolor{blue}{\texttt{\textquotesingle{}\textbf{p}\underline{m}\textbf{p}\textquotesingle{}}}
    &$\mapsto$
    \textcolor{blue}{\texttt{\textquotesingle{}\textbf{p}\underline{1}\textbf{p}\textquotesingle{}}}
    & \qquad\qquad &
    \textcolor{blue}{\texttt{\textquotesingle{}pmm\textquotesingle{}}}
    &$\mapsto$
    \textcolor{blue}{\texttt{\textquotesingle{}pmm\textquotesingle{}}} \\
    \textcolor{blue}{\texttt{\textquotesingle{}mpp\textquotesingle{}}}
    &$\mapsto$
    \textcolor{blue}{\texttt{\textquotesingle{}mpp\textquotesingle{}}}
    & \qquad\qquad &
    \textcolor{blue}{\texttt{\textquotesingle{}\textbf{m}\underline{p}\textbf{m}\textquotesingle{}}}
    &$\mapsto$
    \textcolor{blue}{\texttt{\textquotesingle{}\textbf{m}\underline{0}\textbf{m}\textquotesingle{}}} \\
    \textcolor{blue}{\texttt{\textquotesingle{}mmp\textquotesingle{}}}
    &$\mapsto$
    \textcolor{blue}{\texttt{\textquotesingle{}mmp\textquotesingle{}}}
    & \qquad\qquad &
    \textcolor{blue}{\texttt{\textquotesingle{}\textbf{m}\underline{m}\textbf{m}\textquotesingle{}}}
    &$\mapsto$
    \textcolor{blue}{\texttt{\textquotesingle{}\textbf{m}\underline{1}\textbf{m}\textquotesingle{}}}
\end{tabular}
\end{center}
Since the \lstinline!else id! clause above is often redundant, the syntactic sugar
\lstinline!(pm >> std in {'p_p', 'm_m'})! also performs the mapping above.
Because basis patterns are an extension of basis literals, the requirement that
basis vectors in a basis literal are orthogonal still applies to basis
patterns. For instance, the Qwerty type checker would flag
\lstinline!{'p_p', 'p_0'}! as invalid because \lstinline!'pp'! and
\lstinline!'p0'! are not orthogonal. All basis vectors in a basis pattern must
also have the target symbol \lstinline!'_'! at the same position in each
vector. For instance, \lstinline!{'p_p','mm_'}! and \lstinline!{'p_p', 'mmm'}! are
both invalid as opposed to the valid basis pattern \lstinline!{'p_p','m_m'}! shown
earlier.

The \textit{padding} symbol
\lstinline!'?'! can represent a qubit neither matched on (as in \lstinline!'0'!) nor targeted by the predicated function (as in \lstinline!'_'!). For example, \lstinline!{'0?1','1?0'} >> {'1?0','0?1'}! is a 3-qubit basis translation that leaves the middle qubit unchanged while performing
\discretionary{\lstinline!{'01','10'} >>!}{\lstinline!{'10','01'}!}{\lstinline!{'01','10'} >> {'10','01'}!}
on the outer two qubits.

Quantum phase estimation (QPE) written in Qwerty (Fig.~\ref{fig:qpe}) shows
nontrivial basis patterns in action. The QPE algorithm approximates how much a black-box operation
\lstinline!op! tilts a particular state by constructing a Fourier basis state
(a row of Fig.~\ref{fig:fourier-nested}) and then measuring in the Fourier
basis. It is crucial that for any nonnegative integer $J$, \lstinline!op! can effectively be repeated $2^J$
times efficiently; for instance, line~\lnqpeuserangleexp{} of the example QPE caller
code in Fig.~\ref{fig:qpeuser} satisfies this requirement by adjusting the
rotation that \lstinline!tilt_op! performs to achieve the same
affect as repeating \lstinline!tilt_op! $2^J$ times.
Using Qwerty syntax, Fig.~\ref{fig:qpe-trace} traces through how
lines~\lnqpescanstart{}-\lnqpescanend{} of Fig.~\ref{fig:qpe} construct the
desired Fourier basis state by tilting the \lstinline!'1'! in each of the three
initial \lstinline!'p'! states. The left three qubits at the end of
Fig.~\ref{fig:qpe-trace} are exactly the three-qubit Fourier basis state found
on the third-to-last row of Fig.~\ref{fig:fourier-nested}. Because the third-to-last row has index 5 in the list of basis vectors, the
measurement on line~\lnqpemeas{} of Fig.~\ref{fig:qpe} yields the bits $101_2 =
5_{10}$. Thus, the calculation of $360\times 0.101_2$ on
line~\lnqpeuserfinalcalc{} of Fig.~\ref{fig:qpeuser} produces $225\degree$, which is
exactly the angle originally written on line~\lnqpeuserangle{} of
Fig.~\ref{fig:qpeuser}. (In less ideal cases, the measurement will project the
state to the nearest Fourier basis state with high probability, providing a
reasonable approximation.)

As discussed in Section~\ref{sec:meta}, the compiler will expand the the
\lstinline!for! construct on line~\lnqpeloop{} of Fig.~\ref{fig:qpe} into a
pipeline with \lstinline!j! substituted with $0$, $1$, and $2$ in the first, second, and third
stages of the pipeline, respectively.
The Qwerty QPE code expects that \lstinline!op! is
defined in terms of a dimension variable (say, \lstinline!J!) that determines
that \lstinline!op! should be applied effectively $2^J$ times.
(Lines~\lnqpeuserJ{}-\lnqpeuserangleexp{} of Fig.~\ref{fig:qpeuser} display an example of this.)
This is useful
because the \textit{instantiation} \lstinline!op[[!$\ldots$\lstinline!]]! on
line~\lnqpeinstant{} of Fig.~\ref{fig:qpe} --- also a metaQwerty feature ---
will summon a specialization of \lstinline!op! with its dimension variable
\lstinline!J! set to $\texttt{prec}-1-\texttt{j}$. If \lstinline!op! is
implemented correctly, this will synthesize a version of \lstinline!op! that effectively
runs itself $2^J$ times.

\begin{codeannotations}{qpe}%
\codeannotation[top middle right]{pat}{(62pt,-3pt)}{(228pt,-3pt)}{0}{(20pt,25pt)}{Basis \\ \small pattern}
\codeannotation[bottom right]{fourier}{(40pt,-3pt)}{(169pt,6pt)}{1}{(0pt,-25pt)}{Measurement in \\ \small the Fourier basis}
\end{codeannotations}%

\begin{figure}
\centering
\begin{lstlisting}[name=qpe,style=nums,xleftmargin=5mm,linewidth=0.98\linewidth]
from fractions import Fraction
from qwerty import *

def qpe(prec, get_init_state, op):
  @qpu[[M]]
  def kernel():
    return (
      'p'**prec * get_init_state()
      | (op[[prec-1-j]]
         in '?'**j * '1' * '?'**(prec-1-j)
            * '_'**M
         for j in range(prec))
      | fourier[[prec]].measure
        * discard**M)

  bits = kernel()
  return Fraction(int(bits),
                  2**len(bits))
\end{lstlisting}
\caption{Quantum phase estimation expressed in Qwerty~\cite{cleve_quantum_1998,nielsen_quantum_2010}}\label{fig:qpe}
\end{figure}
\begin{figure}
\centering
\begin{lstlisting}[name=qpeuser,style=nums,xleftmargin=5mm,linewidth=0.98\linewidth]
from qwerty import *
from qpe import qpe

angle_deg = 225.0
precision = 3

@qpu
def init1():
  return '1'

@qpu[[J]]
@reversible
def tilt_op(q):
  return q | '1' >> '1'@(angle_deg*2**J)

print('Expected:', angle)
angle_got = 360*qpe(precision, init1,
                    tilt_op)
print('Actual:', float(angle_got))
\end{lstlisting}
\caption{Example Qwerty code that invokes the quantum phase estimation implementation in Fig.~\ref{fig:qpe}}\label{fig:qpeuser}
\end{figure}

\begin{figure}
\centering
\includegraphics[width=0.85\linewidth]{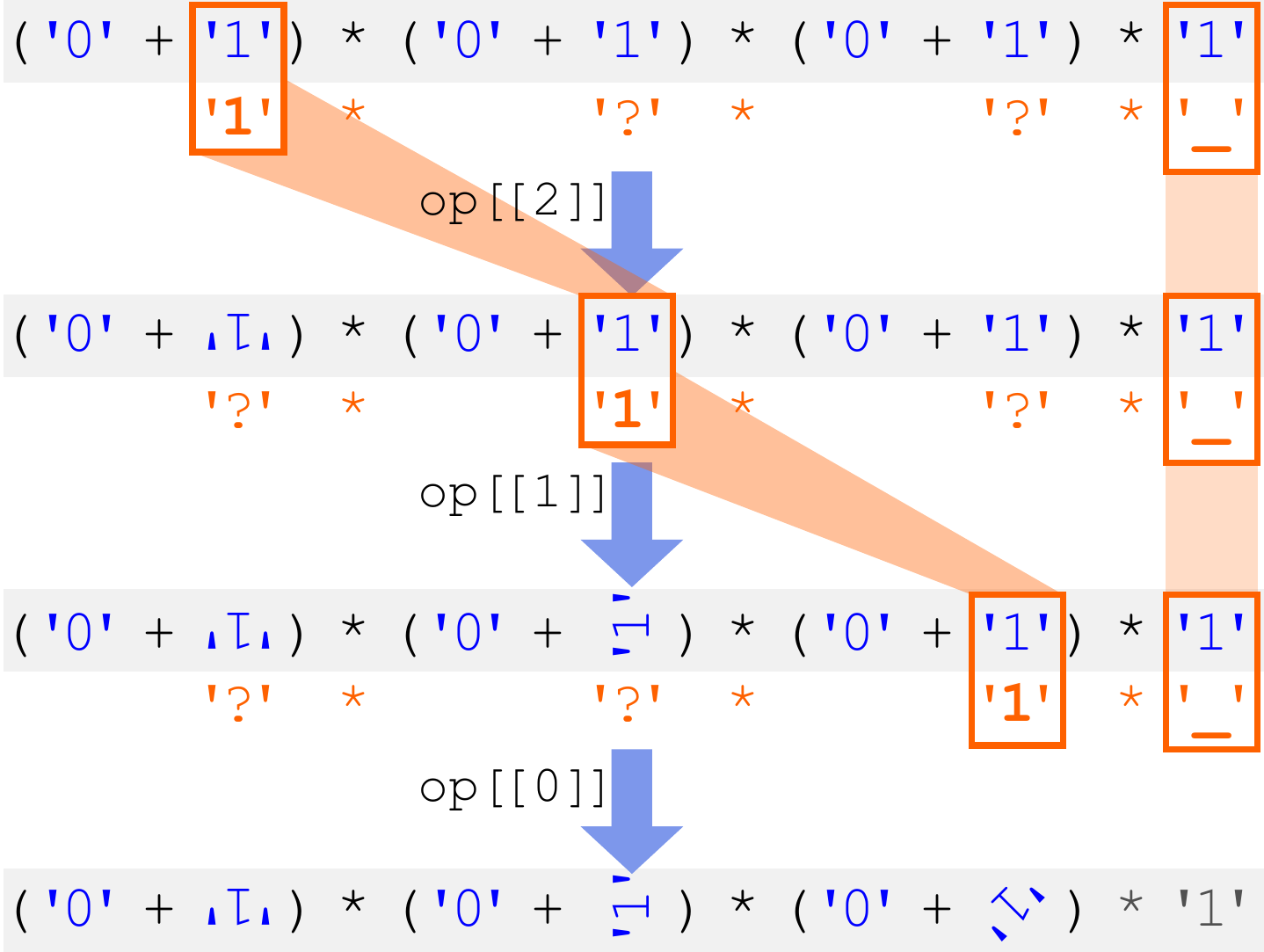}
\caption{A Qwerty-based visualization of how the example quantum phase estimation program in Fig.~\ref{fig:qpe} constructs the desired Fourier basis state
}\label{fig:qpe-trace}
\end{figure}

\section{Classical Operations}\label{sec:classical}

With quantum operations in Qwerty tackled, we turn our attention to classical
operations.
Broadly speaking, there are four categories of classical operations that a
quantum programming language can support~\cite{green_quipper_2013}:
\begin{enumerate}
    \item Compile-time classical operations;
    \item Classical operations that run on qubits;
    \item Classical operations that run on a classical processor during qubit lifetime; and
    \item Classical operations that run on a classical processor outside qubit lifetime.
\end{enumerate}
Category 1 is useful for metaprogramming and is handled by the preprocessing step of expanding metaQwerty to Qwerty (Section~\ref{sec:meta}).
This section describes how Qwerty handles on categories 2, 3, and 4: classical
functions, mid-quantum branching, and Python integration, respectively.

\subsection{Classical Functions}\label{sec:classical-embed}
Quantum algorithms that solve classical problems usually require running
classical code on qubits. The most famous example is perhaps the oracle in
Grover's algorithm, which marks desired states in an unstructured search. With
some notable
exceptions~\cite{green_quipper_2013,abhari_scaffold_2012,seidel_qrisp_2024,de_muelenaere_qgat_2024,bichsel_silq_2020,paradis_unqomp_2021},
traditional quantum programming languages require programmers to express
classical oracles as quantum circuits or by calling into libraries that produce
circuits. Instead, Qwerty allows programmers to write classical logic directly
in a subset of Python, as shown on
lines~\lngroveroraclestart{}-\lngroveroracleend{} of Fig.~\ref{fig:grover}.
The most novel aspect of classical functions in Qwerty is the programmer's
choice of three \textit{embeddings} of a classical function \lstinline!f!:
either \lstinline!f.sign!, \lstinline!f.xor!, or \lstinline!f.inplace!, discussed in Sections~\ref{sec:bv}, \ref{sec:period}, and \ref{sec:shor}, respectively.

\subsection{Mid-Quantum Branching}
Algorithms such as quantum teleportation (Fig.~\ref{fig:teleport}) make use of
classical branching during qubit lifetime. Although circuit-oriented
descriptions may call such branching ``classical controls,'' Qwerty includes
classical conditionals instead, which are more general (and more familiar to
classical programmers). Lines~\lnteleportcondone{} and \lnteleportcondtwostart{}-\lnteleportcondtwoend{} of
Fig.~\ref{fig:teleport} are examples of classical conditional expressions
written in Qwerty. (This syntax is identical to Python conditional expression
syntax.)

\begin{figure}
\centering
\begin{lstlisting}[name=teleport,style=nums,xleftmargin=5mm,linewidth=0.98\linewidth]
@qpu
def teleport(payload: qubit) -> qubit:
  alice, bob = '00' + '11'

  bit_flip, sign_flip = (
    alice * payload | (flip if '_1' else id)
                    | (std * pm).measure)

  teleported_payload = (
    bob | (flip if bit_flip else id)
        | ('1' >> -'1'
           if sign_flip else id))

  return teleported_payload
\end{lstlisting}
\caption{Quantum teleportation written in Qwerty~\cite{mermin_quantum_2007,nielsen_quantum_2010}}\label{fig:teleport}
\end{figure}

\subsection{Python Integration}
Qwerty \textit{kernels}, functions annotated with \lstinline!@qpu! that contain
Qwerty code and run on a quantum accelerator, may be treated as if they were
Python functions
(e.g., the call on line~\lnqpekernelcall{} of Fig.~\ref{fig:qpe}).
Classical pre- or post-processing may be written in
Python and make use of third-party Python libraries (e.g.,
\texttt{numpy}~\cite{harris2020array}) or Qwerty runtime libraries for
continued fractions (\lstinline!cfrac!) or registers of classical bits
(\lstinline!bit[N]!).

\section{Qwerty Examples}\label{sec:ex}
Now that the basics of Qwerty are defined, this section will explain three
Qwerty programs in detail, elaborating on Qwerty features in the process.
\begin{codeannotations}{teleport}%
\codeannotation[bottom right]{cond}{(52pt,-2pt)}{(180pt,6pt)}{1}{(11pt,-5pt)}{Classical \\ \small conditional}
\end{codeannotations}%

\begin{figure}
\centering
\begin{lstlisting}[name=bv,style=nums,xleftmargin=5mm,linewidth=0.98\linewidth]
from qwerty import *

def bv(secret_str):
  @classical[[N]]
  def f(x: bit[N]) -> bit:
    return (secret_str & x).xor_reduce()

  @qpu[[N]]
  def kernel() -> bit[N]:
    return ('p'**N | f.sign
                   | pm**N >> std**N
                   | measure**N)

  return kernel()

secret_str = bit[4](0b1101)
print(bv(secret_str))
\end{lstlisting}
\caption{The Bernstein--Vazirani algorithm written in Qwerty~\cite{mermin_quantum_2007,fallek_transport_2016,bernstein_quantum_1997}}\label{fig:bv}
\end{figure}

\begin{figure}
\centering
\subfloat[Tracing through the Qwerty pipeline on lines~\lnbvpipestart{}-\lnbvpipeend{} of Fig.~\ref{fig:bv} using the four-bit example secret string $1101$ defined on line~\lnbvsecretdef{} of Fig.~\ref{fig:bv}]{%
\includegraphics[width=0.75\linewidth]{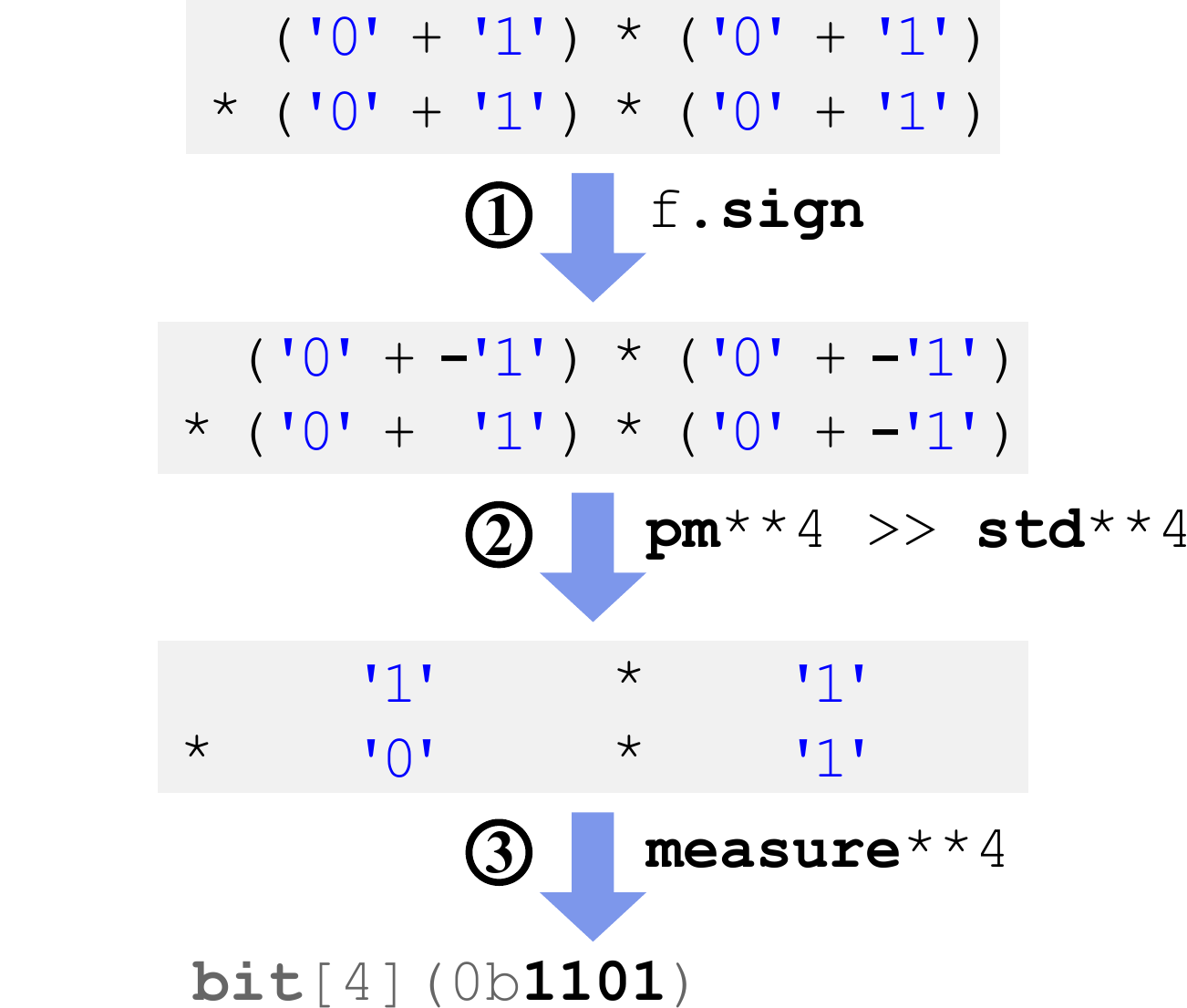}
\label{fig:bv-explain:trace}}

\subfloat[The key insight behind the Bernstein--Vazirani algorithm, which justifies step \circleit{1} in Fig.~\ref{fig:bv-explain:trace}]{%
\includegraphics[width=\linewidth]{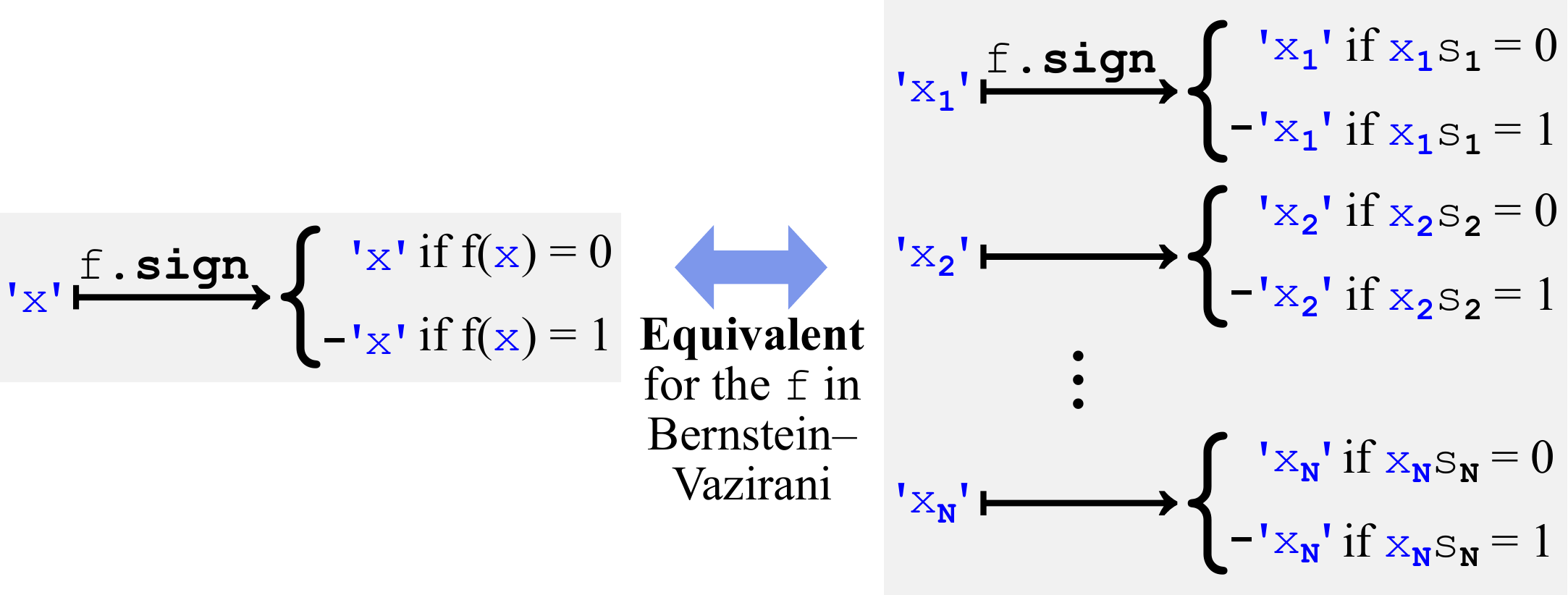}
\label{fig:bv-explain:insight}}
\caption{Explanation of the Bernstein--Vazirani code in Fig.~\ref{fig:bv} using Qwerty syntax}\label{fig:bv-explain}
\end{figure}

\subsection{Bernstein--Vazirani}\label{sec:bv}
We begin with the Bernstein--Vazirani algorithm because its simplicity makes it a popular introductory
algorithm, even if it achieves only a modest speedup~\cite{asfaw_building_2022,wootton_teaching_2021,mermin_quantum_2007,rieffel_quantum_2014}.
The Bernstein--Vazirani algorithm finds an $N$-bit secret string $s$ using a
single quantum invocation of a black-box classical function $f(x) = x_1s_1
\oplus x_2s_2 \oplus \cdots \oplus x_Ns_N$. (Here, $\oplus$ denotes XOR, and
$x_is_i$ denotes ANDing $x_i$ with $s_i$.) For comparison, a classical
algorithm would need $N$ invocations of the black box to determine $s$: first
$f(100\cdots 0)$ to get $s_1$, then $f(010\cdots 0)$ to get $s_2$,
and so on~\cite{mermin_quantum_2007,fallek_transport_2016,bernstein_quantum_1997}.

Fig.~\ref{fig:bv} shows Bernstein--Vazirani implemented in Qwerty.
Line~\lnbvwrapperfunc{} of Fig.~\ref{fig:bv} defines a Python wrapper function \lstinline!bv()!, a
practice common in Qwerty programs.
Such wrapper functions contain any classical pre- or
post-processing as well as the (nested) definitions of any classical functions
or quantum kernels --- in this case, \lstinline!f! and \lstinline!kernel!,
respectively, on lines~\lnbvfdef{} and \lnbvkerneldef{} of Fig.~\ref{fig:bv}.
The function \lstinline!f! is implemented as classical Python code:
the \lstinline!&! operator on line~\lnbvand{} of Fig.~\ref{fig:bv} is typical C
syntax for a bitwise AND, and the \lstinline!x.xor_reduce()! syntax represents
XORing all constituent bits of a classical register \lstinline!x! together into a single result bit, identically to \lstinline!^x! in Verilog. (Similar reduction operations exist for other Boolean operations, e.g., \lstinline!x.and_reduce()!.)

Fig.~\ref{fig:bv-explain:trace} uses Qwerty syntax to trace through the Qwerty pipeline on
lines~\lnbvpipestart{}-\lnbvpipeend{} of Fig.~\ref{fig:bv}.
The crucial step is \circleit{1}, in which applying \lstinline!f.sign!
effectively reveals the answer, albeit in the \lstinline!'p'!/\lstinline!'m'!
basis (hence the later basis translation from \lstinline!pm! to
\lstinline!std!).
The left side of Fig.~\ref{fig:bv-explain:insight} shows the definition of \lstinline!f.sign! for any classical function \lstinline!f!. This embedding of a black-box oracle is common in textbook quantum algorithms, including Grover's algorithm (line~\lngroversign{} of Fig.~\ref{fig:grover}).
The promised structure of the Bernstein--Vazirani black-box function $f(x)$
allows the behavior of \lstinline!f.sign! to be understood as introducing a negative sign to individual bits of the input, as shown on the right side of
Fig.~\ref{fig:bv-explain:insight}. Specifically, each input bit $x_i$ is tilted
by 180 degrees if and only if both $x_i$ and $s_i$ are $1$.

The Qwerty Bernstein--Vazirani code in Fig.~\ref{fig:bv} is highly expressive.
First, the \lstinline!f.sign! construct leaves synthesis of a Bennett embedding
of \lstinline!f! and preparation of the $\ket{-}$
ancilla to the compiler, avoiding the need to explain phase kickback to explain
the algorithm~\cite{bennett_timespace_1989,cleve_quantum_1998}. Second, the basis translation
\lstinline!pm**N >> std**N! explicitly specifies the state change in step
\circleit{2} of Fig.~\ref{fig:bv-explain:trace}. A \texttt{for} loop of
Hadamard gates, for example, is more verbose and less explicit ---
since the Hadamard gate is self-adjoint, a programmer may write it to perform either
\lstinline!pm >> std! or \lstinline!std >> pm!.

\begin{figure} \centering
\begin{lstlisting}[name=period,style=nums,xleftmargin=5mm,linewidth=0.98\linewidth]
import math
from fractions import Fraction
from qwerty import *

def period_finding(f):
  @qpu[[N]]
  def kernel():
    return ('p'**N * '0'**N
            | f.xor
            | id**N * discard**N
            | fourier[[N]].measure)

  def shift_binary_point(bits):
    return Fraction(int(bits),
                    2**len(bits))

  frac1 = shift_binary_point(kernel())
  frac2 = shift_binary_point(kernel())
  return math.lcm(frac1.denominator,
                  frac2.denominator)

@classical
def mod4(x: bit[3]) -> bit[3]:
  return x % 4

period = period_finding(mod4)

if period == 4:
  print('Success!')
else:
  print('Period finding failed')
\end{lstlisting}
\caption{Period finding written in Qwerty~\cite{mermin_quantum_2007,rieffel_quantum_2014}}\label{fig:period}
\end{figure}

\begin{figure} \centering
\includegraphics[width=0.92\linewidth]{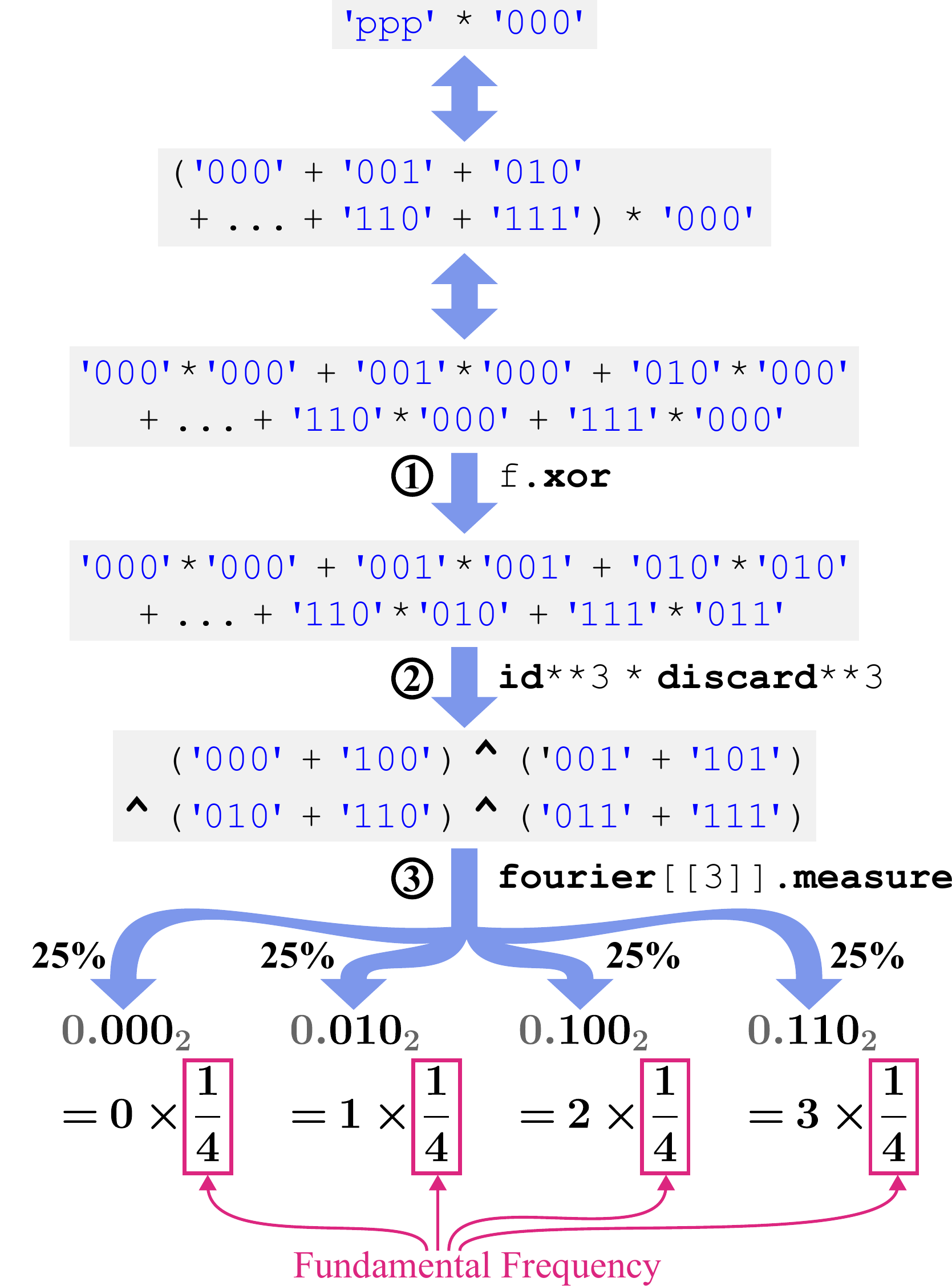}
\caption{Tracing the Qwerty pipeline for period finding on lines~\lnperiodpipestart{}-\lnperiodpipeend{} of Fig.~\ref{fig:period} using Qwerty syntax. (The \lstinline!^! operator written after step \circleit{2} represents an ensemble of possible states as described in Section~\ref{sec:ensemble-lit}. Note also that \lstinline!*! has higher precedence than \lstinline!+!.)}\label{fig:period-trace}
\end{figure}

\subsection{Period Finding}\label{sec:period}
Quantum algorithms involving a black-box implementation of a classical function
$f(x)$ typically reveal some property of $f(x)$. In Bernstein--Vazirani
(Section~\ref{sec:bv}), this property is a secret string used to construct
$f(x)$, an admittedly contrived property. A more useful property of a black box
function $f(x)$ is its \textit{fundamental period}, i.e., the smallest positive
$r$ such that $f(x+r) = f(x)$~\cite{bracewell_fourier_1999}. Remarkably,
there is a quantum algorithm that can determine the period of $f(x)$ with only
$O(1)$ invocations of $f$. The utility of such a period finding algorithm is exemplified by Shor's integer factoring algorithm,
which is built on finding the period of a modular
multiplier~\cite{mermin_quantum_2007,rieffel_quantum_2014,shor_algorithms_1994,shor_polynomial-time_1999}.

Fig.~\ref{fig:period} shows quantum period finding implemented in Qwerty.
Line~\lnperiodblackboxdef{} of Fig.~\ref{fig:period} defines an example classical black box function
named \lstinline!mod4! that returns the input modulo four, giving the example
function a fundamental period of $4$. The Qwerty pipeline that implements the
quantum portion of period finding is written on
lines~\lnperiodpipestart{}-\lnperiodpipeend{} of Fig.~\ref{fig:period}.
Fig.~\ref{fig:period-trace} traces through each step in this pipeline using
Qwerty syntax. The pipeline begins by initializing two registers on
line~\lnperiodpipestart{} of Fig.~\ref{fig:period}: the first an $N$-qubit
register where each qubit is \lstinline!'p'!, and the second an $N$-qubit
register with all qubits initialized to \lstinline!'0'!. As seen at the top of
Fig.~\ref{fig:period-trace}, the first register can be viewed as a
superposition of all possible inputs $x$ to $f(x)$.

Next, line~\lnperiodxor{} of Fig.~\ref{fig:period} (step~\circleit{1} of
Fig.~\ref{fig:period-trace}) creates a superposition of all pairs of inputs and
outputs $(x,f(x))$ by passing the initial superposition state through an
\textit{XOR embedding} of $f$, written with the syntax \lstinline!f.xor!. By
definition, when an XOR embedding is passed two registers in computational
basis states, the first register (the input register) is treated as the input
$x$ and is passed through unchanged, and the result $f(x)$ is XORed into the
second register (the output register)~\cite{bennett_timespace_1989}.
On line~\lnperiodinit{} of Fig.~\ref{fig:period}, the output register is initialized to zero, meaning
the output $f(x)$ is effectively copied into the output register for each of
the respective possible inputs $x$ superposed in the input register.

Line~\lnperioddiscard{} of Fig.~\ref{fig:period} (step~\circleit{2} of
Fig.~\ref{fig:period-trace}) proceeds by measuring the output register and
throwing away the measurement result (which would be an arbitrary choice of an
output of $f(x)$), collapsing the input register to a superposition of all
inputs $x$ that produce the discarded output value $f(x)$. The resulting state
is an ensemble of possible resulting superpositions (for details on the
\lstinline!^! operator, see Section~\ref{sec:ensemble-lit}); the four possible
superpositions shown after step~\circleit{2} in Fig.~\ref{fig:period-trace} are
the pairs of inputs that produce $f(x) = 000_2$, $001_2$, $010_2$, and
$011_2$, respectively.

One can interpret a superposition of computational basis vectors as a signal in
the time domain, whereas period finding warrants analysis in the frequency
domain, which period finding achieves by measuring in the Fourier basis.
Line~\lnperiodmeas{} of Fig.~\ref{fig:period} (step \circleit{3} of
Fig.~\ref{fig:period-trace}), the final quantum step of period finding, can be
understood by imagining the state beforehand (a superposition of $x$ inputs
that produce the same output $f(x)$) as a signal in the time domain. In the
same sense that a time signal is a weighted sum of functions with different frequencies (a discrete
Fourier series)~\cite{smith_scientist_1997,bracewell_fourier_1999}, the superposition of inputs to $f$ that produce the same output is itself a weighted
superposition of Fourier basis states. Measurement in the Fourier basis thus
reveals which Fourier basis states --- frequencies --- are in this
superposition. Because each of these frequencies measured are multiples of the
fundamental frequency, the fundamental frequency can be extracted classically
from sampled measurements. The reciprocal of this fundamental frequency is the
desired fundamental period: for example, $\left(\frac{1}{4}\right)^{-1} = 4$,
the fundamental period of the example function \lstinline!mod4()! defined on
line~\lnperiodblackboxdef{} of Fig.~\ref{fig:period}.

Lines~\lnperiodpoststart{}-\lnperiodpostend{} of Fig.~\ref{fig:period} attempt
to recover the fundamental period of $f$ from the measured frequencies using
Python code. The function \lstinline!shift_binary_point()! interprets a
measurement result $b_1b_2\cdots b_N$ as a fractional binary number
$0.b_1b_2\cdots b_N$ in rational form using the \lstinline!fractions! module in
the Python standard library~\cite{pyfrac}. Line~\lnperiodlcm{} of
Fig.~\ref{fig:period} guesses the denominator of the fundamental frequency
(that is, the fundamental period) by running the quantum kernel twice and
taking the least common multiple of the
denominators~\cite{mermin_quantum_2007,nielsen_quantum_2010}. The hope is that
different measurements will be made and thus different common factors between
numerator and denominator will have canceled out. (More complex post-processing is
needed when the period is not a power of two, as seen in
Section~\ref{sec:shor}, specifically lines~\lnshorpoststart{}-\lnshorpostend{}
of Fig.~\ref{fig:shor}~\cite{cleve_quantum_1998,nielsen_quantum_2010}.)

\begin{figure} \centering
\begin{lstlisting}[name=shor,style=nums,xleftmargin=5mm,linewidth=0.98\linewidth]
import math
from qwerty import *
from qpe import qpe

def order_finding(err_tol, x, modN):
  m = math.ceil(math.log2(modN))
  prec = 2*m + 1 + math.ceil(
    math.log2(2+1/(2*err_tol)))

  @qpu
  def one():
    return '0'**(m-1) * '1'

  @classical[[J]]
  @reversible
  def mult(y: bit[m]) -> bit[m]:
    return x**2**J * y % modN

  op = mult.inplace
  frac1 = qpe(prec, one, op)
  frac2 = qpe(prec, one, op)

  def get_denom(frac):
    cf = cfrac(frac)
    for conv in reversed(cf.convergents()):
      if conv.denominator < modN:
        return conv.denominator

  return math.lcm(get_denom(frac1),
                  get_denom(frac2))

print(order_finding(0.2, 7, 15))
\end{lstlisting}
\caption{Shor's order finding algorithm written in
Qwerty~\cite{cleve_quantum_1998,nielsen_quantum_2010,jacob_shor_conv}}\label{fig:shor}
\end{figure}

\subsection{Shor's Order Finding Algorithm}\label{sec:shor}
The final Qwerty example in this paper implements quantum order finding, the heart of Shor's
celebrated integer factoring algorithm, which promises exponential speedup over classical algorithms for integer factoring~\cite{cleve_quantum_1998,mermin_quantum_2007,nielsen_quantum_2010,rieffel_quantum_2014,shor_algorithms_1994,shor_polynomial-time_1999}.
Since Shor's algorithm is primarily a classical algorithm, this section focuses
on its sole quantum component, the quantum order finding subroutine.
(We omit the remaining classical steps since they could be implemented
in pure Python.)

The order finding algorithm, expressed as Qwerty in Fig.~\ref{fig:shor},
attempts to find the multiplicative order of $x$ mod $N$, that is, the smallest
positive integer $r$ such that $x^r \equiv 1 \pmod N$.
Although an implementation based on period finding (Section~\ref{sec:period}) is possible~\cite{mermin_quantum_2007,rieffel_quantum_2014,shor_algorithms_1994,shor_polynomial-time_1999},
the formulation of order finding
in this section
uses quantum phase estimation to determine how much a multiplier
tilts a particular family of states; the superposition of these states is in
fact equivalent to \lstinline!'1'! as initialized on line~\lnshorinitone{} of
Fig.~\ref{fig:shor}~\cite{cleve_quantum_1998,nielsen_quantum_2010}. (The calculation for the
precision of the angle estimate from QPE written on lines
\lnshorprecstart{}-\lnshorprecend{} of Fig.~\ref{fig:shor} is taken directly
from Nielsen and Chuang~\cite{nielsen_quantum_2010}.)

The multiplier passed
to QPE acts on the state \textit{in-place}, that is, piping \lstinline!'y'!
into the multiplier yields \lstinline!'z'! where $z = xy \pmod N$ for some coprime constants $x$ and $N$. This is possible on a quantum computer because this
multiplier is a reversible operation.
The \lstinline!mult.inplace! expression
on line~\lnshorinplace{} of Fig.~\ref{fig:shor} produces this in-place
embedding of \lstinline!mult! that acts on \lstinline!m! qubits. This
embedding requires the \lstinline!@reversible! annotation on
line~\lnshorreversible{} of Fig.~\ref{fig:shor}, which marks the classical
function \lstinline!mult! (line~\lnshormult{} of Fig.~\ref{fig:shor}) as
reversible, triggering specialized circuit synthesis for reversible classical
code.

The full details of the classical post-processing in the order finding
subroutine are beyond the scope of this
paper~\cite{cleve_quantum_1998,nielsen_quantum_2010,jacob_shor_conv}, but we
highlight that lines~\lnshorcfrac{} and \lnshorconv{} of Fig.~\ref{fig:shor}
make use of the continued fraction functionality included in the Qwerty
runtime for Python. The overall conclusion to draw from the implementation of order finding in
Qwerty (Fig.~\ref{fig:shor}) is that quantum--classical algorithms are
comfortable to express in Qwerty no matter what their mix of classical versus
quantum operations may be, even if different sorts of classical computation are
involved (see Section~\ref{sec:classical}).

\section{Related Work}\label{sec:relwork}
Prior quantum programming languages are mostly rooted in quantum circuit--level
programming~\cite{qcl,green_quipper_2013,ross_algebraic_2015,abhari_scaffold_2012,litteken_updated_2020,javadi-abhari_quantum_2024,steiger_projectq_2018,cirq,wecker_liqui_2014,svore_q_2018,lubinski_advancing_2022,diaz-caro_realizability_2019,bergholm_pennylane_2022,ittah_catalyst_2024,mintz_qcor_2020,mccaskey_extending_2021,cross_openqasm_2022,paykin_qwire_2017,amy_sized_2019,bichsel_silq_2020,paradis_unqomp_2021,seidel_qrisp_2024,voichick_qunity_2023,yuan_twist_2022,li_verified_2022,duncan_introducing_2024,koch_guppy_2024,vax_qmod_2025}.
Exceptions exist to the prevalence of gates in quantum programming languages,
though, such as QML~\cite{altenkirch_functional_2005},
Quantum$\Pi$~\cite{carette_quantum_2023}, QPPL~\cite{inoue_quantum_2024},
Q is for Quantum~\cite{rudolph_q_2017},
quAPL~\cite{nunez-corrales_quapl_2023},
CQ~\cite{binkowski_cq_2025},
Qualtran~\cite{harrigan_expressing_2024}, Tower~\cite{yuan_tower_2022},
\textsc{Oqimp}~\cite{li_verified_2022}, Qrisp~\cite{seidel_qrisp_2024},
Neko~\cite{pinto_neko_2023},
NchooseK~\cite{khetawat_implementing_2019,wilson_mapping_2021,wilson_combining_2022},
QGAT~\cite{de_muelenaere_qgat_2024}, Aleph~\cite{aleph},
and Qutes~\cite{faro_qutes_2025}. The theoretical work QML and Quantum$\Pi$ explore
novel quantum language designs, but example algorithms implemented in both
papers are expressed as
gates~\cite{altenkirch_functional_2005,carette_quantum_2023}. The QPPL language
views quantum programming as a generalization of
probabilistic programming, but writing programs that impart phases other than $-1$
and $1$ may be nontrivial. quAPL is an ambitious, deeply expressive language
embedded in APL, but its state evolution primitives are still gates.
CQ has clever primitives for preparing a superposition of computational basis states that satisfy some classical predicate, but a programmer cannot trace a CQ program in CQ syntax.
Q is for Quantum is a remarkably approachable idea for explaining quantum
computing concepts, but its lack of
generality makes it unsound~\cite{economou_teaching_2020}, and it has not been
realized as a programming language. Qualtran can
express large-scale algorithms as compositions of quantum building blocks, but
ultimately its primitives are also gates. (Q\#~\cite{svore_q_2018} and
PennyLane~\cite{bergholm_pennylane_2022} have similarly rich libraries of
pre-written circuits yet gates also remain their core primitives.)

The
gate-free constructions in the remaining languages are designed for particular
purposes: Tower focuses on a reversible version of C useful for manipulating pointer-based data structures; \textsc{Oqimp} is designed
specifically for synthesizing oracles (Quipper~\cite{green_quipper_2013} and
Scaffold~\cite{abhari_scaffold_2012} feature similar functionality); Qrisp
has expressive quantum data types and can efficiently generate circuits for classical logic, but quantum operations not provided by its standard library require circuit-level programming; Neko requires that the algorithm implemented is
compatible with a map-filter-reduce structure; NchooseK has elegant primitives, but it is designed specifically for constraint satisfaction problems; QGAT provides a high level of abstraction
specifically for a particular oracle--Fourier transform program structure;
Aleph is impressively beginner-friendly but synthesizes only amplitude
amplification and thus cannot achieve exponential speedup for factoring, for
instance; and Qutes includes constructs similar to Qwerty's qubit literals and
superposition literals but relies on gates for state evolution.

\section{Conclusion}
Qwerty is focused on algorithms for quantum information processing as opposed
to quantum physical system modeling (e.g., Hamiltonian
simulation)~\cite{manin_radio,feynman_simulating_1982}.
Qwerty also primarily targets future fault-tolerant hardware, where fidelity is
less dependent on hand-tuning low-level circuitry.

Our goal in creating Qwerty is to help non-experts reason about quantum
computation while abstracting away the complexity of gate
engineering. Qwerty achieves this through
abstractions such as qubit literals based on a string analogy,
batteries-included quantum embeddings of classical functions, the intuitive
superposition operator, rich basis pattern matching, and the basis translation.
These constructs provide a programmer with a rich suite of primitives to
realize algorithms as code. Embedding Qwerty in Python achieves both
approachability and convenience for the classical component of
quantum--classical programs. This is demonstrated through expressive Qwerty
implementations of notable quantum algorithms such as Grover's and period finding.

\section*{Acknowledgment}
The authors thank Elton Pinto and Eugene Dumitrescu for helpful discussions on
quantum programming language design~\cite{dumitrescu_integrating_2023}. We also thank Pulkit Gupta for helpful
discussions on the draft. We acknowledge support for this work from NSF
planning grant \#2016666, ``Enabling Quantum Computer Science and
Engineering'' and through the ORNL STAQCS project. Support for this work also
came from the U.S. Department of Energy, Office of Science, Advanced Scientific
Computing Research Accelerated Research in Quantum Computing Program under
field work proposal ERKJ332. This research was supported in part through
research infrastructure and services provided by the Rogues Gallery
testbed~\cite{powell_wrangling_2019,young_experimental_2019} hosted by the
Center for Research into Novel Computing Hierarchies (CRNCH) at Georgia Tech.
The Rogues Gallery testbed is primarily supported by the NSF under NSF Award
Number \#2016701. Any opinions, findings and conclusions, or recommendations
expressed in this material are those of the author(s), and do not necessarily
reflect those of the NSF.

\appendices
\def\thesubsection{\thesection.\arabic{subsection}}
\def\thesubsectiondis{\thesection.\arabic{subsection}.}

\section{\formname{} Language}\label{app:form}

While the main text motivates the Qwerty language and describes some example
programs, it does not define Qwerty rigorously.
To address this gap,
this appendix defines \textit{\formname{}}, a formalized subset of Qwerty, including its syntax (Section~\ref{sec:mini-syntax}), type system (Section~\ref{sec:mini-types}), semantics (Section~\ref{sec:mini-semantics}), and soundness (Section~\ref{sec:mini-props}).
The formalization presented draws from Selinger and Valiron's quantum
$\lambda$-calculus~\cite{selinger_lambda_2006}, the $\mu$Q language defined by
Yuan et al.~\cite{yuan_twist_2022}, the formalization of Q\# by Singhal et
al.~\cite{singhal_q_2022}, Pierce's simply typed lambda
calculus~\cite{pierce_types_2002}, and Wadler's linear lambda
calculus~\cite{wadler_is_1991}.

\begin{figure}[b]\centering
$\textstyle
\begin{alignedat}{2}
&\desc{(Types)} &\  \typevar{} {}::={}&
    \unittype{}
    \alt \regvar\broadcasttok{\nvar}
    \alt \arrowtok{\typevar_1}{\funckindvar}{\typevar_2} \\
&\desc{(Reg. Kind)} &\  \regvar{} {}::={}&
    \bittype{}
    \alt \qubittype{}
    \alt \basistype{} \\
&\desc{(Func. Kind)} &\  \funckindvar{} {}::={}&
    \revtok{}
    \alt \irrevtok{} \\
&\desc{(Terms)} &\  \termvar{} {}::={}&
    \exprvar{}
    \alt \basisvar{} \\
&\desc{(Expr.)} &\  \exprvar{} {}::={}&
    \varvar{}
    \alt \qlitvar{}
    \alt \zerotok{}
    \alt \onetok{}
    \alt \unittok{}
    \alt \qrefvar{\indexvar}
    \alt \bitensortok{\exprvar_1}{\exprvar_2}
    \\ &&\mid&\
    \exprvar_1 \pipetok{} \exprvar_2
    \alt \measuretok{\basisvar}
    \alt \discardtok{}
    \\ &&\mid&\
    \unpacktok{\mvar}{\exprvar_1}{\exprvar_2}
    \\ &&\mid&\
    \transtok{\basisvar_1}{\basisvar_2}
    \alt \lambdatok{\varvar}{\typevar}{\exprvar}
    \alt \adjtok{\exprvar{}}
    \\ &&\mid&\
    \ifelsetok{\exprvar_1}{\basisvar}{\exprvar_2}
    \\
&\desc{(Qubit Lit.)} &\  \qlitvar{} {}::={}&
    \unittok{}
    \alt \qavar
    \alt \bitensortok{\qlitvar_1}{\qlitvar_2} \\
&\desc{(Qubit Atom)} &\  \qavar {}::={}&
    \qlittok{0}
    \alt \qlittok{1}
    \alt \tilttok{\qlitvar}{\tiltvar}
    \alt \superpostok{\qlitvar_1}{\qlitvar_2}
\end{alignedat}$
\caption{Mini-Qwerty types and expressions}
\label{fig:mini-syntax-expr}
\end{figure}

\begin{figure}\centering
$\textstyle
\begin{alignedat}{2}
&\desc{(Value)} &\  \valvar{} {}::={}&
    \unittok{}
    \alt \propervalvar{} \\
&\desc{(Proper Val.)} &\  \propervalvar{} {}::={}&
    \regelemvar{}
    \alt \funcvalvar{}
    \alt \bitensortok{\propervalvar_1}{\propervalvar_2} \\
&\desc{(Func. Val.)} &\  \funcvalvar{} {}::={}&
    \revfuncvalvar{}
    \alt \lambdatok{\varvar}{\typevar}{\exprvar}
    \alt \measuretok{\basisvar}
    \\ &&\mid&\
    \discardtok{} \\
&\desc{(Rev. Func. Val.)} &\  \revfuncvalvar{} {}::={}&
    \adjtok{\revfuncvalvar{}}
    \alt \transtok{\basisvar_1}{\basisvar_2}
    \alt \ifelsetok{\valvar_1}{\basisvar}{\valvar_2} \\
&\desc{(Reg. Elem.)} &\  \regelemvar{} {}::={}&
    \zerotok{}
    \alt \onetok{}
    \alt \qrefvar{\indexvar}
\end{alignedat}$
\caption{Mini-Qwerty values}
\label{fig:mini-syntax-values}
\end{figure}

\begin{figure}\centering
$\textstyle
\begin{alignedat}{2}
&\desc{(Basis)} &\  \basisvar{} {}::={}&
    \basislittok{\mvar{}}
    \alt \bitensortok{\basisvar_1}{\basisvar_2}
    \alt \unittok{} \\
&\desc{(Basis Vec.)} &\  \bvvar{} {}::={}&
    \unittok{}
    \alt \vavar{}
    \alt \tilttok{\bvvar}{\tiltvar}
    \alt \superpostok{\bvvar_1}{\bvvar_2}
    \\ &&\mid&\
    \bitensortok{\bvvar_1}{\bvvar_2} \\
&\desc{(Vec. Atom)} &\  \vavar{} {}::={}&
    \qlittok{0}
    \alt \qlittok{1}
    \alt \qlittok{?}
    \alt \qlittok{\_}
\end{alignedat}$
\caption{Mini-Qwerty syntax for bases}
\label{fig:mini-syntax-bases}
\end{figure}

The following definitions are used throughout this appendix.
The variables $\nvar$ and $\mvar$ are integers, where $\nvar{} \ge 0$ and $\mvar{} > 0$. Real numbers are represented by $\tiltvar{}$.
We write $A \qctxunion B$ to denote the union of disjoint sets $A$ and $B$~\cite{yuan_twist_2022}.
We assume $\rawpermutation$ is a permutation, and $\permutationof{i}$ applies the permutation $\rawpermutation$ to $i$~\cite{keller_applied_2017}.
Writing $\intbit{j}{m}{y}$ means the $j$th bit of the $m$-bit binary representation of $y$, i.e., $y = (\intbit{1}{m}{y}\intbit{2}{m}{y}\cdots\intbit{m}{m}{y})_2$.
The state space of one qubit, i.e., $\vspan(\ket{0},\ket{1})$, is written as $\hilbertspace{}$.
The notation $(\cdot)_{i=1}^{\nvar}$ denotes a comma-separated list of length $\nvar$.
The empty list is written as $\emptyveclist$, a dot.
The number of elements in a list $\ell$ is written as $\vert \ell \vert$.
If a list $\ell$ contains only integers, then $\ell^{+\nvar}$ denotes the list
where each entry is incremented by $\nvar$, i.e.,
$\ell_1{+}\nvar{},\ell_2{+}\nvar{},\ldots,\ell_\mvar{+}\nvar{}$.
We define the tensor product of vector lists $\left(\ket{\psi_i}\right)_{i=1}^{\nvar_1} \otimes \left(\ket{\phi_j}\right)_{j=1}^{\nvar_2}$ as $\left(\ket{\psi_i}\otimes\ket{\phi_j}\right)\!{\substack{i=1,2,\ldots,\nvar_1 \\ j=1,2,\ldots,\nvar_2}}$, that is,
\begin{align*}
    &\ket{\psi_1}\otimes\ket{\phi_1},\ket{\psi_1}\otimes\ket{\phi_2},\ldots,\ket{\psi_1}\otimes\ket{\phi_{\nvar_2}},\ldots, \\
    &\ket{\psi_2}\otimes\ket{\phi_1},\ket{\psi_2}\otimes\ket{\phi_2},\ldots,\ket{\psi_2}\otimes\ket{\phi_{\nvar_2}},\ldots, \\
    &\ldots, \\
    &\ket{\psi_{\nvar_1}}\otimes\ket{\phi_1},\ket{\psi_{\nvar_1}}\otimes\ket{\phi_2},\ldots,\ket{\psi_{\nvar_1}}\otimes\ket{\phi_{\nvar_2}}\!.
\end{align*}
For convenience, $\ket{}$ is defined such that $\braket{}{} = 1$ and
$\ket{\psi}\otimes\ket{} = \ket{}\otimes\ket{\psi} = \ket{\psi}$.
(One can imagine $\ket{}$ as the $1{\times}1$ matrix $\left[\:1\:\right]$.)
For any vector space $V$, we define $V^{\otimes 0}$ as $\{\ket{}\}$.

\subsection{Mini-Qwerty Syntax}\label{sec:mini-syntax}
We begin by defining types, syntax, and values in \formname{}.
Fig.~\ref{fig:mini-syntax-expr} shows \formname{} types.
The $\broadcasted{\regvar}{\nvar}$ type is a \textit{register}
of qubits or bits.
(For simplicity, the basis type is also represented as a register type.)
The distinction between the function types
$\arrowtok{\typevar_1}{\irrevtok}{\typevar_2}$ and
$\arrowtok{\typevar_1}{\revtok}{\typevar_2}$ exists because some functions
(e.g., measurement) are not reversible and thus cannot be adjointed or
predicated, whereas other functions (e.g., basis translations) are
reversible~\cite{singhal_q_2022}.
As in prior work~\cite{wadler_is_1991,paykin_qwire_2017,yuan_twist_2022},
\formname{} has both linear types and nonlinear types. In particular,
functions, bit registers, and $\unittype$ are nonlinear, but nonempty registers of
qubits ($\broadcasted{\qubittype}{\mvar}$) are linear.

Fig.~\ref{fig:mini-syntax-expr} also defines \formname{} terms. Terms are
either expressions $\exprvar$, which are reduced to values by execution, or
bases $\basisvar$, which syntactically encode a constant basis (described below).
Variables are shown as $\varvar$ as in typical lambda calculus.
The unit literal is written $\unittok$ since $\unittype$s function as empty
registers in \formname{}.
The syntax $\qrefvar{\indexvar}$ is a reference to the qubit at index
$\indexvar$ of quantum memory and is intended to be included only in
intermediate computations~\cite{yuan_twist_2022}.
To introduce quantum registers,
programmers write qubit literals $\qlitvar{}$ instead. Qubit literals are constructed as tensor products of \textit{qubit atoms} ($\qavar$), the smallest syntactic units describing quantum states to prepare. For example, $\bitensortok{\qlittok{0}}{\qlittok{1}}$ is built with two qubit atoms, $\qlittok{0}$ and $\qlittok{1}$.
(As in Qwerty, the operator $\rawbitensortok{}$ is a tensor product
and can concatenate both registers and functions.) The $\zerotok{}$ and $\onetok{}$ bit constants
 act similarly to qubit literals except
that they introduce $\bittype$ registers instead of $\qubittype$
registers. The \texttt{\textbf{let}} expression unpacks a register,
allowing programmers to operate on individual qubits or bits.
(In contrast to \formname{}, Qwerty omits the \texttt{\textbf{let}} and \texttt{\textbf{in}} keywords to align with Python syntax.)

Whereas the lambda calculus requires a function call (an application) to be written as $\exprvar_2 \exprvar_1$, \formname{} expects $\exprvar_1 \pipetok \exprvar_2$ to align with Qwerty.
The pipe operator is written as ${\pipetok}$ rather than as the literal pipe character $\texttt{\textbf{|}}$ used in Qwerty to avoid readers confusing it with the the ``or'' symbol $\vert$ used to construct the grammar in Fig.~\ref{fig:mini-syntax-expr}.
Programmers may define custom functions with
$\lambdatok{\varvar}{\typevar}{\exprvar}$, but \formname{} contains built-in functions as well.
These built in functions correspond exactly to operations described in the main text: $\discardtok{}$ in Section~\ref{sec:qubit-type}; $\transtok{\basisvar_1}{\basisvar_2}$ in Section~\ref{sec:evol}; $\measuretok{\basisvar}$ in Section~\ref{sec:meas};  and $\ifelsetok{\exprvar_1}{\basisvar}{\exprvar_1}$ in Section~\ref{sec:pred}. The syntax $\adjtok{\exprvar}$ yields the adjoint of a function $\exprvar$. (This syntax also exists in Qwerty, although the main text omits discussion of it in the interest of space.)

Fig.~\ref{fig:mini-syntax-values} defines the subset of \formname{} expressions that are considered values, i.e., results of evaluation. The
distinction between a value ($\valvar$) and a \textit{proper value}
($\propervalvar$) exists because $\unittok{}$ represents an empty tensor
product, so it does not make sense to include it in a fully-evaluated tensor product. (For example, the unit literal $\unittok{}$ in the expression $\bitensortok{\qlittok{0}}{\unittok{}}$ only serves to clutter the expression and should be removed by reduction.) \textit{Register elements} ($\regelemvar$) represent individual elements of a register
value. For example, $\bitensortok{\qrefvar{2}}{\qrefvar{7}}$ is a qubit
register value holding two qubit register elements with references to qubits at
indices 2 and 7, respectively.

The syntax for \formname{} expressions $\exprvar$ is largely orthogonal
to the syntax for \formname{} bases $\basisvar$ shown in Fig.~\ref{fig:mini-syntax-bases}.
The \formname{} basis literal syntax $\basislittok{\mvar}$ is identical to the Qwerty basis literal syntax discussed in Section~\ref{sec:blit}.
\textit{Basis vector atoms} ($\vavar$) are similar to the qubit atoms mentioned earlier except for two differences. First, each $\vavar$ corresponds to only one qubit. Second, in addition to $\qlittok{0}$ and $\qlittok{1}$, a $\vavar$ may also be the padding atom $\qlittok{?}$ and the target atom $\qlittok{\_}$, both described in Section~\ref{sec:pred}.

\newcommand{\rulename}[1]{{\textsc{\footnotesize#1}}}
\newcommand{\ruledef}[3]{\infer[\rulename{#1}]{#3}{#2}}
\newcommand{\rulegap}{\quad}
\newcommand{\rulebrk}{\\[0.7em]}
\newcommand{\evalrulebrk}{\\[0.7em]}
\newcommand{\assumgap}{\quad}
\newcommand{\assumbrk}{\\[-0.3em]}

\subsection{Mini-Qwerty Type System}\label{sec:mini-types}

\begin{figure}
    \centering$\textstyle
    \begin{gathered}
        \ruledef{T-Var}
                {}
                {\typerel{\typectxbind{\varvar}{\typevar}}{\emptyqctx}{\varvar}{\typevar}}
        \rulegap
        \ruledef{T-Exch}
                {\typerel{\tyctx_1,\typectxbind{\varvar_1}{\typevar_1},\typectxbind{\varvar_2}{\typevar_2},\tyctx_2}
                         {\qctx}
                         {\termvar}
                         {\typevar'}}
                {\typerel{\tyctx_1,\typectxbind{\varvar_2}{\typevar_2},\typectxbind{\varvar_1}{\typevar_1},\tyctx_2}
                         {\qctx}
                         {\termvar}
                         {\typevar'}}
        \rulebrk{}
        \ruledef{T-Contract}
                {\typerel{\tyctx,\typectxbind{\varvar_1}{\typevar},\typectxbind{\varvar_2}{\typevar}}{\qctx}{\termvar}{\typevar'}
                 \assumgap{}
                 \typevar \neq \broadcasted{\qubittype}{\mvar}}
                {\typerel{\tyctx,\typectxbind{\varvar_1}{\typevar}}{\qctx}{\substitute{\varvar_2}{\varvar_1}{\termvar}}{\typevar'}}
        \rulebrk{}
        \ruledef{T-Weaken}
                {\typerel{\tyctx}{\qctx}{\termvar}{\typevar}
                 \assumgap{}
                 \typevar \neq \broadcasted{\qubittype}{\mvar}}
                {\typerel{\tyctx,\typectxbind{\varvar}{\typevar'}}{\qctx}{\termvar}{\typevar}}
    \end{gathered}
    $
\caption{\formname{} structural rules for linear typing}
\label{fig:mini-types-linear}
\end{figure}

Now that \formname{} syntax is defined, this section introduces the \formname{} type system.
The typing relation used is $\typerel{\tyctx}{\qctx}{\termvar}{\typevar}$,
where $\tyctx$ is an ordered list of type bindings
$\typectxbind{\varvar}{\typevar}$ in which all variables $\varvar$ are distinct; $\qctx$ is a
set holding all valid qubit indices; $\termvar$ is a term; and $\typevar$ is a
type~\cite{yuan_twist_2022}. Linear typing is achieved with the structural
rules in Fig.~\ref{fig:mini-types-linear}. \rulename{T-Contract} allows
duplicating nonlinear values, and \rulename{T-Weaken} allows discarding
nonlinear values~\cite{wadler_is_1991}. The former rule uses the notation
$\rangesubstitute{\varvar_i}{\termvar_i}{i=1}{\mvar}{\termvar'}$,
which simultaneously replaces every $\varvar_i$ found in $\termvar'$ with a
respective new term $\termvar_i$. (It is assumed that all $x_i$ are distinct.)

\begin{figure}
    \centering$\textstyle
    \begin{gathered}
        \ruledef{T-B0}
                {}
                {\typerel{\emptytyctx}{\emptyqctx}{\zerotok}{\broadcasted{\bittype}{1}}}
        \rulegap{}
        \ruledef{T-B1}
                {}
                {\typerel{\emptytyctx}{\emptyqctx}{\onetok}{\broadcasted{\bittype}{1}}}
        \rulebrk{}
        \ruledef{T-Unit}
                {}
                {\typerel{\emptytyctx}{\emptyqctx}{\unittok}{\unittype}}
        \rulegap{}
        \ruledef{T-Q}
                {}
                {\typerel{\emptytyctx}{\{\indexvar{}\}}{\qrefvar{\indexvar}}{\broadcasted{\qubittype}{1}}}
    \end{gathered}
    $
\caption{\formname{} type rules for values}
\label{fig:mini-types-vals}
\end{figure}

\begin{figure}
    \centering$\textstyle
    \begin{gathered}
        \ruledef{S-Rev}
                {}
                {\subtyperel{\arrowtok{\typevar_1}{\revtok}{\typevar_2}}{\arrowtok{\typevar_1}{\irrevtok}{\typevar_2}}}
        \rulegap{}
        \ruledef{S-Unit}
                {}
                {\subtyperel{\unittype}{\broadcasted{\regvar}{0}}}
        \rulebrk{}
        \ruledef{S-Func}
                {\subtyperel{\typevar_1'}{\typevar_1} \assumgap \subtyperel{\typevar_2}{\typevar_2'}}
                {\subtyperel{\arrowtok{\typevar_1}{\funckindvar}{\typevar_2}}{\arrowtok{\typevar_1'}{\funckindvar}{\typevar_2'}}}
        \rulegap{}
        \ruledef{S-Trans}
                {\subtyperel{\typevar_1}{\typevar_2} \assumgap \subtyperel{\typevar_2}{\typevar_3}}
                {\subtyperel{\typevar_1}{\typevar_3}}
        \rulebrk{}
        \ruledef{S-Refl}
                {}
                {\subtyperel{\typevar}{\typevar}}
        \rulegap{}
        \ruledef{T-Sub}
                {\typerel{\tyctx}{\qctx}{\termvar}{\typevar_1} \assumgap{} \subtyperel{\typevar_1}{\typevar_2}}
                {\typerel{\tyctx}{\qctx}{\termvar}{\typevar_2}}
    \end{gathered}
    $
\caption{\formname{} subtyping rules}
\label{fig:mini-types-subtyping}
\end{figure}

The remaining type rules are logical rules.
Fig.~\ref{fig:mini-types-vals} defines straightforward type rules for values.
Fig.~\ref{fig:mini-types-subtyping} defines subtyping rules that
allow \formname{} programmers to both use reversible functions as irreversible
functions and $\unittok{}$ as an empty basis or empty register. The first two
rules, \rulename{S-Rev} and \rulename{S-Unit}, handle both of these respective
cases.
The \rulename{S-Func} rule makes function types respect both of these cases
too. The remaining three rules in Fig.~\ref{fig:mini-types-subtyping} establish
${\rawsubtyperel}$ as a preorder (reflexive and transitive) and integrate
subtyping into typing judgments~\cite{pierce_types_2002}. In this appendix, we
assume without loss of generality that redundant instances of \rulename{S-Refl}
are not included in typing derivations.

\begin{figure}
    \centering$\textstyle
    \begin{gathered}
        \ruledef{T-Q0}
                {}
                {\typerel{\emptytyctx}{\emptyqctx}{\qlittok{0}}{\broadcasted{\qubittype}{1}}}
        \rulegap{}
        \ruledef{T-Q1}
                {}
                {\typerel{\emptytyctx}{\emptyqctx}{\qlittok{1}}{\broadcasted{\qubittype}{1}}}
        \rulebrk{}
        \ruledef{T-Sup}
                {\begin{gathered}
                 \orthorel{\qlitvar_1}{\qlitvar_2}
                 \assumbrk{}
                 \typerel{\emptytyctx}{\emptyqctx}{\qlitvar_1}{\broadcasted{\qubittype}{\mvar}}
                 \assumgap{} \typerel{\emptytyctx}{\emptyqctx}{\qlitvar_2}{\broadcasted{\qubittype}{\mvar}}
                 \end{gathered}}
                {\typerel{\emptytyctx}{\emptyqctx}{\superpostok{\qlitvar_1}{\qlitvar_2}}{\broadcasted{\qubittype}{\mvar}}}
        \rulebrk{}
        \ruledef{T-Tilt}
                {\typerel{\emptytyctx}{\emptyqctx}{\qlitvar}{\broadcasted{\qubittype}{\mvar}}}
                {\typerel{\emptytyctx}{\emptyqctx}{\tilttok{\qlitvar}{\tiltvar}}{\broadcasted{\qubittype}{\mvar}}}
        \rulebrk{}
        \ruledef{T-BLit}
                {\begin{gathered}
                 \forall_{i \in 1..\mvar_1} \qdim{\bvvar_i} = \mvar_2
                 \assumbrk{}
                 \forall_{i,j \in 1..\mvar_1}^{i \ne j} \orthorel{\bvvar_i}{\bvvar_j}
                 \end{gathered}}
                {\typerel{\emptytyctx}{\emptyqctx}{\basislittok{\mvar_1}}{\broadcasted{\basistype}{\mvar_2}}}
    \end{gathered}
    $
\caption{\formname{} type rules for qubit and basis literals}
\label{fig:mini-types-qblit}
\end{figure}

\begin{figure}
    \centering$\textstyle
    \begin{aligned}
        \vecf{\qlittok{0}} &\defas \ket{0} \\
        \vecf{\qlittok{1}} &\defas \ket{1} \\
        \vecf{\qlittok{?}} &\defas \ket{} \\
        \vecf{\qlittok{\_}} &\defas \ket{} \\
        \vecf{\unittok{}} &\defas \ket{} \\
        \vecf{\bitensortok{\bvvar_1}{\bvvar_2}} &\defas \vecf{\bvvar_1} \otimes \vecf{\bvvar_2} \\
        \vecf{\tilttok{\bvvar}{\tiltvar}} &\defas e^{i(2\pi\cdot\tiltvar/360\degree)}\vecf{\bvvar} \\
        \vecf{\superpostok{\bvvar_1}{\bvvar_2}} &\defas \frac{1}{\sqrt{2}}\left(\vecf{\bvvar_1} + \vecf{\bvvar_2}\right) \\
                                                &\quad\ \ \text{if }\braket{\bvvar_1}{\bvvar_2}\!\!=\!0
    \end{aligned}
    $
\caption{Definition of the states represented by \formname{} basis vectors}
\label{fig:mini-vec}
\end{figure}

\begin{figure}
    \centering$\textstyle
    \begin{aligned}
        \qdim{\qlittok{0}} &\defas 1 \\
        \qdim{\qlittok{1}} &\defas 1 \\
        \qdim{\qlittok{?}} &\defas 1 \\
        \qdim{\qlittok{\_}} &\defas 1 \\
        \qdim{\unittok{}} &\defas 0 \\
        \qdim{\bitensortok{\bvvar_1}{\bvvar_2}} &\defas \qdim{\bvvar_1} + \qdim{\bvvar_2} \\
        \qdim{\tilttok{\bvvar}{\tiltvar}} &\defas \qdim{\bvvar} \\
        \qdim{\superpostok{\bvvar_1}{\bvvar_2}} &\defas \qdim{\bvvar_1} \quad \text{if }\qdim{\bvvar_1} \!=\! \qdim{\bvvar_2}
    \end{aligned}
    $
\caption{Definition of the dimension of a \formname{} basis vector}
\label{fig:mini-qdim-vec}
\end{figure}

Fig.~\ref{fig:mini-types-qblit} shows typing for qubit literals and basis
literals. The check for orthogonality $\orthorel{\qlitvar_1}{\qlitvar_2}$ in
\rulename{T-Sup} ensures that the resulting superposition is a unit vector, and
the similar check in \rulename{T-BLit} ensures that every well-typed basis
literal is an orthonormal basis.
These assumptions make use of the definition of state vectors
$\babyket{\bvvar}$ represented by individual basis vector constants $\bvvar$
shown in Fig.~\ref{fig:mini-vec}. (This definition also applies to qubit
literals because every $\qlitvar$ is a $\bvvar$.)
The \rulename{T-BLit} rule also makes use of the notation $\qdim{\bvvar}$ defined in Fig.~\ref{fig:mini-qdim-vec}, which denotes the number of qubits a basis vector $\bvvar$ describes.

\begin{figure}
    \centering$\textstyle
    \begin{gathered}
        \ruledef{T-Tensor}
                {\typerel{\tyctx_1}{\qctx_1}{\termvar_1}{\broadcasted{\regvar}{\nvar_1}}
                 \assumgap{} \typerel{\tyctx_2}{\qctx_2}{\termvar_2}{\broadcasted{\regvar}{\nvar_2}}}
                {\typerel{\tyctx_1,\tyctx_2}{\qctx_1 \qctxunion \qctx_2}{\bitensortok{\termvar_1}{\termvar_2}}{\broadcasted{\regvar}{\nvar_1 + \nvar_2}}}
        \rulebrk{}
        \ruledef{T-TensF}
                {\begin{gathered}
                 \typerel{\tyctx_1}{\qctx_1}{\exprvar_1}{\arrowtok{\broadcasted{\regvar_1}{\nvar_1}}{\funckindvar}{\broadcasted{\regvar_2}{\nvar_2}}}
                 \assumbrk{}
                 \typerel{\tyctx_2}{\qctx_2}{\exprvar_2}{\arrowtok{\broadcasted{\regvar_1}{\nvar_3}}{\funckindvar}{\broadcasted{\regvar_2}{\nvar_4}}}
                 \end{gathered}}
                {\typerelwrap{\tyctx_1,\tyctx_2}{\qctx_1 \qctxunion \qctx_2}{\bitensortok{\exprvar_1}{\exprvar_2}}{\arrowtok{\broadcasted{\regvar_1}{\nvar_1 + \nvar_3}}{\funckindvar}{\broadcasted{\regvar_2}{\nvar_2 + \nvar_4}}}}
        \rulebrk{}
        \ruledef{T-Unpack}
                {\begin{gathered}
                 \typerel{\tyctx_1}{\qctx_1}{\exprvar_1}{\broadcasted{\regvar}{\mvar}}
                 \assumbrk{}
                 \typerel{\tyctx_2,\typectxbind{\varvar_1}{\broadcasted{\regvar}{1}},\typectxbind{\varvar_2}{\broadcasted{\regvar}{1}},\ldots,\typectxbind{\varvar_{\mvar}}{\broadcasted{\regvar}{1}}}{\qctx_2}{\exprvar_2}{\typevar}
                 \end{gathered}}
                {\typerelaltwrap{\tyctx_1,\tyctx_2}{\qctx_1 \qctxunion \qctx_2}{\unpacktok{\mvar}{\exprvar_1}{\exprvar_2}}{\typevar}}
    \end{gathered}
    $
\caption{\formname{} type rules for tensor products}
\label{fig:mini-types-tensor}
\end{figure}

More complex expressions in \formname{} are built with tensor products;
the rules in Fig.~\ref{fig:mini-types-tensor} define the typing for these tensor
product expressions. \rulename{T-Tensor} allows combining bases and
registers with the tensor product. \rulename{T-Unpack} allows unpacking
registers into individual bits or qubits. \rulename{T-TensF} facilitates
Qwerty's characteristic pipeline-like program structure by allowing tensor
products of functions to behave as functions themselves.

\begin{figure}
    \centering$\textstyle
    \begin{gathered}
        \ruledef{T-BTrans}
                {\begin{gathered}
                 \typerel{\emptytyctx}{\emptyqctx}{\basisvar_1}{\broadcasted{\basistype}{\mvar}}
                 \assumgap{} \typerel{\emptytyctx}{\emptyqctx}{\basisvar_2}{\broadcasted{\basistype}{\mvar}}
                 \assumbrk{}
                 \eqrel{\atompos{\_}{\basisvar_1}}{\emptyindexlist}
                 \assumgap{}
                 \eqrel{\atompos{\_}{\basisvar_2}}{\emptyindexlist}
                 \assumbrk{}
                 \eqrel{\atompos{?}{\basisvar_1}}{\atompos{?}{\basisvar_2}}
                 \assumgap{}
                 \bspan{\basisvar_1} = \bspan{\basisvar_2}
                 \end{gathered}}
                {\typerel{\emptytyctx}{\emptyqctx}{\transtok{\basisvar_1}{\basisvar_2}}{\arrowtok{\broadcasted{\qubittype}{\mvar}}{\revtok}{\broadcasted{\qubittype}{\mvar}}}}
        \rulebrk{}
        \ruledef{T-Meas}
                {\begin{gathered}
                 \typerel{\emptytyctx}{\emptyqctx}{\basisvar}{\broadcasted{\basistype}{\mvar}}
                 \assumgap{}
                 \bspan{\basisvar} = \hilbertspace^{\otimes \mvar}
                 \assumbrk{}
                 \eqrel{\atompos{?}{\basisvar}}{\emptyindexlist}
                 \assumgap{}
                 \eqrel{\atompos{\_}{\basisvar}}{\emptyindexlist}
                 \end{gathered}}
                {\typerel{\emptytyctx}{\emptyqctx}{\measuretok{\basisvar}}{\arrowtok{\broadcasted{\qubittype}{\mvar}}{\irrevtok}{\broadcasted{\bittype}{\mvar}}}}
        \rulebrk{}
        \ruledef{T-Discard}
                {}
                {\typerel{\emptytyctx}{\emptyqctx}{\discardtok}{\arrowtok{\broadcasted{\qubittype}{1}}{\irrevtok}{\unittype}}}
    \end{gathered}
    $
\caption{\formname{} type rules for core quantum primitives}
\label{fig:mini-types-quantum-expr}
\end{figure}

The aforementioned type rules handle initial state preparation and the tensor
product, but primitives for state evolution and measurement are necessary for
useful quantum programs. Typing for these core quantum primitives is shown in
Fig.~\ref{fig:mini-types-quantum-expr}. These core rules make use of the span
of a basis $\basisvar$, defined as $\bspan{\basisvar} \defas
\vspan\left(\vecsf{\basisvar}\right)$, where the list of state vectors
$\vecsf{\basisvar}$ represented by a basis $\basisvar$ is defined in
Fig.~\ref{fig:mini-basis-vecs}.
Because these rules also require knowledge of where single-qubit vector atoms
($\vavar$) reside in larger vectors, Fig.~\ref{fig:mini-atompos} defines
$\atomposany{\vavar}{\basisvar}$ as a list of qubit indices (starting from 1)
where $\vavar$ occurs in the basis $\basisvar$. For example,
$\atomposany{\qlittok{?}}{\basislittokvecs{\bitensortok{\qlittok{1}}{\bitensortok{\qlittok{?}}{\qlittok{?}}}}}
= 2,3$.

\begin{figure}
    \centering$\textstyle
    \begin{aligned}
        \vecsf{\bitensortok{\basisvar_1}{\basisvar_2}} &\defas \vecsf{\basisvar_1} \otimes \vecsf{\basisvar_2} \\
        \vecsf{\unittok} &\defas \emptyveclist \\
        \vecsf{\basislittok{\mvar}} &\defas \vecf{\bvvar_1}\!,\vecf{\bvvar_2}\!,\ldots,\vecf{\bvvar_{\mvar}}
    \end{aligned}
    $
\caption{Definition of the correspondence between \formname{} bases and lists of state vectors}
\label{fig:mini-basis-vecs}
\end{figure}

\begin{figure}
    \centering$\textstyle
    \begin{aligned}
        \atomposany{\vavar}{\unittok} &\defas \emptyindexlist \\
        \atomposany{\vavar}{\vavar} &\defas 1 \\
        \atomposany{\vavar}{\vavar'} &\defas \emptyindexlist \quad\text{if }\vavar \ne \vavar' \\
        \atomposany{\vavar}{\tilttok{\bvvar}{\tiltvar}} &\defas \atomposany{\vavar}{\bvvar} \\
        \atomposany{\vavar}{\superpostok{\bvvar_1}{\bvvar_2}} &\defas \atomposany{\vavar}{\bvvar_1} \\
                            &\quad\ \ \text{if } \atomposany{\vavar}{\bvvar_1}=\atomposany{\vavar}{\bvvar_2} \\
        \atomposany{\vavar}{\bitensortok{\bvvar_1}{\bvvar_2}} &\defas
                            \atomposany{\vavar}{\bvvar_1},\atomposanysup{\vavar}{+\qdim{\bvvar_1}}{\bvvar_2} \\
        \atomposany{\vavar}{\basislittok{\mvar}} &\defas
                            \atomposany{\vavar}{\bvvar_1} \\
                            &\quad\ \ \text{if }  \forall_{i=1}^{\mvar}, \atomposany{\vavar}{\bvvar_1} \!=\! \atomposany{\vavar}{\bvvar_i} \\
        \atomposany{\vavar}{\bitensortok{\basisvar_1}{\basisvar_2}} &\defas
                            \atomposany{\vavar}{\basisvar_1},\atomposanysup{\vavar}{+\qdim{\basisvar_1}}{\basisvar_2}
    \end{aligned}
    $
\caption{Definition of indices of basis vector atoms in \formname{} bases}
\label{fig:mini-atompos}
\end{figure}

Basis translation type checking is
defined by \rulename{T-BTrans}, which bans the predication-specific target atom
$\qlittok{\_}$ (see Section~\ref{sec:pred}) and enforces some consistency
checks between both bases. These checks are that both bases have matching spans
(see Section~\ref{sec:btrans-typecheck}) and have padding $\qlittok{?}$ atoms at the
same positions. The padding check ensures that both bases indicate the same
qubits should be left unmodified.
The \rulename{T-Meas} rule defines typing for the other key quantum primitive,
measurement. The rule checks that measurement is performed in a basis that
spans the full $\mvar$-qubit space and has no target $\qlittok{\_}$ or padding
$\qlittok{?}$ vector atoms. The former is because $\qlittok{\_}$ is specific to
predication, and the latter is because (for simplicity in \formname{})
programmers cannot measure only a subset of qubits as $\qlittok{?}$ would
imply.

\begin{figure}
    \centering$\textstyle
    \begin{gathered}
        \ruledef{T-Pipe}
                {\typerel{\tyctx_1}{\qctx_1}{\exprvar_1}{\typevar_1}
                 \assumgap{} \typerel{\tyctx_2}{\qctx_2}{\exprvar_2}{\arrowtok{\typevar_1}{\funckindvar}{\typevar_2}}}
                {\typerel{\tyctx_1,\tyctx_2}{\qctx_1 \qctxunion \qctx_2}{\exprvar_1 \pipetok \exprvar_2}{\typevar_2}}
        \rulebrk{}
        \ruledef{T-Lam}
                {\begin{gathered}
                 \typerel{\tyctx,\typectxbind{\varvar}{\typevar_1}}{\qctx}{\exprvar}{\typevar_2}
                 \assumbrk{}
                 \forall_{\varvar' \in \freevars{\exprvar}} \: \typectxbind{\varvar'}{\broadcasted{\qubittype}{\mvar}} \notin \tyctx
                 \end{gathered}}
                {\typerel{\tyctx}{\qctx}{\lambdatok{\varvar}{\typevar_1}{\exprvar}}{\arrowtok{\typevar_1}{\irrevtok}{\typevar_2}}}
    \end{gathered}
    $
\caption{\formname{} type rules for functions}
\label{fig:mini-types-funcs}
\end{figure}

These quantum primitives act as functions and thus must be called to change the
state of the quantum system. Fig.~\ref{fig:mini-types-funcs} shows the \formname{} type rules
for defining and calling functions.
\rulename{T-Pipe} defines the typing for the ${\pipetok}$ operator used to call
functions. The \rulename{T-Lam} rule prevents lambdas, whose type is nonlinear, from
having free variables that reference nonempty registers of qubits, a linear
type~\cite{selinger_lambda_2006}.

\begin{figure}
    \centering$\textstyle
    \begin{gathered}
        \ruledef{T-Adj}
                {\typerel{\tyctx}{\qctx}{\exprvar}{\arrowtok{\broadcasted{\qubittype}{\mvar}}{\revtok}{\broadcasted{\qubittype}{\mvar}}}}
                {\typerel{\tyctx}{\qctx}{\adjtok{\exprvar}}{\arrowtok{\broadcasted{\qubittype}{\mvar}}{\revtok}{\broadcasted{\qubittype}{\mvar}}}}
        \rulebrk{}
        \ruledef{T-Pred}
                {\begin{gathered}
                 \typerel{\emptytyctx}{\emptyqctx}{\basisvar}{\broadcasted{\basistype}{\mvar_2}}
                 \assumbrk{}
                 \typerel{\tyctx_1}{\qctx_1}{\exprvar_1}{\arrowtok{\broadcasted{\qubittype}{\mvar_1}}{\revtok}{\broadcasted{\qubittype}{\mvar_1}}}
                 \assumbrk{}
                 \typerel{\tyctx_2}{\qctx_2}{\exprvar_2}{\arrowtok{\broadcasted{\qubittype}{\mvar_1}}{\revtok}{\broadcasted{\qubittype}{\mvar_1}}}
                 \assumbrk{}
                 \eqrel{\atomposcard{\_}{\basisvar}}{\mvar_1}
                 \assumgap{}
                 \neqrel{\atomposnegneg{\_}{?}{\basisvar}}{\emptyindexlist}
                 \end{gathered}}
                {\typerelwrap{\tyctx_1,\tyctx_2}{\qctx_1 \qctxunion \qctx_2}{\ifelsetok{\exprvar_1}{\basisvar}{\exprvar_2}}{\arrowtok{\broadcasted{\qubittype}{\mvar_2}}{\revtok}{\broadcasted{\qubittype}{\mvar_2}}}}
    \end{gathered}
    $
\caption{\formname{} type rules for adjointing and predication}
\label{fig:mini-types-adjpred}
\end{figure}

\begin{figure}
    \centering$\textstyle
    \begin{aligned}
        \qdim{\unittok{}} &\defas 0 \\
        \qdim{\basislittok{\mvar}} &\defas \qdim{\bvvar_1} \quad \text{if } \forall_{i=1}^{\mvar}, \qdim{\bvvar_1} \!=\! \qdim{\bvvar_i} \\
        \qdim{\bitensortok{\basisvar_1}{\basisvar_2}} &\defas \qdim{\basisvar_1} + \qdim{\basisvar_2}
    \end{aligned}
    $
\caption{Definition of the dimension of a \formname{} basis}
\label{fig:mini-qdims-basis}
\end{figure}

Finally, the \formname{} type system must also account for taking the adjoint of a
function or predicating it. Typing for these operations on reversible functions
is handled by the rules in Fig.~\ref{fig:mini-types-adjpred}. The
\rulename{T-Pred} rule verifies that the number of target qubits
($\qlittok{\_}$ vector atoms) matches the number of qubits expected by both
function operands. \rulename{T-Pred} also checks that the predicating basis
$\basisvar$ is nontrivial (contains vector atoms other than $\qlittok{\_}$ and
$\qlittok{?}$), since predication in \formname{} must apply the two functions
supplied to disjoint subspaces. A trivial basis such as
$\basislittokvecs{\qlittok{\_}}$ fails to identify any subspace at all, so it
cannot be a predicate.
The notation $\atomposanynegmany{\mvar}{\basisvar}$ used in this check
means the indices of all vector atoms in $\basisvar$ that are equal to none of
the $\vavar_i$s, i.e., all indices in the list
$1,2,\ldots,\qdim{\basisvar}$ except those in the list $\atomposany{\vavar_1}{\basisvar},\atomposany{\vavar_2}{\basisvar},\ldots,\atomposany{\vavar_{\mvar}}{\basisvar}$.
The syntax $\qdim{\basisvar}$ represents the number of qubits for a basis
$\basisvar$ and is defined in Fig.~\ref{fig:mini-qdims-basis}.

\subsection{Mini-Qwerty Semantics}\label{sec:mini-semantics}
This section presents the last portion of the definition of \formname{}:
operational semantics. Unlike typical classical languages, the state of an
abstract machine executing \formname{} cannot be described only with an
expression $\exprvar$, since nontrivial Qwerty programs manipulate quantum
memory as well. Thus, we characterize the state of an abstract \formname{}
machine with $\statepair{\qstatevar}{\exprvar}$, a pair of a quantum state
$\qstatevar$ and an expression $\exprvar$. We then define a small step of
evaluation with a binary relation
$\smallstepprob{\qstatevar}{\exprvar}{\probvar}{\qstatevaralt}{\exprvar'}$,
where $\probvar$ is a positive real
probability~\cite{selinger_lambda_2006,yuan_twist_2022}. Omitting $\probvar$ as
in $\smallstep{\qstatevar}{\exprvar}{\qstatevaralt}{\exprvar'}$ implies that
$\probvar=1$.

\begin{figure}
    \centering$\textstyle
    \begin{gathered}
        \ruledef{E-QAtom}
                {}
                {\smallstep{\qstatevar}{\qavar}{\qstatevar \otimes \babyket{\qavar}\!}{\bigplus_{j=1}^{\qdim{\qavar}} \qrefvar{\nvar + j}}}
        \evalrulebrk{}
        \ruledef{E-Discard}
                {}
                {\smallstep{\qstatevar}{\qrefvar{\indexvar} \pipetok{} \discardtok{}}{\qstatevar}{\unittok{}}}
        \evalrulebrk{}
        \ruledef{E-BTrans}
                {}
                {\smallstepwrap{\qstatevar}{\bigplus_{j=1}^{\qdim{\basisvar_1}} \qrefvar{\indexveccoord{j}} \pipetok{} \transtok{\basisvar_1}{\basisvar_2}}{\btransunitary{\basisvar_1}{\basisvar_2}^{\indexvec}\qstatevar}{\bigplus_{j=1}^{\qdim{\basisvar_1}} \qrefvar{\indexveccoord{j}}}}
        \evalrulebrk{}
        \ruledef{E-Meas}
                {}
                {\smallstepprobwrap{\qstatevar}{\bigplus_{j=1}^{\qdim{\basisvar}} \qrefvar{\indexveccoord{j}} \pipetok{} \measuretok{\basisvar}}{\probof{y}}{\qstatevarhat{y}}{\bigplus_{j=1}^{\qdim{\basisvar}} \intbit{j}{\qdim{\basisvar}}{y}}}
    \end{gathered}
    $
\caption{Core quantum evaluation rules for \formname{}}
\label{fig:mini-eval-core-quantum}
\end{figure}

Evaluation rules for \formname{}'s core quantum primitives are shown in
Fig.~\ref{fig:mini-eval-core-quantum}. The \rulename{E-QAtom} rule handles
introducing qubits and assumes that $\qstatevar \in \hilbertspace^{\otimes
\nvar}$.
The \rulename{E-QAtom} rule and many others use the notation
$\bigplus_{j=1}^{\nvar} \termvar_j$, which represents an $\nvar$-fold tensor
product. An empty tensor product is a unit literal, i.e., $\bigplus_{j=1}^{0}
\termvar_j \defas \unittok{}$, and a one-fold tensor product is the term
itself, that is, $\bigplus_{j=1}^{1} \termvar_j \defas \termvar_1$. Otherwise,
$\bigplus_{j=1}^{\nvar} \termvar_j$ with $\nvar > 1$ is defined as
$\bitensortok{\termvar_1}{\bitensortok{\termvar_2}{\bitensortok{\cdots}{\termvar_{\nvar}}}}$
with any associativity.

Conversely to \rulename{E-QAtom}, the rule \rulename{E-Discard} eliminates qubits.
Because we choose to avoid mixed states in our description of \formname{}
semantics, the \rulename{E-Discard} rule discards a qubit with
index $\indexvar$ by merely removing the last classical reference
$\qrefvar{\indexvar}$ to it. By the principle of implicit measurement, this is
equivalent to a measurement~\cite[\S4.4]{nielsen_quantum_2010}.

Basis translations accomplish state evolution using the \rulename{E-BTrans}
rule.
A crucial convenience in this rule is defining $T^{\indexvec{}}$ for a linear
operator $T$ as a version of $T$ such that $T$ is performed on the qubits at
the indices held in $\indexvec{}$. The identity operator is applied to qubits
at indices not in $\indexvec{}$~\cite{mermin_quantum_2007}. It is assumed that
$T : \hilbertspace^{\otimes \nvar} \rightarrow \hilbertspace^{\otimes \nvar}$
and $\vert \indexvec{} \vert = \nvar$. The total number of qubits is greater than or equal to $\nvar$ and is implied from context.
Using this notation, we can define the unitary applied by a basis translation as
\begin{gather*}
    \textstyle\btransunitary{\basisvar_1}{\basisvar_2} = \left(\sum_i \basisvecidx{\basisvar_2}{i}\basisvecidxbra{\basisvar_1}{i} + \sum_k \ket{e_k}\!\bra{e_k}\right)^{\atomposneg{?}{\basisvar_1}}.
\end{gather*}
This definition assumes that both nonempty bases have no target vector atoms and the same
padding vector atoms, or more formally that $\qdim{\basisvar_1} = \qdim{\basisvar_2} > 0$;
$\atompos{?}{\basisvar_1} = \atompos{?}{\basisvar_2}$; and $\atompos{\_}{\basisvar_1} = \atompos{\_}{\basisvar_2} = \emptyindexlist$. The vectors $\ket{e_k}$
are defined as an orthonormal basis of $\vspan(\basisvar_1)^{\perp}$,
assuming that $\vspan(\basisvar_1) = \vspan(\basisvar_2)$.

In the measurement evaluation rule \rulename{E-Meas}, $y$ is a nonnegative
integer with $\log_2(y) < \qdim{\basisvar}$. The vector $\qstatevarproj{y}$ to which
$\qstatevar$ is projected by the measurement operator for $y$ is defined as
\begin{gather*}
\qstatevarproj{y} \defas \left(\proj{y}\right)^{\indexvec}\btransunitary{\basisvar}{S_{\qdim{\basisvar}}}^{\indexvec}\qstatevar.
\end{gather*}
Here, the $\nvar$-qubit standard basis $S_{\nvar}$ is defined as $\bigplus_{j=1}^{\nvar}\basislittokstd{}$.
The measurement outcome $\qstatevarhat{y}$ is $\qstatevarproj{y}$ normalized,
i.e., $\qstatevarproj{y}/\!\sqrt{\probof{y}}$.
These definitions assume that $\qdim{\basisvar} > 0$, that $\basisvar$ spans the whole $\qdim{\basisvar}$-qubit space, and that $\basisvar$ contains no $\qlittok{?}$ or $\qlittok{\_}$ atoms.

\begin{figure}
    \centering$\textstyle
    \begin{gathered}
        \ruledef{E-LUnit}
                {}
                {\smallstep{\qstatevar}{\bitensortok{\unittok}{\exprvar}}{\qstatevar}{\exprvar}}
        \evalrulebrk{}
        \ruledef{E-RUnit}
                {}
                {\smallstep{\qstatevar}{\bitensortok{\valvar}{\unittok}}{\qstatevar}{\valvar}}
        \evalrulebrk{}
        \ruledef{E-Tens1}
                {\smallstepprob{\qstatevar}{\exprvar_1}{\probvar}{\qstatevaralt}{\exprvar_1'}}
                {\smallstepprob{\qstatevar}{\bitensortok{\exprvar_1}{\exprvar_2}}{\probvar}{\qstatevaralt}{\bitensortok{\exprvar_1'}{\exprvar_2}}}
        \evalrulebrk{}
        \ruledef{E-Tens2}
                {\smallstepprob{\qstatevar}{\exprvar}{\probvar}{\qstatevaralt}{\exprvar'}}
                {\smallstepprob{\qstatevar}{\bitensortok{\propervalvar}{\exprvar}}{\probvar}{\qstatevaralt}{\bitensortok{\propervalvar}{\exprvar'}}}
        \evalrulebrk{}
        \ruledef{E-Unpack}
                {\smallstepprob{\qstatevar}{\exprvar_1}{\probvar}{\qstatevaralt}{\exprvar_1'}}
                {\smallstepprobwrap{\qstatevar}{\unpacktok{\mvar}{\exprvar_1}{\exprvar_2}}
                                   {\probvar}
                                   {\qstatevaralt}{\unpacktok{\mvar}{\exprvar_1'}{\exprvar_2}}}
        \evalrulebrk{}
        \ruledef{E-UnpackV}
                {}
                {\smallstepwrap{\qstatevar}{\unpacktok{\mvar}{\!\bigplus_{j=1}^{\mvar} \regelemvar_j}{\exprvar}}
                               {\qstatevar}{\rangesubstitute{\varvar_j}{\regelemvar_j}{j=1}{\mvar}{\exprvar}}}
        \evalrulebrk{}
        \ruledef{E-TensPipe}
                {\smallstepprob{\qstatevar}{\bigplus_{k=1}^{\nvar_1} \regelemvar_k \pipetok{} \funcvalvar_1}
                               {\probvar}
                               {\qstatevaralt}{\exprvar}}
                {\smallstepprobwrap{\qstatevar}{\bigplus_{k=1}^{\nvar_1 + \nvar_2} \regelemvar_k \pipetok{} \!\bigplus_{j=1}^{m+1} \funcvalvar_j}
                                   {\probvar}
                                   {\qstatevaralt}{\bitensortok{\exprvar}{\!\!\left(\bigplus_{k=1+\nvar_1}^{\nvar_1 + \nvar_2} \regelemvar_k \pipetok{} \!\bigplus_{j=2}^{m+1} \funcvalvar_j\right)}}}
    \end{gathered}
    $
\caption{Tensor product rules for \formname{}}
\label{fig:mini-eval-tensor}
\end{figure}

To support the tensor product operations vital for useful \formname{} programs,
Fig.~\ref{fig:mini-eval-tensor} defines how tensor products are executed.
Unit literals $\unittok{}$ are eliminated
from tensor products by the \rulename{E-LUnit} and \rulename{E-RUnit} rules because
the definition of \formname{} values does not include unit
literals in tensor products.
The \rulename{E-Tens1} and \rulename{E-Tens2} rules ensure that expressions in
tensor products are evaluated left-to-right. The latter rule specifies a proper value $\propervalvar$ instead of a value $\valvar$ to ensure that unit literals are removed from the left-hand side of the tensor product before evaluation proceeds.
The \rulename{E-UnpackV} rule unpacks tensor products of bits or qubits but
only after \rulename{E-Unpack} reduces the desired register to a value.
The \rulename{E-TensPipe} rule implements support for piping a register into a
tensor product of functions, a particularly distinct feature of Qwerty.
\rulename{E-TensPipe} effectively peels off the first function on the
right-hand side of the pipe and executes it on its chosen number of register
elements.

\begin{figure}
    \centering$\textstyle
    \begin{gathered}
        \ruledef{E-Pipe1}
                {\smallstepprob{\qstatevar{}}{\exprvar_2}{\probvar}{\qstatevaralt{}}{\exprvar_2'}}
                {\smallstepprob{\qstatevar{}}{\exprvar_1 \pipetok{} \exprvar_2}{\probvar}{\qstatevaralt{}}{\exprvar_1 \pipetok{} \exprvar_2'}}
        \evalrulebrk{}
        \ruledef{E-Pipe2}
                {\smallstepprob{\qstatevar{}}{\exprvar}{\probvar}{\qstatevaralt{}}{\exprvar'}}
                {\smallstepprob{\qstatevar{}}{\exprvar \pipetok{} \valvar}{\probvar}{\qstatevaralt{}}{\exprvar' \pipetok{} \valvar}}
        \evalrulebrk{}
        \ruledef{E-LamReg}
                {}
                {\smallstepwrap{\qstatevar}{\!\bigplus_{j=1}^{\nvar} \regelemvar_j \pipetok{} \lambdatok{\varvar}{\broadcasted{\regvar}{\nvar}}{\exprvar}}
                               {\qstatevar}{\substitute{\varvar}{\!\bigplus_{j=1}^{\nvar} \regelemvar_j}{\exprvar}}}
        \evalrulebrk{}
        \ruledef{E-LamUnit}
                {}
                {\smallstepwrap{\qstatevar}{\unittok{} \pipetok{} \lambdatok{\varvar}{\unittype}{\exprvar}}
                               {\qstatevar}{\substitute{\varvar}{\unittok}{\exprvar}}}
        \evalrulebrk{}
        \ruledef{E-LamFunc}
                {}
                {\smallstepwrap{\qstatevar}{\!\bigplus_{j=1}^{\mvar} \funcvalvar_j \pipetok{} \lambdatok{\varvar}{\arrowtok{\typevar_1}{\funckindvar}{\typevar_2}}{\exprvar}}
                               {\qstatevar}{\substitute{\varvar}{\!\bigplus_{j=1}^{\mvar} \funcvalvar_j}{\exprvar}}}
    \end{gathered}
    $
\caption{Function evaluation rules for \formname{}}
\label{fig:mini-eval-func}
\end{figure}

The semantics for pipes when tensor products of functions are not explicitly involved is
more straightforward, as shown in Fig.~\ref{fig:mini-eval-func}.
Together, \rulename{E-Pipe1} and \rulename{E-Pipe2} implement call-by-value
semantics for \formname{}. The reason for three separate evaluation rules for
calling lambdas is that the assumption for the \rulename{E-TensPipe} rule
effectively requires that any function that expects a register cannot
take a step of evaluation with the wrong number of register elements.
That is, the \rulename{E-LamReg} rule cannot apply if the number of register
element values ($\regelemvar_j$s) does not match the $\nvar$ in the explicit
register type annotation $\broadcasted{\regvar}{\nvar}$ in the callee lambda
$\lambdatok{\varvar}{\broadcasted{\regvar}{\nvar}}{\exprvar}$.
The other two rules \rulename{E-LamUnit} and \rulename{E-LamFunc} implement calling lambdas with explicit argument types that are not registers. Both of these latter rules could in principle be written as one rule but are separate for readability.

\begin{figure}
    \centering$\textstyle
    \begin{gathered}
        \ruledef{E-Adj}
                {\smallstepprob{\qstatevar}{\exprvar}
                               {\probvar}
                               {\qstatevaralt}{\exprvar'}}
                {\smallstepprob{\qstatevar}{\adjtok{\exprvar}}
                               {\probvar}
                               {\qstatevaralt}{\adjtok{\exprvar'}}}
        \evalrulebrk{}
        \ruledef{E-AdjV}
                {}
                {\smallstep{\qstatevar}{\adjtok{\!\left(\bitensortok{\propervalvar_1}{\propervalvar_2}\right)}}
                           {\qstatevar}{\bitensortok{\adjtok{\propervalvar_1}}{\adjtok{\propervalvar_2}}}}
        \evalrulebrk{}
        \ruledef{E-Pred1}
                {\smallstepprob{\qstatevar}{\exprvar_1}
                               {\probvar}
                               {\qstatevaralt}{\exprvar_1'}}
                {\smallstepprobwrap{\qstatevar}{\ifelsetok{\exprvar_1}{\basisvar}{\exprvar_2}}
                                   {\probvar}
                                   {\qstatevar}{\ifelsetok{\exprvar_1'}{\basisvar}{\exprvar_2}}}
        \evalrulebrk{}
        \ruledef{E-Pred2}
                {\smallstepprob{\qstatevar}{\exprvar}
                               {\probvar}
                               {\qstatevaralt}{\exprvar'}}
                {\smallstepprobwrap{\qstatevar}{\ifelsetok{\valvar}{\basisvar}{\exprvar}}
                                   {\probvar}
                                   {\qstatevar}{\ifelsetok{\valvar}{\basisvar}{\exprvar'}}}
        \evalrulebrk{}
        \ruledef{E-AdjPipe}
                {\begin{gathered}
                 \forall{\qstatevaralt \in \hilbertspace^{\otimes \mvar}}, \\
                 \textstyle\multismallstepwrap{\qstatevaralt}{\bigplus_{j=1}^{\mvar} \qrefvar{j} \pipetok{} \revfuncvalvar{}}
                                              {U\qstatevaralt{}}{\bigplus_{j=1}^{\mvar} \qrefvar{\permutationof{j}}}
                 \end{gathered}}
                {\smallstepwrap{\qstatevar}{\bigplus_{j=1}^{\mvar} \qrefvar{\indexveccoord{j}} \pipetok{} \adjtok{\revfuncvalvar}}
                               {\left(U^\dagger\right)^{\indexvec}\qstatevar}{\bigplus_{j=1}^{\mvar} \qrefvar{\invpermutationof{\indexveccoord{j}}}}}
        \evalrulebrk{}
        \ruledef{E-PredPipe}
                {\begin{gathered}
                 \forall{\qstatevaralt \in \hilbertspace^{\otimes \mvar}}, \\
                 \textstyle\multismallstepwrap{\qstatevaralt}{\bigplus_{j=1}^{\mvar} \qrefvar{j} \pipetok{} \valvar_1}{U{'}\qstatevaralt{}}{\bigplus_{j=1}^{\mvar} \qrefvar{\altpermutationof{j}}} \\
                 \textstyle\multismallstepwrap{\qstatevaralt}{\bigplus_{j=1}^{\mvar} \qrefvar{j} \pipetok{} \valvar_2}{U{''}\qstatevaralt{}}{\bigplus_{j=1}^{\mvar} \qrefvar{\altaltpermutationof{j}}}
                 \end{gathered}}
                {\smallstepwrap{\qstatevar}{\bigplus_{j=1}^{\qdim{\basisvar}} \qrefvar{\indexveccoord{j}} \pipetok{} \ifelsetok{\valvar_1}{\basisvar}{\valvar_2}}
                               {\left(\predunitary{\basisvar}{\altpermunitary{} U'}{\altaltpermunitary{} U''}\right)^{\indexvec}\qstatevar}{\bigplus_{j=1}^{\qdim{\basisvar}} \qrefvar{\indexveccoord{j}}}}
    \end{gathered}
    $
\caption{Reversible function evaluation rules for \formname{}}
\label{fig:mini-eval-revfunc}
\end{figure}

The final execution rules in Fig.~\ref{fig:mini-eval-revfunc} handle adjointing
and predication. The rules \rulename{E-Adj}, \rulename{E-Pred1}, and
\rulename{E-Pred2} ensure that the adjoint and predication operators act only
on values. The \rulename{E-AdjV} rule distributes the adjoint operator
$\rawadjtok$ into a tensor product; this is done to make execution of $\exprvar_1 \pipetok \adjtok{\exprvar_2}$ (discussed shortly) not depend on \rulename{E-TensPipe} to accept the right number of register elements, since \rulename{E-TensPipe} itself requires this.
The \rulename{E-AdjPipe} and \rulename{E-PredPipe} rules perform adjointed or predicated execution and thus assume the adjointed or predicated functions act as a unitary $U$. In
\rulename{E-PredPipe}, $\permunitary{}$ is defined as a unitary that permutes
qubits $1,2,\ldots,m$ according to the permutation $\rawpermutation{}$.
If $U$ and $W$ are unitaries acting on the space $\hilbertspace^{\otimes\atomposcard{\_}{\basisvar}}$, then the unitary $\predunitary{\basisvar}{U}{W}$ performed by predication is defined as
\begin{align*}
    \textstyle\predunitary{\basisvar}{U}{W} \defas
        &\textstyle \left(\sum_i \basisvecidx{\basisvar}{i}\basisvecidxbra{\basisvar}{i}\right)^{\atomposnegneg{?}{\_}{\basisvar}}U^{\atompos{\_}{\basisvar}} \\
        &\textstyle {}+ \left(\sum_k \ket{e_k}\!\bra{e_k} \right)^{\atomposnegneg{?}{\_}{\basisvar}}W^{\atompos{\_}{\basisvar}}
\end{align*}
where $\ket{e_k}$ is an orthonormal basis of
$\vspan(\basisvar)^{\perp}$. The unitary $\predunitary{\basisvar}{U}{W}$ is undefined for empty or trivial bases $\basisvar$, that is, if $\qdim{\basisvar} = 0$ or $\atomposnegneg{?}{\_}{\basisvar}$ is empty, respectively.

\subsection{Mini-Qwerty Properties}\label{sec:mini-props}
With \formname{} now defined, we argue in this section that the language is safe and
universal. Safety will be shown by proving the progress and preservation
theorems~\cite{pierce_types_2002}, beginning with the progress theorem.
Because the proof for progress reasons about which evaluation rules may be applied
to expressions, it is useful to have guarantees about the structure of values
with particular types, which the first lemma provides.

\begin{lemma}[Canonical forms of values]\label{lem:canon-forms}
The following properties of \formname{} values $\valvar$ hold:
\begin{enumerate}
    \item If $\typerel{\tyctx}{\qctx}{\valvar}{\unittype}$, then $\eqrel{\valvar}{\unittok{}}$.
    \item If $\typerel{\tyctx}{\qctx}{\valvar}{\broadcasted{\basistype}{0}}$, then $\eqrel{\valvar}{\unittok{}}$.
    \item If $\typerel{\tyctx}{\qctx}{\valvar}{\broadcasted{\qubittype}{0}}$, then $\eqrel{\valvar}{\unittok{}}$.
    \item If $\typerel{\tyctx}{\qctx}{\valvar}{\broadcasted{\bittype}{0}}$, then $\eqrel{\valvar}{\unittok{}}$.
    \item If $\typerel{\tyctx}{\qctx}{\valvar}{\broadcasted{\bittype}{\mvar}}$, then $\eqrel{\valvar}{\!\bigplus_{j=1}^{\mvar} \mathbb{B}_j}$ where each $\mathbb{B}_j \in \{\zerotok{},\onetok{}\}$.
    \item If $\typerel{\tyctx}{\qctx}{\valvar}{\broadcasted{\qubittype}{\mvar}}$, then $\eqrel{\valvar}{\!\bigplus_{j=1}^{\mvar} \qrefvar{\indexveccoord{j}}}$ where $\indexvec{} \subseteq \qctx$.
    \item If $\typerel{\tyctx}{\qctx}{\valvar}{\arrowtok{\typevar_1}{\irrevtok}{\typevar_2}}$, then $\eqrel{\valvar}{\!\bigplus_{j=1}^{\mvar}\funcvalvar_j}$.
    \item If $\typerel{\tyctx}{\qctx}{\valvar}{\arrowtok{\typevar_1}{\revtok}{\typevar_2}}$, then $\eqrel{\valvar}{\!\bigplus_{j=1}^{\mvar}\revfuncvalvar_j}$.
\end{enumerate}
\end{lemma}
\begin{proof}
\begin{LaTeXdescription}
Propositions 1-4 follow from the definition of typing rules and values.
Propositions 5-8 can be proven by structural induction on proper values $\propervalvar$.
\end{LaTeXdescription}
\end{proof}

Unlike other evaluation rules, the \rulename{E-AdjPipe} and \rulename{E-PredPipe} rules
require that a particular value evolves the state of quantum memory by a
unitary operator. The next lemma guarantees this for reversible values.

\begin{lemma}[Unitary progress]\label{lem:unitary-progress}
For any value $\valvar$ such that $\typerel{\tyctx}{\qctx}{\valvar}{\arrowtok{\broadcasted{\qubittype}{\mvar}}{\revtok}{\broadcasted{\qubittype}{\mvar}}}$, then for
all values $\!\bigplus_{j=1}^{\mvar} \qrefvar{\indexveccoord{j}}$ and
all $\qstatevar \in \hilbertspace^{\otimes \mvar}$ such that $\{1,2,\ldots,\mvar\} \subseteq \qctx$,
execution is unitary up to a permutation of input qubits: $\multismallstep{\qstatevar}{\!\bigplus_{j=1}^{\mvar} \qrefvar{\indexveccoord{j}} \pipetok \valvar}{U\qstatevar}{\!\bigplus_{j=1}^{\mvar} \qrefvar{\permutationof{\indexveccoord{j}}}}$.
\end{lemma}
\begin{proof}
Structural induction on operational semantics derivations informed by Lemma~\ref{lem:canon-forms}.
\end{proof}

In the present definition of \formname{}, the permutation $\rawpermutation$ in
the statement of Lemma~\ref{lem:unitary-progress} is always the identity
permutation. This is because the operational semantics for no
$\revfuncvalvar$s permute qubits nontrivially, meaning that by
Lemma~\ref{lem:canon-forms}, all functions with a reversible type permute
their input qubits only trivially. Nevertheless, we involve permutations in this appendix (in the definition of rules \rulename{E-AdjPipe} and \rulename{E-PredPipe}, for instance) for the
sake of future extensions of \formname{} that support reversible lambdas,
e.g., $\revlambdatok{\varvar}{\typevar}{\exprvar}$.

With these lemmas established, we can prove that well-typed \formname{} terms
can take a step of evaluation.

\begin{theorem}[Progress]
For any expression $\exprvar$ such that $\typerel{\emptytyctx}{\qctx}{\exprvar}{\typevar}$, either $\exprvar$ is a value or there exists $\exprvar'$ such that for all $\qstatevar \in \hilbertspace^{\otimes \nvar}$ where $\qctx \subseteq \{1,2,\ldots,\nvar\}$, $\smallstepprob{\qstatevar}{\exprvar}{\probvar}{\qstatevaralt}{\exprvar'}$.
\end{theorem}
\begin{proof}
By induction on typing derivations.

\begin{LaTeXdescription}
    \item[Case \rulename{T-Pipe}:] Consider the following subcases:
        \begin{LaTeXdescription}
            \item[Subcase 1: $\exprvar_2$ is not a value:] Apply \rulename{E-Pipe1} by the induction hypothesis (IH).
            \item[Subcase 2: $\exprvar_2$ is a value but $\exprvar_1$ is not a value:] Apply \rulename{E-Pipe2} by the IH.
            \item[Subcase 3: $\exprvar_2$ and $\exprvar_1$ are values:] Consider the following subsubcases:
                \begin{LaTeXdescription}
                    \item[Subsubcase 1: $\exprvar_2$ is $\lambdatok{\varvar}{\broadcasted{\regvar}{\nvar}}{\exprvar}$:] By Lemma~\ref{lem:canon-forms}, $\exprvar_1$ is $\!\bigplus_{j=1}^{\nvar} \regelemvar_j$. Apply \rulename{E-LamReg}.
                    \item[Subsubcase 2: $\exprvar_2$ is $\adjtok{\revfuncvalvar}$:] By Lemma~\ref{lem:unitary-progress}, apply \rulename{E-AdjPipe} .
                    \item[Subsubcase 3: $\exprvar_2$ is $\ifelsetok{\valvar_1}{\basisvar}{\valvar_2}$:] By Lemma~\ref{lem:unitary-progress}, apply \rulename{E-PredPipe}.
                \end{LaTeXdescription}
                The other subsubcases are similar.
        \end{LaTeXdescription}

    \item[Case \rulename{T-Adj}:] Consider the following subcases:
        \begin{LaTeXdescription}
            \item[Subcase 1: $\exprvar$ is not a value:] Apply \rulename{E-Adj} by the IH.
            \item[Subcase 2: $\exprvar$ is value:] Consider the following subsubcases:
                \begin{LaTeXdescription}
                    \item[Subsubcase 1: $\exprvar$ is a reversible function value $\revfuncvalvar$:] Value.
                    \item[Subsubcase 2: $\exprvar$ is not a reversible function value $\revfuncvalvar$:] By Lemma~\ref{lem:canon-forms}, $\exprvar$ is a tensor product of proper values. Apply \rulename{E-AdjV}.
                \end{LaTeXdescription}
        \end{LaTeXdescription}
\end{LaTeXdescription}
The other cases are straightforward.
\end{proof}

With the progress theorem tackled, we proceed to the preservation theorem.
First, though, we establish that replacing a variable with an expression of the
same type does not produce an ill-typed expression.

\begin{lemma}[Substitution]\label{lem:subst}
If $\typerel{\typectxbind{\varvar_1}{\typevar_1},\typectxbind{\varvar_2}{\typevar_2},\ldots,\typectxbind{\varvar_{\mvar}}{\typevar_{\mvar}}}{\qctx'}{\exprvar'}{\typevar'}$ and $\typerel{\emptytyctx}{\qctx_i}{\exprvar_i}{\typevar_i}$ for all $i \in \{1,2,\ldots,\mvar\}$ and all $\qctx_i$ are disjoint (including with $\qctx'$), then $\typerel{\emptytyctx}{\qctx''}{\rangesubstitute{\varvar_i}{\exprvar_i}{i=1}{\mvar}{\exprvar'}}{\typevar'}$ where $\qctx'' = \left(\bigqctxunion_{i=1}^{\mvar} \qctx_i\right) \qctxunion \qctx'$.
\end{lemma}
\begin{proof}
A derivation of $\typerel{\emptytyctx}{\qctx''}{\rangesubstitute{\varvar_i}{\exprvar_i}{i=1}{\mvar}{\exprvar'}}{\typevar'}$ can be built by removing
structural rules (Fig.~\ref{fig:mini-types-linear})
in the derivation of
$\typerel{\typectxbind{\varvar_1}{\typevar_1},\typectxbind{\varvar_2}{\typevar_2},\ldots,\typectxbind{\varvar_{\mvar}}{\typevar_{\mvar}}}{\qctx}{\exprvar'}{\typevar'}$
or replacing them with the respective derivations of $\typerel{\emptytyctx}{\qctx_i}{\exprvar_i}{\typevar_i}$.
\end{proof}

Next, whereas the typing rules identify the number of qubits for a qubit
literal by the $\nvar$ in its $\broadcasted{\qubittype}{\nvar}$ type, the
evaluation rules use $\qdim{\qlitvar}$. The next lemma shows that these
definitions align. The statement of the lemma treats $\unittok{}$ specially
because $\unittok{}$ is a syntactically valid qubit literal yet could have
types such as $\unittype$ or $\broadcasted{\bittype}{0}$ rather than
exclusively $\broadcasted{\qubittype}{0}$.

\begin{lemma}[Type of qubit literal]\label{lem:qlit-dim}
If $\typerel{\emptytyctx}{\qctx}{\qlitvar}{\typevar}$, then either $\eqrel{\qlitvar}{\unittok}$ or $\eqrel{\typevar}{\broadcasted{\qubittype}{\qdim{\qlitvar}}}$.
\end{lemma}
\begin{proof}
By induction on typing derivations.
\end{proof}

The following lemma is similar to the previous one, arguing that the type rules
for qubit literals align with the definition of the $\ket{\qlitvar}$ notation
used in the evaluation rules.

\begin{lemma}[State for qubit literal]\label{lem:qlit-ket}
If $\typerel{\emptytyctx}{\qctx}{\qlitvar}{\broadcasted{\qubittype}{\nvar}}$, then $\ket{\qlitvar} \in \hilbertspace^{\otimes \nvar}$.
\end{lemma}
\begin{proof}
By induction on typing derivations.
\end{proof}

The discussion of \rulename{E-LamReg} in the previous section
mentioned that the construction of the \rulename{E-PredPipe} rule practically
mandates that function calls can take a step of execution only if
the input register is the correct length.
The lemma below argues that the operational semantics of \formname{} satisfy
this requirement.

\begin{lemma}[Function input size]\label{lem:func-size}
If $\typerel{\emptytyctx}{\qctx}{\funcvalvar}{\arrowtok{\broadcasted{\regvar_1}{\nvar_1}}{\funckindvar}{\broadcasted{\regvar_2}{\nvar_2}}}$,
then there exists an expression $\exprvar$ such that $\smallstepprob{\qstatevar}{\!\bigplus_{j=1}^{\nvar_3} \regelemvar_j \pipetok \funcvalvar}{\probvar}{\qstatevaralt}{\exprvar}$ only if $\nvar_3 = \nvar_1$.
\end{lemma}
\begin{proof}
By contraposition. Assume $\nvar_3 \ne \nvar_1$ and show that there is no $\exprvar$ such that $\smallstepprob{\qstatevar}{\!\bigplus_{j=1}^{\nvar_3} \regelemvar_j \pipetok \funcvalvar}{\probvar}{\qstatevaralt}{\exprvar}$.
Proceed by structural induction on $\valvar$.
(Induction on $\valvar$ instead of $\funcvalvar$ is used due to the $\valvar_1$ and $\valvar_2$ subterms of the function value $\ifelsetok{\valvar_1}{\basisvar}{\valvar_2}$.)

\begin{LaTeXdescription}
    \item[Case $\lambdatok{\varvar}{\broadcasted{\regvar}{\nvar}}{\exprvar}$:] The only applicable evaluation rule is \rulename{E-LamReg}, which requires that $\nvar_3 = \nvar_1$.

    \item[Case $\adjtok{\revfuncvalvar}$:] By the IH.

    \item[Case $\ifelsetok{\valvar_1}{\basisvar}{\valvar_2}$:] The only applicable evaluation rule is \rulename{E-PredPipe}, which requires that $\nvar_3 = \nvar_1 = \qdim{\basisvar}$.

    \item[Cases where $\valvar$ is not a $\funcvalvar$:] Claim vacuously holds.
\end{LaTeXdescription}
The other cases are similar.
\end{proof}

The final lemma used in the proof of preservation states that new qubits
introduced by evaluation do not conflict with existing qubits.

\begin{lemma}[Evaluation cannot decrease qubit count]\label{lem:growth}
If $\typerel{\tyctx}{\qctx}{\exprvar}{\typevar}$ and $\smallstepprob{\qstatevar}{\exprvar}{\probvar}{\qstatevaralt}{\exprvar'}$ and $\typerel{\tyctx'}{\qctx'}{\exprvar'}{\typevar'}$, where $\qstatevar \in \hilbertspace^{\otimes \nvar}$ with $\qctx \subseteq \{1,2,\ldots,\nvar\}$ and $\qstatevaralt \in \hilbertspace^{\otimes \nvar'}$ with $\qctx' \subseteq \{1,2,\ldots,\nvar'\}$, then $\nvar' \ge \nvar$ and $\qctx' - \qctx \subseteq \{\nvar+1,\nvar+2,\ldots,\nvar'\}$.
\end{lemma}
\begin{proof}
First, observe that no evaluation rule shrinks the number of qubits in quantum
memory $\qstatevar$, so it is impossible for $\nvar' < \nvar$. (Note that
neither \rulename{E-Discard} nor \rulename{E-Meas} remove qubits from quantum memory; they only
discard classical references to qubit indices.) Second, the only evaluation
rule that can cause the qubit context to grow ($\qctx' - \qctx \neq
\ouremptyset$) is \rulename{E-QAtom}, which is defined such that all
indices inserted are greater than $\nvar$ and thus all indices in the original qubit context
$\qctx$.
\end{proof}

We proceed to proving that each step of execution in \formname{} preserves
the type of an expression and does not introduce any invalid qubit pointers.

\newcommand{\gvar}{g}
\newcommand{\zvar}{z}
\newcommand{\svar}{\rho}
\newcommand{\dvar}{d}
\begin{theorem}[Preservation]
Suppose that $\dvar$ and $\dvar'$ are nonnegative integers; that $\gvar$ and
$\gvar'$ are expressions; and $\gamma$ is a type context.
If $\typerel{\emptytyctx}{\qctx}{\gvar}{\typevar}$ and $\smallstepprob{\qstatevar}{\gvar}{\probvar}{\qstatevar}{\gvar'}$, where $\qstatevar \in \hilbertspace^{\otimes \dvar}$ and $\qctx \subseteq \{1,2,\ldots,\dvar\}$, then $\typerel{\emptytyctx}{\qctx'}{\gvar'}{\typevar}$ where $\qstatevaralt \in \hilbertspace^{\otimes \dvar'}$ and $\qctx' \subseteq \{1,2,\ldots,\dvar'\}$.
\end{theorem}
\begin{proof}
By induction on evaluation derivations.
\begin{LaTeXdescription}
    \item[Case \rulename{E-QAtom}:]
        Suppose that $\typerel{\emptytyctx}{\qctx}{\qavar}{\typevar}$.
        By Lemma~\ref{lem:qlit-dim}, then either $\eqrel{\qavar}{\unittok}$ or $\eqrel{\typevar}{\broadcasted{\qubittype}{\qdim{\qavar}}}$. The former is not possible because $\unittok$ is a value.
        The rules \rulename{T-Tensor} and \rulename{T-Q} can be used to show that the resulting term $\bigplus_{j=1}^{\qdim{\qavar}} \qrefvar{\nvar + j}$ has the same type $\broadcasted{\qubittype}{\qdim{\qavar}}$.
        By Lemma~\ref{lem:qlit-ket}, the qubit indices $\nvar{+}j$ are indeed less than or equal to $\dvar'$.

    \item[Case \rulename{E-UnpackV}:]
        A typing derivation for the judgment $\typerel{\emptytyctx}{\qctx_1\qctxunion\qctx_2}{\unpacktok{\mvar}{\!\bigplus_{j=1}^{\mvar} \regelemvar_j}{\exprvar}}{\typevar}$ ends in \rulename{T-Unpack}, which has an assumption that
$\typerel{\typectxbind{\varvar_1}{\broadcasted{\regvar}{1}},\typectxbind{\varvar_2}{\broadcasted{\regvar}{1}},\ldots,\typectxbind{\varvar_{\mvar}}{\broadcasted{\regvar}{1}}}{\qctx_2}{\exprvar_2}{\typevar}$.
The judgments $\typerel{\emptytyctx}{\qctx_j'}{\regelemvar_j}{\broadcasted{\regvar}{1}}$ are trivial.
Applying Lemma~\ref{lem:subst} yields $\typerel{\emptytyctx}{\qctx_1 \qctxunion \qctx_2}{\rangesubstitute{\varvar_j}{\regelemvar_j}{j=1}{\mvar}{\exprvar}}{\typevar}$ as needed.

    \item[Case \rulename{E-TensPipe}:]
        Suppose that
        \begin{gather*}
            \textstyle\typerel{\emptytyctx}{\qctx_1 \qctxunion \qctx_2}{\!\bigplus_{k=1}^{\nvar_1 + \nvar_2} \regelemvar_k \pipetok{} \!\bigplus_{j=1}^{m+1} \funcvalvar_j}{\typevar}.
        \end{gather*}
        Notice that by definition of $\regelemvar$ and relevant typing rules, it must be true that $\typerel{\emptytyctx}{\qctx_1}{\!\bigplus_{k=1}^{\nvar_1 + \nvar_2} \regelemvar_k}{\broadcasted{\regvar_1}{\nvar_1 + \nvar_2}}$.
        By \rulename{T-Pipe}, it follows that
        \begin{gather*}
            \textstyle\typerel{\emptytyctx}{\qctx_2}{\!\bigplus_{j=1}^{m+1} \funcvalvar_j}{\arrowtok{\broadcasted{\regvar_1}{\nvar_1 + \nvar_2}}{\funckindvar'}{\broadcasted{\regvar_2}{\nvar_3}}}.
        \end{gather*}
        Thus, the type $\typevar$ of the original expression is $\broadcasted{\regvar_2}{\nvar_3}$.
        Furthermore, $\funcvalvar_1$ specifically must satisfy
        \begin{gather}
            \textstyle\typerel{\emptytyctx}{\qctx_2'}{\funcvalvar_1}{\arrowtok{\broadcasted{\regvar_1}{\nvar_5}}{\funckindvar}{\broadcasted{\regvar_2}{\nvar_4}}} \label{eq:fv1welltyped}
        \end{gather}
        where the expected input register size $\nvar_5$ of $\funcvalvar_1$ is no larger than the size of the input register $\nvar_1 + \nvar_2$; the output register size $\nvar_4$ of $\funcvalvar_1$ is no more than the size $\nvar_3$ of the register produced by the function call; and $\qctx_2'$ is the appropriate subset of the overall qubit context $\qctx_2$ (that is, $\qctx_2 = \qctx_2' \qctxunion \qctx_2''$).

        Recall that the \rulename{E-TensPipe} rule has an assumption that $\funcvalvar_1$ accepts a register of size $\nvar_1$:
        \begin{gather}
            \textstyle\smallstepprob{\qstatevar}{\bigplus_{k=1}^{\nvar_1} \regelemvar_k \pipetok{} \funcvalvar_1}{\probvar}{\qstatevaralt}{\exprvar} \label{eq:fv1steps}
        \end{gather}
        By Lemma~\ref{lem:func-size}, Equations (\ref{eq:fv1welltyped}) and
        (\ref{eq:fv1steps}) imply that $\funcvalvar_1$ has a type that expects $\nvar_1$ register elements as input:
        \begin{gather}
        \textstyle\typerel{\emptytyctx}{\qctx_2'}{\funcvalvar_1}{\arrowtok{\broadcasted{\regvar_1}{\nvar_1}}{\funckindvar}{\broadcasted{\regvar_2}{\nvar_4}}} \label{eq:fv1wantsn1}
        \end{gather}
        Using \rulename{T-Tensor}, it can be shown that the first $\nvar_1$ values in the original input register have type $\broadcasted{\regvar_1}{\nvar_1}$ as the type of $\funcvalvar_1$ expects, i.e.,
        \begin{gather}
        \textstyle\typerel{\emptytyctx}{\qctx_1'}{\!\bigplus_{k=1}^{\nvar_1} \regelemvar_k}{\broadcasted{\regvar_1}{\nvar_1}} \label{eq:firstn1aren1}
        \end{gather}
        where $\qctx_1 = \qctx_1' \qctxunion \qctx_1''$.
        This means that piping the first $\nvar_1$ values to $\funcvalvar_1$ is
        well-typed. This is because Equations (\ref{eq:fv1wantsn1}) and
        (\ref{eq:firstn1aren1}) can be combined with \rulename{T-Pipe} to show that
        \begin{gather}
            \textstyle\typerel{\emptytyctx}{\qctx_1' \qctxunion \qctx_2'}{\!\bigplus_{k=1}^{\nvar_1} \regelemvar_k \pipetok \funcvalvar_1}{\broadcasted{\regvar_2}{\nvar_4}}. \label{eq:pipeislegit}
        \end{gather}
        This establishes that the expression in Equation (\ref{eq:pipeislegit}), which is also present in the assumption of \rulename{E-TensPipe}, is well-typed, allowing the IH to be applied. 
        The IH reveals that $\typerel{\emptytyctx}{\qctx'''}{\exprvar}{\broadcasted{\regvar_2}{\nvar_4}}$, i.e., that the new expression $\exprvar$ shares the same type
        $\broadcasted{\regvar_2}{\nvar_4}$.

        We now turn our focus to the final portion of the consequent of \rulename{E-TensPipe},
        the tensor product of remaining functions $\!\bigplus_{j=2}^{m+1} \funcvalvar_j$
        and the tensor product of remaining inputs
        $\!\bigplus_{k=1+\nvar_1}^{\nvar_1+\nvar_2} \regelemvar_k$.
        It is obvious that for the original expression to be well-typed, the
        remaining functions must consume the remaining register elements $\nvar_2$:
        \begin{gather*}
            \textstyle\typerel{\emptytyctx}{\qctx_2''}{\!\bigplus_{j=2}^{m+1} \funcvalvar_j}{\arrowtok{\broadcasted{\regvar_1}{\nvar_2}}{\funckindvar''}{\broadcasted{\regvar_2}{\nvar_3-\nvar_4}}}
        \end{gather*}
        Furthermore, the remaining register elements are of the input type expected by the remaining functions:
        \begin{gather*}
            \textstyle\typerel{\emptytyctx}{\qctx_1''}{\!\bigplus_{k=1+\nvar_1}^{\nvar_1+\nvar_2} \regelemvar_k}{\broadcasted{\regvar_1}{\nvar_2}}
        \end{gather*}
        Per \rulename{T-Pipe}, piping these remaining register elements into the remaining functions yields type $\broadcasted{\regvar_2}{\nvar_3-\nvar_4}$:
        \begin{gather*}
            \textstyle\typerel{\emptytyctx}{\qctx_1'' \qctxunion \qctx_2''}{\!\bigplus_{k=1+\nvar_1}^{\nvar_1+\nvar_2} \regelemvar_k \pipetok \!\bigplus_{j=2}^{m+1} \funcvalvar_j}{\broadcasted{\regvar_2}{\nvar_3-\nvar_4}}
        \end{gather*}
        Finally, because $\nvar_4 + (\nvar_3{-}\nvar_4) = \nvar_3$ and the original overall expression type $\typevar$ is $\broadcasted{\regvar_2}{\nvar_3}$, applying \rulename{T-Tensor} would give
        \begin{gather*}
            \textstyle\typerel{\emptytyctx}{\qctx''' \qctxunion \qctx_1'' \qctxunion \qctx_2''}{\bitensortok{\exprvar}{\left(\!\bigplus_{k=1+\nvar_1}^{\nvar_1+\nvar_2} \regelemvar_k \pipetok \!\bigplus_{j=2}^{m+1} \funcvalvar_j\right)}}{\typevar}
        \end{gather*}
        as desired.

        However, we must briefly justify why the disjoint union $\qctx'''
        \qctxunion \qctx_1'' \qctxunion \qctx_2''$ in this typing judgment is
        defined. First, by construction we have $\qctx = (\qctx_1' \qctxunion
        \qctx_1'') \qctxunion (\qctx_2' \qctxunion \qctx_2'')$; consequently, the contexts
        $\qctx_1' \qctxunion \qctx_2'$ and $\qctx_1'' \qctxunion \qctx_2''$ are
        necessarily disjoint.

        When applied to the earlier usage of the inductive hypothesis, where
        $\qctx'''$ was introduced, Lemma~\ref{lem:growth} shows that any
        indices in $\qctx''' - (\qctx_1' \qctxunion \qctx_2')$
        are at least $\dvar+1$. Thus, these new indices cannot be in the qubit context
        $\qctx_1'' \qctxunion \qctx_2'' \subseteq \{1,2,\ldots,\dvar\}$. It
        follows that $\qctx'''$ and $\qctx_1'' \qctxunion \qctx_2''$ are
        disjoint as required for the desired typing judgment to be valid.

\end{LaTeXdescription}
The remaining cases are straightforward.
\end{proof}

The final theorem in this appendix demonstrates that \formname{} can simulate a
universal quantum gate set.

\begin{theorem}[Universality]
\formname{} is universal for quantum computation.
\end{theorem}
\begin{proof}
The following basis translation performs a unitary transformation equal
to an $R_z(\tiltvar)$ gate up to a global phase:
\begin{gather*}
    \transtok{\basislittokvecs{\qlittok{1}}}{\basislittokvecs{\tilttok{\qlittok{1}}{\tiltvar'}}}
\end{gather*}
where henceforth $\tiltvar' = \tiltvar \cdot 360\degree/2\pi$.

Similarly, the following acts as an $R_y(\tiltvar)$ up to a global phase:
\begin{align*}
    &\basislittokvecs{\superpostok{\qlittok{0}}{\left(\tilttok{\qlittok{1}}{90}\right)}} \\
    \rawtranstok{}\, &\basislittokvecs{\superpostok{\qlittok{0}}{\left(\tilttok{\left(\tilttok{\qlittok{1}}{90}\right)}{\tiltvar'}\right)}}
\end{align*}
Above, the basis vector $(\superpostok{\qlittok{0}}{\left(\tilttok{\qlittok{1}}{90}\right)})$ represents the plus eigenstate of $Y$, $\frac{1}{\sqrt{2}}\left(\ket{0} + i\ket{1}\right)$.

The following applies a global phase of $\theta$:
\begin{align*}
    &\basislittokvecs{\qlittok{0},\qlittok{1}} \\
    \rawtranstok{}\, &\basislittokvecs{\tilttok{\qlittok{0}}{\tiltvar'},\tilttok{\qlittok{1}}{\tiltvar'}}
\end{align*}
Thus, using a ZYZ decomposition~\cite[\S4.2]{nielsen_quantum_2010}, \formname{}
is capable of executing any one-qubit unitary by applying these
three functions with different choices of $\tiltvar$.

Furthermore, a CNOT can be performed in \formname{} with the following basis translation:
\begin{gather*}
    \ifelsetok{\texttt{flip}}{\basislittokvecs{\bitensortok{\qlittok{1}}{\qlittok{\_}}}}{\texttt{id}}
\end{gather*}
where $\texttt{flip} \defas \transtok{\basislittokvecs{\qlittok{0},\qlittok{1}}}{\basislittokvecs{\qlittok{1},\qlittok{0}}}$ and $\texttt{id} \defas \transtok{\basislittokvecs{\qlittok{?}}}{\basislittokvecs{\qlittok{?}}}$.

Thus, by the universality of single qubit gates and
CNOTs~\cite[\S4.5.2]{nielsen_quantum_2010}, \formname{} is universal.
\end{proof}

\subsection{Conclusion and Future Work}

In this appendix, we presented \formname{}, a formal definition of a subset of
the Qwerty language. Proofs were presented of its safety and universality.
Yet as implied by name \textit{Mini}-Qwerty itself,
crucial aspects of Qwerty were excluded from this appendix in
the interest of space and time, namely Meta-Qwerty (Section~\ref{sec:meta}),
embeddings of classical functions (Section~\ref{sec:classical-embed}), and reversible function definitions (e.g., line~\lnqpeuserrevdec{} of Fig.~\ref{fig:qpeuser}). Future
work should expand \formname{} to cover more features of the Qwerty language
and compiler.
The authors also emphasize that faithfulness to the Qwerty domain-specific
language embedded in Python was prioritized over theoretical beauty. Future
work can eschew Qwerty implementation details in favor of a simpler, more
elegant basis-oriented quantum--classical lambda calculus.

\bibliographystyle{IEEEtran}
\bibliography{zotero,handcrafted}

\begin{thebibliography}{10}
\providecommand{\url}[1]{#1}
\csname url@samestyle\endcsname
\providecommand{\newblock}{\relax}
\providecommand{\bibinfo}[2]{#2}
\providecommand{\BIBentrySTDinterwordspacing}{\spaceskip=0pt\relax}
\providecommand{\BIBentryALTinterwordstretchfactor}{4}
\providecommand{\BIBentryALTinterwordspacing}{\spaceskip=\fontdimen2\font plus
\BIBentryALTinterwordstretchfactor\fontdimen3\font minus
  \fontdimen4\font\relax}
\providecommand{\BIBforeignlanguage}[2]{{%
\expandafter\ifx\csname l@#1\endcsname\relax
\typeout{** WARNING: IEEEtran.bst: No hyphenation pattern has been}%
\typeout{** loaded for the language `#1'. Using the pattern for}%
\typeout{** the default language instead.}%
\else
\language=\csname l@#1\endcsname
\fi
#2}}
\providecommand{\BIBdecl}{\relax}
\BIBdecl

\bibitem{grover_fast_1996}
L.~K. Grover, ``A fast quantum mechanical algorithm for database search,'' in
  \emph{Proceedings of the twenty-eighth annual {ACM} symposium on {Theory} of
  {Computing}}, ser. {STOC} '96.\hskip 1em plus 0.5em minus 0.4em\relax New
  York, NY, USA: Association for Computing Machinery, Jul. 1996, pp. 212--219.

\bibitem{shor_polynomial-time_1999}
P.~W. Shor, ``Polynomial-{Time} {Algorithms} for {Prime} {Factorization} and
  {Discrete} {Logarithms} on a {Quantum} {Computer},'' \emph{SIAM Review},
  vol.~41, no.~2, pp. 303--332, Jan. 1999, publisher: Society for Industrial
  and Applied Mathematics.

\bibitem{shor_why_2003}
------, ``Why haven't more quantum algorithms been found?'' \emph{Journal of
  the ACM}, vol.~50, no.~1, pp. 87--90, Jan. 2003.

\bibitem{aaronson_how_2022}
\BIBentryALTinterwordspacing
S.~Aaronson, ``\BIBforeignlanguage{en}{How {Much} {Structure} {Is} {Needed} for
  {Huge} {Quantum} {Speedups}?}'' arXiv, Tech. Rep. arXiv:2209.06930, Sep.
  2022. [Online]. Available: \url{http://arxiv.org/abs/2209.06930}
\BIBentrySTDinterwordspacing

\bibitem{cobb_towards_2022}
A.~Cobb, J.-G. Schneider, and K.~Lee, ``Towards {Higher}-{Level} {Abstractions}
  for {Quantum} {Computing},'' in \emph{Proceedings of the 2022 {Australasian}
  {Computer} {Science} {Week}}, ser. {ACSW} '22.\hskip 1em plus 0.5em minus
  0.4em\relax New York, NY, USA: Association for Computing Machinery, Mar.
  2022, pp. 115--124.

\bibitem{di_matteo_abstraction_2024}
O.~Di~Matteo, S.~Núñez-Corrales, M.~Stechly, S.~P. Reinhardt, and T.~Mattson,
  ``An {Abstraction} {Hierarchy} {Toward} {Productive} {Quantum}
  {Programming},'' in \emph{2024 {IEEE} {International} {Conference} on
  {Quantum} {Computing} and {Engineering} ({QCE})}, vol.~01, Sep. 2024, pp.
  979--989.

\bibitem{furntratt_towards_2024}
H.~Fürntratt, P.~Schnabl, F.~Krebs, R.~Unterberger, and H.~Zeiner,
  ``\BIBforeignlanguage{en}{Towards {Higher} {Abstraction} {Levels}
  in {Quantum} {Computing}},'' in
  \emph{\BIBforeignlanguage{en}{Service-{Oriented} {Computing} – {ICSOC} 2023
  {Workshops}}}, F.~Monti, P.~Plebani, N.~Moha, H.-y. Paik, J.~Barzen,
  G.~Ramachandran, D.~Bianchini, D.~A. Tamburri, and M.~Mecella, Eds.\hskip 1em
  plus 0.5em minus 0.4em\relax Singapore: Springer Nature, 2024, pp. 162--173.

\bibitem{meyer_todays_2022}
J.~C. Meyer, G.~Passante, S.~J. Pollock, and B.~R. Wilcox, ``Today's
  interdisciplinary quantum information classroom: {Themes} from a survey of
  quantum information science instructors,'' \emph{Physical Review Physics
  Education Research}, vol.~18, no.~1, p. 010150, Jun. 2022.

\bibitem{johansson_shut_2018}
A.~Johansson, S.~Andersson, M.~Salminen-Karlsson, and M.~Elmgren,
  ``\BIBforeignlanguage{en}{“{Shut} up and calculate”: the available
  discursive positions in quantum physics courses},''
  \emph{\BIBforeignlanguage{en}{Cultural Studies of Science Education}},
  vol.~13, no.~1, pp. 205--226, Mar. 2018.

\bibitem{singh_review_2015}
C.~Singh and E.~Marshman, ``Review of student difficulties in upper-level
  quantum mechanics,'' \emph{Physical Review Special Topics - Physics Education
  Research}, vol.~11, no.~2, p. 020117, Sep. 2015, publisher: American Physical
  Society.

\bibitem{grover_quantum_1997}
L.~K. Grover, ``Quantum {Mechanics} {Helps} in {Searching} for a {Needle} in a
  {Haystack},'' \emph{Physical Review Letters}, vol.~79, no.~2, pp. 325--328,
  Jul. 1997.

\bibitem{nielsen_quantum_2010}
M.~A. Nielsen and I.~L. Chuang, \emph{\BIBforeignlanguage{English}{Quantum
  {Computation} and {Quantum} {Information}: 10th {Anniversary} {Edition}}},
  1st~ed.\hskip 1em plus 0.5em minus 0.4em\relax Cambridge ; New York:
  Cambridge University Press, Dec. 2010.

\bibitem{rieffel_quantum_2014}
E.~G. Rieffel and W.~H. Polak, \emph{\BIBforeignlanguage{English}{Quantum
  {Computing}: {A} {Gentle} {Introduction}}}.\hskip 1em plus 0.5em minus
  0.4em\relax Cambridge, Massachusetts London, England: The MIT Press, Aug.
  2014.

\bibitem{mermin_quantum_2007}
N.~D. Mermin, \emph{\BIBforeignlanguage{en}{Quantum {Computer} {Science}: {An}
  {Introduction}}}.\hskip 1em plus 0.5em minus 0.4em\relax Cambridge University
  Press, Aug. 2007.

\bibitem{green_quipper_2013}
A.~S. Green, P.~L. Lumsdaine, N.~J. Ross, P.~Selinger, and B.~Valiron,
  ``Quipper: a scalable quantum programming language,'' in \emph{Proceedings of
  the 34th {ACM} {SIGPLAN} {Conference} on {Programming} {Language} {Design}
  and {Implementation}}, ser. {PLDI} '13.\hskip 1em plus 0.5em minus
  0.4em\relax New York, NY, USA: Association for Computing Machinery, Jun.
  2013, pp. 333--342.

\bibitem{abhari_scaffold_2012}
A.~J. Abhari, A.~Faruque, M.~J. Dousti, L.~Svec, O.~Catu, A.~Chakrabati, C.-F.
  Chiang, S.~Vanderwilt, J.~Black, F.~Chong, M.~Martonosi, M.~Suchara,
  K.~Brown, M.~Pedram, and T.~Brun, ``\BIBforeignlanguage{en}{Scaffold:
  {Quantum} {Programming} {Language}},'' Princeton University Department of
  Computer Science, Tech. Rep., Jul. 2012.

\bibitem{li_verified_2022}
L.~Li, F.~Voichick, K.~Hietala, Y.~Peng, X.~Wu, and M.~Hicks, ``Verified
  compilation of {Quantum} oracles,'' \emph{Proceedings of the ACM on
  Programming Languages}, vol.~6, no. OOPSLA2, pp. 146:589--146:615, Oct. 2022.

\bibitem{seidel_qrisp_2024}
\BIBentryALTinterwordspacing
R.~Seidel, S.~Bock, R.~Zander, M.~Petrič, N.~Steinmann, N.~Tcholtchev, and
  M.~Hauswirth, ``Qrisp: {A} {Framework} for {Compilable} {High}-{Level}
  {Programming} of {Gate}-{Based} {Quantum} {Computers},'' Jun. 2024. [Online].
  Available: \url{http://arxiv.org/abs/2406.14792}
\BIBentrySTDinterwordspacing

\bibitem{selinger_lambda_2006}
P.~Selinger and B.~Valiron, ``\BIBforeignlanguage{en}{A lambda calculus for
  quantum computation with classical control},''
  \emph{\BIBforeignlanguage{en}{Mathematical Structures in Computer Science}},
  vol.~16, no.~3, pp. 527--552, Jun. 2006.

\bibitem{paykin_qwire_2017}
J.~Paykin, R.~Rand, and S.~Zdancewic, ``{QWIRE}: a core language for quantum
  circuits,'' \emph{ACM SIGPLAN Notices}, vol.~52, no.~1, pp. 846--858, Jan.
  2017.

\bibitem{yuan_twist_2022}
C.~Yuan, C.~McNally, and M.~Carbin, ``Twist: sound reasoning for purity and
  entanglement in {Quantum} programs,'' \emph{Proceedings of the ACM on
  Programming Languages}, vol.~6, no. POPL, pp. 30:1--30:32, Jan. 2022.

\bibitem{axler_linear_2023}
S.~Axler, \emph{\BIBforeignlanguage{English}{Linear {Algebra} {Done} {Right}}},
  4th~ed.\hskip 1em plus 0.5em minus 0.4em\relax Cham, Switzerland: Springer,
  Dec. 2023.

\bibitem{qcl}
B.~\"Omer, ``Quantum programming in {QCL},'' Master's thesis, Technical
  University of Vienna, 2000.

\bibitem{amy_sized_2019}
M.~Amy, ``\BIBforeignlanguage{en}{Sized {Types} for {Low}-{Level} {Quantum}
  {Metaprogramming}},'' in \emph{\BIBforeignlanguage{en}{Reversible
  {Computation}}}, M.~K. Thomsen and M.~Soeken, Eds.\hskip 1em plus 0.5em minus
  0.4em\relax Cham: Springer International Publishing, 2019, pp. 87--107.

\bibitem{cleve_quantum_1998}
R.~Cleve, A.~Ekert, C.~Macchiavello, and M.~Mosca, ``Quantum algorithms
  revisited,'' \emph{Proceedings of the Royal Society of London. Series A:
  Mathematical, Physical and Engineering Sciences}, vol. 454, no. 1969, pp.
  339--354, Jan. 1998.

\bibitem{de_muelenaere_qgat_2024}
\BIBentryALTinterwordspacing
U.~De~Muelenaere, ``{QGAT}: {A} {Generate}-and-{Test} {Paradigm} for {Quantum}
  {Circuits},'' in \emph{Fourth {International} {Workshop} on {Programming}
  {Languages} for {Quantum} {Computing} ({PLanQC} 2024)}, London, UK, Jan.
  2024. [Online]. Available:
  \url{https://popl24.sigplan.org/details/planqc-2024-papers/14/QGAT-A-Generate-and-Test-Paradigm-for-Quantum-Circuits}
\BIBentrySTDinterwordspacing

\bibitem{bichsel_silq_2020}
B.~Bichsel, M.~Baader, T.~Gehr, and M.~Vechev, ``Silq: a high-level quantum
  language with safe uncomputation and intuitive semantics,'' in
  \emph{Proceedings of the 41st {ACM} {SIGPLAN} {Conference} on {Programming}
  {Language} {Design} and {Implementation}}, ser. {PLDI} 2020.\hskip 1em plus
  0.5em minus 0.4em\relax New York, NY, USA: Association for Computing
  Machinery, Jun. 2020, pp. 286--300.

\bibitem{paradis_unqomp_2021}
A.~Paradis, B.~Bichsel, S.~Steffen, and M.~Vechev, ``Unqomp: synthesizing
  uncomputation in {Quantum} circuits,'' in \emph{Proceedings of the 42nd {ACM}
  {SIGPLAN} {International} {Conference} on {Programming} {Language} {Design}
  and {Implementation}}, ser. {PLDI} 2021.\hskip 1em plus 0.5em minus
  0.4em\relax New York, NY, USA: Association for Computing Machinery, Jun.
  2021, pp. 222--236.

\bibitem{harris2020array}
\BIBentryALTinterwordspacing
C.~R. Harris, K.~J. Millman, S.~J. van~der Walt, R.~Gommers \emph{et~al.},
  ``Array programming with {NumPy},'' \emph{Nature}, vol. 585, no. 7825, pp.
  357--362, Sep. 2020. [Online]. Available:
  \url{https://doi.org/10.1038/s41586-020-2649-2}
\BIBentrySTDinterwordspacing

\bibitem{fallek_transport_2016}
S.~D. Fallek, C.~D. Herold, B.~J. McMahon, K.~M. Maller, K.~R. Brown, and J.~M.
  Amini, ``\BIBforeignlanguage{en}{Transport implementation of the
  {Bernstein}–{Vazirani} algorithm with ion qubits},''
  \emph{\BIBforeignlanguage{en}{New Journal of Physics}}, vol.~18, no.~8, p.
  083030, Aug. 2016.

\bibitem{bernstein_quantum_1997}
E.~Bernstein and U.~Vazirani, ``Quantum {Complexity} {Theory},'' \emph{SIAM
  Journal on Computing}, vol.~26, no.~5, pp. 1411--1473, Oct. 1997.

\bibitem{asfaw_building_2022}
A.~Asfaw, A.~Blais, K.~R. Brown, J.~Candelaria \emph{et~al.}, ``Building a
  {Quantum} {Engineering} {Undergraduate} {Program},'' \emph{IEEE Transactions
  on Education}, vol.~65, no.~2, pp. 220--242, May 2022.

\bibitem{wootton_teaching_2021}
J.~R. Wootton, F.~Harkins, N.~T. Bronn, A.~C. Vazquez, A.~Phan, and A.~T.
  Asfaw, ``Teaching quantum computing with an interactive textbook,'' in
  \emph{2021 {IEEE} {International} {Conference} on {Quantum} {Computing} and
  {Engineering} ({QCE})}, Oct. 2021, pp. 385--391.

\bibitem{bennett_timespace_1989}
C.~H. Bennett, ``Time/{Space} {Trade}-{Offs} for {Reversible} {Computation},''
  \emph{SIAM Journal on Computing}, vol.~18, no.~4, pp. 766--776, Aug. 1989.

\bibitem{bracewell_fourier_1999}
R.~Bracewell, \emph{\BIBforeignlanguage{English}{The {Fourier} {Transform} \&
  {Its} {Applications}}}, 3rd~ed.\hskip 1em plus 0.5em minus 0.4em\relax
  Boston: McGraw-Hill Science/Engineering/Math, Jun. 1999.

\bibitem{shor_algorithms_1994}
P.~Shor, ``Algorithms for quantum computation: discrete logarithms and
  factoring,'' in \emph{Proceedings 35th {Annual} {Symposium} on {Foundations}
  of {Computer} {Science}}, Nov. 1994, pp. 124--134.

\bibitem{smith_scientist_1997}
S.~W. Smith, \emph{\BIBforeignlanguage{English}{The {Scientist} \& {Engineer}'s
  {Guide} to {Digital} {Signal} {Processing}}}, 1st~ed.\hskip 1em plus 0.5em
  minus 0.4em\relax San Diego, Calif: California Technical Pub, Jan. 1997.

\bibitem{pyfrac}
\BIBentryALTinterwordspacing
{Python Software Foundation}, ``fractions --- rational numbers,'' 2024.
  [Online]. Available: \url{https://docs.python.org/3/library/fractions.html}
\BIBentrySTDinterwordspacing

\bibitem{jacob_shor_conv}
\BIBentryALTinterwordspacing
J.~Watkins, ``Continued fractions with {Shor's} algorithm: which convergent?''
  Quantum Computing Stack Exchange, Apr. 2023. [Online]. Available:
  \url{https://quantumcomputing.stackexchange.com/a/32182/}
\BIBentrySTDinterwordspacing

\bibitem{ross_algebraic_2015}
N.~J. Ross, ``\BIBforeignlanguage{en}{Algebraic and {Logical} {Methods} in
  {Quantum} {Computation}},'' Ph.D. dissertation, Dalhousie University,
  Halifax, Nova Scotia, Canada, Aug. 2015.

\bibitem{litteken_updated_2020}
A.~Litteken, Y.-C. Fan, D.~Singh, M.~Martonosi, and F.~T. Chong,
  ``\BIBforeignlanguage{en}{An updated {LLVM}-based quantum research compiler
  with further {OpenQASM} support},'' \emph{\BIBforeignlanguage{en}{Quantum
  Science and Technology}}, vol.~5, no.~3, p. 034013, May 2020.

\bibitem{javadi-abhari_quantum_2024}
\BIBentryALTinterwordspacing
A.~Javadi-Abhari, M.~Treinish, K.~Krsulich, C.~J. Wood, J.~Lishman, J.~Gacon,
  S.~Martiel, P.~D. Nation, L.~S. Bishop, A.~W. Cross, B.~R. Johnson, and J.~M.
  Gambetta, ``Quantum computing with {Qiskit},'' Jun. 2024. [Online].
  Available: \url{http://arxiv.org/abs/2405.08810}
\BIBentrySTDinterwordspacing

\bibitem{steiger_projectq_2018}
D.~S. Steiger, T.~Häner, and M.~Troyer,
  ``\BIBforeignlanguage{en-GB}{{ProjectQ}: an open source software framework
  for quantum computing},'' \emph{\BIBforeignlanguage{en-GB}{Quantum}}, vol.~2,
  p.~49, Jan. 2018.

\bibitem{cirq}
\BIBentryALTinterwordspacing
{Cirq Developers}, \emph{Cirq}.\hskip 1em plus 0.5em minus 0.4em\relax Zenodo,
  May 2024. [Online]. Available:
  \url{https://zenodo.org/doi/10.5281/zenodo.4062499}
\BIBentrySTDinterwordspacing

\bibitem{wecker_liqui_2014}
\BIBentryALTinterwordspacing
D.~Wecker and K.~M. Svore, ``{LIQUi}{\textbar}{\textgreater}: {A} {Software}
  {Design} {Architecture} and {Domain}-{Specific} {Language} for {Quantum}
  {Computing},'' Feb. 2014. [Online]. Available:
  \url{http://arxiv.org/abs/1402.4467}
\BIBentrySTDinterwordspacing

\bibitem{svore_q_2018}
K.~Svore, A.~Geller, M.~Troyer, J.~Azariah, C.~Granade, B.~Heim,
  V.~Kliuchnikov, M.~Mykhailova, A.~Paz, and M.~Roetteler, ``Q\#: {Enabling}
  {Scalable} {Quantum} {Computing} and {Development} with a {High}-level
  {DSL},'' in \emph{Proceedings of the {Real} {World} {Domain} {Specific}
  {Languages} {Workshop} 2018}, ser. {RWDSL2018}.\hskip 1em plus 0.5em minus
  0.4em\relax New York, NY, USA: Association for Computing Machinery, Feb.
  2018, pp. 1--10.

\bibitem{lubinski_advancing_2022}
T.~Lubinski, C.~Granade, A.~Anderson, A.~Geller, M.~Roetteler, A.~Petrenko, and
  B.~Heim, ``Advancing hybrid quantum–classical computation with real-time
  execution,'' \emph{Frontiers in Physics}, vol.~10, p. 940293, 2022.

\bibitem{diaz-caro_realizability_2019}
A.~Díaz-Caro, M.~Guillermo, A.~Miquel, and B.~Valiron, ``Realizability in the
  {Unitary} {Sphere},'' in \emph{2019 34th {Annual} {ACM}/{IEEE} {Symposium} on
  {Logic} in {Computer} {Science} ({LICS})}, Jun. 2019, pp. 1--13.

\bibitem{bergholm_pennylane_2022}
\BIBentryALTinterwordspacing
V.~Bergholm, J.~Izaac, M.~Schuld, C.~Gogolin \emph{et~al.}, ``{PennyLane}:
  {Automatic} differentiation of hybrid quantum-classical computations,'' Jul.
  2022. [Online]. Available: \url{http://arxiv.org/abs/1811.04968}
\BIBentrySTDinterwordspacing

\bibitem{ittah_catalyst_2024}
D.~Ittah, A.~Asadi, E.~O. Lopez, S.~Mironov, S.~Banning, R.~Moyard, M.~J. Peng,
  and J.~Izaac, ``\BIBforeignlanguage{en}{Catalyst: a {Python} {JIT} compiler
  for auto-differentiable hybrid quantum programs},''
  \emph{\BIBforeignlanguage{en}{Journal of Open Source Software}}, vol.~9,
  no.~99, p. 6720, Jul. 2024.

\bibitem{mintz_qcor_2020}
T.~M. Mintz, A.~J. McCaskey, E.~F. Dumitrescu, S.~V. Moore, S.~Powers, and
  P.~Lougovski, ``{QCOR}: {A} {Language} {Extension} {Specification} for the
  {Heterogeneous} {Quantum}-{Classical} {Model} of {Computation},'' \emph{ACM
  Journal on Emerging Technologies in Computing Systems}, vol.~16, no.~2, pp.
  22:1--22:17, Mar. 2020.

\bibitem{mccaskey_extending_2021}
A.~Mccaskey, T.~Nguyen, A.~Santana, D.~Claudino, T.~Kharazi, and H.~Finkel,
  ``Extending {C}++ for {Heterogeneous} {Quantum}-{Classical} {Computing},''
  \emph{ACM Transactions on Quantum Computing}, vol.~2, no.~2, pp. 6:1--6:36,
  Jul. 2021.

\bibitem{cross_openqasm_2022}
A.~Cross, A.~Javadi-Abhari, T.~Alexander, N.~De~Beaudrap, L.~S. Bishop,
  S.~Heidel, C.~A. Ryan, P.~Sivarajah, J.~Smolin, J.~M. Gambetta, and B.~R.
  Johnson, ``{OpenQASM} 3: {A} {Broader} and {Deeper} {Quantum} {Assembly}
  {Language},'' \emph{ACM Transactions on Quantum Computing}, vol.~3, no.~3,
  pp. 12:1--12:50, Sep. 2022.

\bibitem{voichick_qunity_2023}
\BIBentryALTinterwordspacing
F.~Voichick, L.~Li, R.~Rand, and M.~Hicks, ``Qunity: {A} {Unified} {Language}
  for {Quantum} and {Classical} {Computing} ({Extended} {Version}),''
  \emph{Proceedings of the ACM on Programming Languages}, vol.~7, no. POPL, pp.
  921--951, Jan. 2023, arXiv:2204.12384 [quant-ph]. [Online]. Available:
  \url{http://arxiv.org/abs/2204.12384}
\BIBentrySTDinterwordspacing

\bibitem{duncan_introducing_2024}
\BIBentryALTinterwordspacing
R.~Duncan, M.~Koch, A.~Lawrence, C.~McBride, and C.~Roy,
  ``\BIBforeignlanguage{en}{Introducing {BRAT}},'' in
  \emph{\BIBforeignlanguage{en}{Fourth {International} {Workshop} on
  {Programming} {Languages} for {Quantum} {Computing} ({PLanQC} 2024)}},
  London, UK, Jan. 2024. [Online]. Available:
  \url{https://popl24.sigplan.org/details/planqc-2024-papers/9/Introducing-BRAT}
\BIBentrySTDinterwordspacing

\bibitem{koch_guppy_2024}
\BIBentryALTinterwordspacing
M.~Koch, A.~Lawrence, K.~Singhal, S.~Sivarajah, and R.~Duncan, ``{GUPPY}:
  {Pythonic} {Quantum}-{Classical} {Programming},'' in \emph{Fourth
  {International} {Workshop} on {Programming} {Languages} for {Quantum}
  {Computing} ({PLanQC} 2024)}, London, UK, Jan. 2024. [Online]. Available:
  \url{https://popl24.sigplan.org/details/planqc-2024-papers/8/GUPPY-Pythonic-Quantum-Classical-Programming}
\BIBentrySTDinterwordspacing

\bibitem{vax_qmod_2025}
\BIBentryALTinterwordspacing
M.~Vax, P.~Emanuel, E.~Cornfeld, I.~Reichental, O.~Opher, O.~Roth, T.~Michaeli,
  L.~Preminger, L.~Gazit, A.~Naveh, and Y.~Naveh, ``Qmod: {Expressive}
  {High}-{Level} {Quantum} {Modeling},'' Feb. 2025. [Online]. Available:
  \url{http://arxiv.org/abs/2502.19368}
\BIBentrySTDinterwordspacing

\bibitem{altenkirch_functional_2005}
T.~Altenkirch and J.~Grattage, ``A functional quantum programming language,''
  in \emph{20th {Annual} {IEEE} {Symposium} on {Logic} in {Computer} {Science}
  ({LICS}' 05)}, Jun. 2005, pp. 249--258, iSSN: 1043-6871.

\bibitem{carette_quantum_2023}
\BIBentryALTinterwordspacing
J.~Carette, C.~Heunen, R.~Kaarsgaard, and A.~Sabry, ``The {Quantum} {Effect}:
  {A} {Recipe} for {QuantumPi},'' May 2023, arXiv:2302.01885 [quant-ph].
  [Online]. Available: \url{http://arxiv.org/abs/2302.01885}
\BIBentrySTDinterwordspacing

\bibitem{inoue_quantum_2024}
\BIBentryALTinterwordspacing
J.~Inoue, ``\BIBforeignlanguage{en}{Quantum {Programming} {Without} the
  {Quantum} {Physics}},'' Aug. 2024. [Online]. Available:
  \url{http://arxiv.org/abs/2408.16234}
\BIBentrySTDinterwordspacing

\bibitem{rudolph_q_2017}
T.~Rudolph, \emph{\BIBforeignlanguage{English}{Q is for {Quantum}}}.\hskip 1em
  plus 0.5em minus 0.4em\relax Wroclaw: Terence Rudolph, Jul. 2017.

\bibitem{nunez-corrales_quapl_2023}
S.~Núñez-Corrales, M.~Frenkel, and B.~Abreu, ``{quAPL}: {Modeling} {Quantum}
  {Computation} in an {Array} {Programming} {Language},'' in \emph{2023 {IEEE}
  {International} {Conference} on {Quantum} {Computing} and {Engineering}
  ({QCE})}, vol.~01, Sep. 2023, pp. 1001--1012.

\bibitem{binkowski_cq_2025}
L.~Binkowski, ``\BIBforeignlanguage{en}{{CQ}: {A} high-level imperative
  classical-quantum programming language},'' Jan. 2025, {Bachelor's} thesis,
  Institut für Theoretische Informatik Leibniz Universität Hannover.

\bibitem{harrigan_expressing_2024}
\BIBentryALTinterwordspacing
M.~P. Harrigan, T.~Khattar, C.~Yuan, A.~Peduri, N.~Yosri, F.~D. Malone,
  R.~Babbush, and N.~C. Rubin, ``Expressing and {Analyzing} {Quantum}
  {Algorithms} with {Qualtran},'' Sep. 2024, arXiv:2409.04643 [quant-ph].
  [Online]. Available: \url{http://arxiv.org/abs/2409.04643}
\BIBentrySTDinterwordspacing

\bibitem{yuan_tower_2022}
C.~Yuan and M.~Carbin, ``Tower: data structures in {Quantum} superposition,''
  \emph{Proceedings of the ACM on Programming Languages}, vol.~6, no. OOPSLA2,
  pp. 134:259--134:288, Oct. 2022.

\bibitem{pinto_neko_2023}
\BIBentryALTinterwordspacing
E.~Pinto, ``\BIBforeignlanguage{en}{Neko: {A} quantum map-filter-reduce
  programming language},'' in \emph{\BIBforeignlanguage{en}{Student {Research}
  {Competition} ({SRC}), {POPL}'23}}, Boston, MA, Jan. 2023. [Online].
  Available: \url{https://www.eltonpinto.me/assets/work/neko-popl23src.pdf}
\BIBentrySTDinterwordspacing

\bibitem{khetawat_implementing_2019}
H.~Khetawat, A.~Atrey, G.~Li, F.~Mueller, and S.~Pakin,
  ``\BIBforeignlanguage{en}{Implementing {NChooseK} on {IBM} {Q} {Quantum}
  {Computer} {Systems}},'' in \emph{\BIBforeignlanguage{en}{Reversible
  {Computation}}}, M.~K. Thomsen and M.~Soeken, Eds.\hskip 1em plus 0.5em minus
  0.4em\relax Cham: Springer International Publishing, 2019, pp. 209--223.

\bibitem{wilson_mapping_2021}
E.~Wilson, F.~Mueller, and S.~Pakin, ``Mapping {Constraint} {Problems} onto
  {Quantum} {Gate} and {Annealing} {Devices},'' in \emph{2021 {IEEE}/{ACM}
  {Second} {International} {Workshop} on {Quantum} {Computing} {Software}
  ({QCS})}, Nov. 2021, pp. 110--117.

\bibitem{wilson_combining_2022}
------, ``Combining {Hard} and {Soft} {Constraints} in {Quantum}
  {Constraint}-{Satisfaction} {Systems},'' in \emph{{SC22}: {International}
  {Conference} for {High} {Performance} {Computing}, {Networking}, {Storage}
  and {Analysis}}, Nov. 2022, pp. 1--14, iSSN: 2167-4337.

\bibitem{aleph}
\BIBentryALTinterwordspacing
A.~Paz, ``Aleph,'' 2023. [Online]. Available:
  \url{https://github.com/anpaz/aleph}
\BIBentrySTDinterwordspacing

\bibitem{faro_qutes_2025}
\BIBentryALTinterwordspacing
S.~Faro, F.~P. Marino, and G.~Messina, ``Qutes: {A} {High}-{Level} {Quantum}
  {Programming} {Language} for {Simplified} {Quantum} {Computing},'' Mar. 2025.
  [Online]. Available: \url{http://arxiv.org/abs/2503.13084}
\BIBentrySTDinterwordspacing

\bibitem{economou_teaching_2020}
\BIBentryALTinterwordspacing
S.~E. Economou, T.~Rudolph, and E.~Barnes, ``Teaching quantum information
  science to high-school and early undergraduate students,'' Aug. 2020,
  arXiv:2005.07874. [Online]. Available: \url{http://arxiv.org/abs/2005.07874}
\BIBentrySTDinterwordspacing

\bibitem{manin_radio}
Y.~Manin, ``Computable and uncomputable,'' Soviet Radio Publishing House, 1980.

\bibitem{feynman_simulating_1982}
R.~P. Feynman, ``\BIBforeignlanguage{en}{Simulating physics with computers},''
  \emph{\BIBforeignlanguage{en}{International Journal of Theoretical Physics}},
  vol.~21, no.~6, pp. 467--488, Jun. 1982.

\bibitem{dumitrescu_integrating_2023}
\BIBentryALTinterwordspacing
E.~Dumitrescu, ``\BIBforeignlanguage{en}{Integrating {Across} {Application},
  {Model}, {Algorithm}, {Compilation}, and {Error} {Correction} {Chasms} {With}
  {Quantum} {Type} {Theory}},'' Apr. 2023. [Online]. Available:
  \url{http://arxiv.org/abs/2305.00144}
\BIBentrySTDinterwordspacing

\bibitem{powell_wrangling_2019}
W.~Powell, J.~Riedy, J.~S. Young, and T.~M. Conte, ``Wrangling {Rogues}: {A}
  {Case} {Study} on {Managing} {Experimental} {Post}-{Moore} {Architectures},''
  in \emph{Practice and {Experience} in {Advanced} {Research} {Computing} 2019:
  {Rise} of the {Machines} (learning)}, ser. {PEARC} '19.\hskip 1em plus 0.5em
  minus 0.4em\relax New York, NY, USA: Association for Computing Machinery,
  Jul. 2019, pp. 1--8.

\bibitem{young_experimental_2019}
J.~S. Young, J.~Riedy, T.~M. Conte, V.~Sarkar, P.~Chatarasi, and S.~Srikanth,
  ``Experimental {Insights} from the {Rogues} {Gallery},'' in \emph{2019 {IEEE}
  {International} {Conference} on {Rebooting} {Computing} ({ICRC})}, Nov. 2019,
  pp. 1--8.

\bibitem{singhal_q_2022}
K.~Singhal, K.~Hietala, S.~Marshall, and R.~Rand, ``\BIBforeignlanguage{en}{Q\#
  as a {Quantum} {Algorithmic} {Language}},'' arXiv, Tech. Rep.
  arXiv:2206.03532, Jun. 2022.

\bibitem{pierce_types_2002}
B.~C. Pierce, \emph{\BIBforeignlanguage{English}{Types and {Programming}
  {Languages}}}, 1st~ed.\hskip 1em plus 0.5em minus 0.4em\relax Cambridge,
  Mass: The MIT Press, Feb. 2002.

\bibitem{wadler_is_1991}
P.~Wadler, ``Is there a use for linear logic?'' \emph{ACM SIGPLAN Notices},
  vol.~26, no.~9, pp. 255--273, May 1991.

\bibitem{keller_applied_2017}
M.~T. Keller and W.~T. Trotter, \emph{\BIBforeignlanguage{English}{Applied
  {Combinatorics}}}.\hskip 1em plus 0.5em minus 0.4em\relax Sioux City, Iowa:
  CreateSpace Independent Publishing Platform, 2017.

\end{thebibliography}

\end{document}